\documentclass[11pt, a4paper, logo, internal]{deepmind}

\pdftrailerid{redacted}

\usepackage{dsfont}
\usepackage{dm-colors}
\usepackage{multirow}
\usepackage{changes}
\usepackage{soul}
\usepackage{svg}
\usepackage{adjustbox}
\svgsetup{inkscapelatex=false}
\usepackage{algpseudocode}

\usepackage{tocloft}
\usepackage{etoc}
\usepackage{setspace}
\usepackage{multicol}
\usepackage{mathtools}
\usepackage{dsfont}
\usepackage[dvipsnames]{xcolor}
\usepackage{todonotes}
\usepackage{booktabs}
\usepackage{xfrac}
\usepackage{bbm}

\usepackage{algorithm}
\usepackage{algorithmicx}

\usepackage[most]{tcolorbox}
\usepackage{xparse}
\usepackage{lipsum}
\usepackage{changepage}
\usepackage{enumitem}

\usepackage{ulem}

\usepackage{tocloft}
\usepackage[toc]{multitoc}

\setuptodonotes{size=footnotesize}

\newcommand{\squishlist}{
   \begin{list}{$\bullet$}
    { \setlength{\itemsep}{0pt}      \setlength{\parsep}{3pt}
      \setlength{\topsep}{3pt}       \setlength{\partopsep}{0pt}
      \setlength{\leftmargin}{1.5em} \setlength{\labelwidth}{1em}
      \setlength{\labelsep}{0.5em} } }

\newcommand{\squishlisttwo}{
   \begin{list}{$\bullet$}
    { \setlength{\itemsep}{0pt}    \setlength{\parsep}{0pt}
      \setlength{\topsep}{0pt}     \setlength{\partopsep}{0pt}
      \setlength{\leftmargin}{2em} \setlength{\labelwidth}{1.5em}
      \setlength{\labelsep}{0.5em} } }

\newcommand{\squishend}{
    \end{list}  }

\definecolor{deepblue}{rgb}{0,0,0.5}
\definecolor{deepred}{rgb}{0.6,0,0}
\definecolor{deepgreen}{rgb}{0,0.5,0}
\definecolor{ipython_cyan}{RGB}{64, 128, 128}

\newcommand{\CC}{C\nolinebreak\hspace{-.05em}\raisebox{.4ex}{\tiny\bf +}\nolinebreak\hspace{-.10em}\raisebox{.4ex}{\tiny\bf +}}

\newcommand{\OurApproachShort}{AlphaCode}
\newcommand{\OurDatasetShort}{CodeContests}

\newcommand{\HeadlineRanking}{54.3}
\newcommand{\HeadlineElo}{1238}
\newcommand{\HeadlineEloRanking}{28}
\newcommand{\RankingUnlimitedAttempt}{48.8}
\newcommand{\AverageSubmissionsTenAttempt}{2.4}
\newcommand{\AverageSubmissionsUnlimitedAttempt}{28.8}

\newcommand{\secref}[1]{Section \ref{#1}}
\newcommand{\figref}[1]{Figure \ref{#1}}
\newcommand{\tabref}[1]{Table \ref{#1}}
\newcommand{\appendixref}[1]{Appendix \ref{#1}}

\usepackage[authoryear, sort&compress, round]{natbib} 

\graphicspath{{figures/}}

\title{Competition-Level Code Generation with AlphaCode}

\correspondingauthor{yujiali@deepmind.com, davidhchoi@deepmind.com, vinyals@deepmind.com}

\newcommand{\noaffil}{ \hspace{-1ex}}

\author[*]{Yujia Li}
\author[*]{David Choi}
\author[*]{Junyoung Chung}
\author[*]{Nate Kushman}
\author[*]{Julian Schrittwieser}
\author[*]{Rémi Leblond}
\author[*]{Tom Eccles}
\author[*]{James Keeling}
\author[*]{Felix Gimeno}
\author[*]{Agustin Dal Lago}
\author[*]{Thomas Hubert}
\author[*]{Peter Choy}
\author[*]{Cyprien de Masson d'Autume}
\author[\noaffil{}]{Igor Babuschkin}
\author[\noaffil{}]{Xinyun Chen}
\author[\noaffil{}]{Po-Sen Huang}
\author[\noaffil{}]{Johannes Welbl}
\author[\noaffil{}]{Sven Gowal}
\author[\noaffil{}]{Alexey Cherepanov}
\author[\noaffil{}]{James Molloy}
\author[\noaffil{}]{Daniel J. Mankowitz}
\author[\noaffil{}]{Esme Sutherland Robson}
\author[\noaffil{}]{Pushmeet Kohli}
\author[\noaffil{}]{Nando de Freitas}
\author[\noaffil{}]{Koray Kavukcuoglu}
\author[\noaffil{}]{Oriol Vinyals} 

\affil[*]{Joint first authors}

\AtBeginDocument{\addtocontents{toc}{\protect\thispagestyle{firststyle}}} 

\begin{abstract}
\vspace{-0.2cm}

Programming is a powerful and ubiquitous problem-solving tool. Developing systems that can assist programmers or even generate programs independently could make programming more productive and accessible, yet so far incorporating innovations in AI has proven challenging.
Recent large-scale language models have demonstrated an impressive ability to generate code, and are now able to complete simple programming tasks. However, these models still perform poorly when evaluated on more complex, unseen problems that require problem-solving skills beyond simply translating instructions into code. 
For example, competitive programming problems which require an understanding of algorithms and complex natural language remain extremely challenging.
To address this gap, we introduce \OurApproachShort{}, a system for code generation that can create novel solutions to these problems that require deeper reasoning. In simulated evaluations on recent programming competitions on the Codeforces platform, \OurApproachShort{} achieved on average a ranking of top \HeadlineRanking{}\% in competitions with more than 5,000 participants. 
We found that three key components were critical to achieve good and reliable performance: (1) an extensive and clean competitive programming dataset for training and evaluation, (2) large and efficient-to-sample transformer-based architectures, and (3) large-scale model sampling to explore the search space, followed by filtering based on program behavior to a small set of submissions.

\vspace{-0.3cm}
\end{abstract}

\begin{document}

\maketitle

\titleformat{\part}
 {}
 {}
 {0pt}
 {\color{black}}
\vspace{-1.6cm} 
\part{}
{
\setlength{\columnseprule}{0.5pt}
\begin{multicols}{2}
\etocsettocstyle{}{}
\etocsetnexttocdepth{subsection}
\localtableofcontents
\end{multicols}
}

\section{Introduction}

Computer programming has emerged as a general-purpose problem-solving tool throughout science, industry, and daily life. As part of this growth, there has been continuously increasing demand for tools that can make programmers more productive~\citep{matsakis2014rust}, or make programming and programming education more accessible~\citep{resnick2009scratch}. Developing AI systems that can effectively model and understand code can transform these tools and the way we interact with them. Systems that can generate code are not only useful, but also stepping stones that can lead to greater understanding of AI and how it relates to programming.

Generating code that solves a specified task requires searching in the huge structured space of possible programs, with a very sparse reward signal. Single character edits can completely change program behaviour even if they don't cause crashes, solutions can look dramatically different even for the same problem, and judging if a partial or incorrect program is useful is a difficult challenge. Therefore, most prior work has been limited to either restricted domain-specific programming languages~\citep{gulwani2011automating} or short code snippets~\citep{raychev2014code,bruch2009learning}.

Recent large-scale transformer-based \citep{vaswani2017attention} language models, used to achieve impressive performance generating text \citep{brown2020language}, have successfully generated code that solves simple programming problems in Python \citep{chen2021codex,austin2021program}. A stripped-down version of our model, without the modifications described in \secref{sec:approach}, performs similarly to Codex (\tabref{tab:human-eval}). However,
problems used in the Codex paper and similar work consist of mostly simple task descriptions with short solutions -- far from the full complexity of real-world programming. Generating an entire program in a general-purpose programming language such as \CC{} or Python, starting from a long natural language task description, has remained an open problem.  The difference in difficulty between generating short code snippets and entire programs can be %
analogous to that of imperative versus declarative problem solving. Generating short code snippets typically amounts to translating the task specification directly into code, and sometimes reduces to invoking the correct API calls. In contrast, generating entire programs often relies on understanding the task and figuring out how to accomplish it, which requires deeper algorithmic reasoning.

Competitive programming problems represent a significant step forward in all these aspects.
Solving such problems requires understanding complex natural language descriptions, reasoning about previously unseen problems, mastering a wide range of algorithms and data structures, and precisely implementing solutions that can span hundreds of lines. 
Solutions are evaluated by executing them on an exhaustive suite of unknown tests, checking for correct behaviour on edge cases as well as execution speed.  The fact that the test cases used for evaluation are hidden is an important part of the challenge. %
These complex problems are newly created for each competition, with the understanding that competitors can draw on solutions to previous contests (either implicitly, by remembering old problems, or explicitly, by searching for them). %
Moreover, competitive programming is very popular; events like the International Collegiate Programming Competition \citep{icpc} and the International Olympiad in Informatics \citep{ioi} are widely recognized as some of the most prestigious competitions in computer science, drawing hundreds of thousands of participants from around the world. Using problems that humans find challenging from such battle-tested competitions ensures robustness against shortcuts and provides a meaningful benchmark for many aspects of intelligence.

Early work using program synthesis for competitive programming has shown that large transformer models can achieve low single-digit solve rates \citep{hendrycks2021measuring,chen2021codex}, but could not yet reliably generate solutions for the vast majority of problems.  Furthermore, as we show in \secref{sec:false-positives}, the lack of sufficient test cases in existing competitive programming datasets makes the metrics defined on them prone to high false positive rates (with 30\% or more programs which pass all tests but are not actually correct), and therefore unreliable for measuring research progress. %

In this paper we present \OurApproachShort{}, a code generation system applied to solving competitive programming problems. We use large transformer language models to generate code, pre-training them on selected GitHub code and fine-tuning on our curated set of competitive programming problems. For each unseen problem we generate a large set of program samples, filter them based on execution results on example tests from the problem description, then cluster the remaining samples to obtain a small set of candidates to be submitted for %
evaluation. We describe \OurApproachShort{} in detail in \secref{sec:approach}.

A core part of developing our system was ensuring that submissions are rigorously evaluated and that evaluation problems are truly unseen during training, so difficult problems cannot be solved by copying from the training set. Towards this goal, we release a new training and evaluation competitive programming dataset, \OurDatasetShort{}\footnote{The dataset is located at \url{https://github.com/deepmind/code_contests}.} (\secref{sec:dataset}). This dataset combines data from various sources, splits temporally so all training data predates all evaluation problems, adds additional generated tests to ensure correctness, and evaluates submissions in a setting that mirrors that of competitive programming.  In our evaluation (\secref{sec:false-positives}), \OurDatasetShort{} reduces the false positive rate from 30-60\% in existing datasets to just 4\%.
Our best model solves 34.2\% of held-out competitive programming problems in this dataset, using at most 10 submissions per problem (comparable to humans), as opposed to previously reported solve rates of around 1-5\% on existing datasets (see %
\secref{sec:public-benchmark-results}). %

\begin{figure}[t]
    \centering
    \begin{tabular}{cc}
        \raisebox{0.3cm}{
        \includegraphics[width=0.60\textwidth]{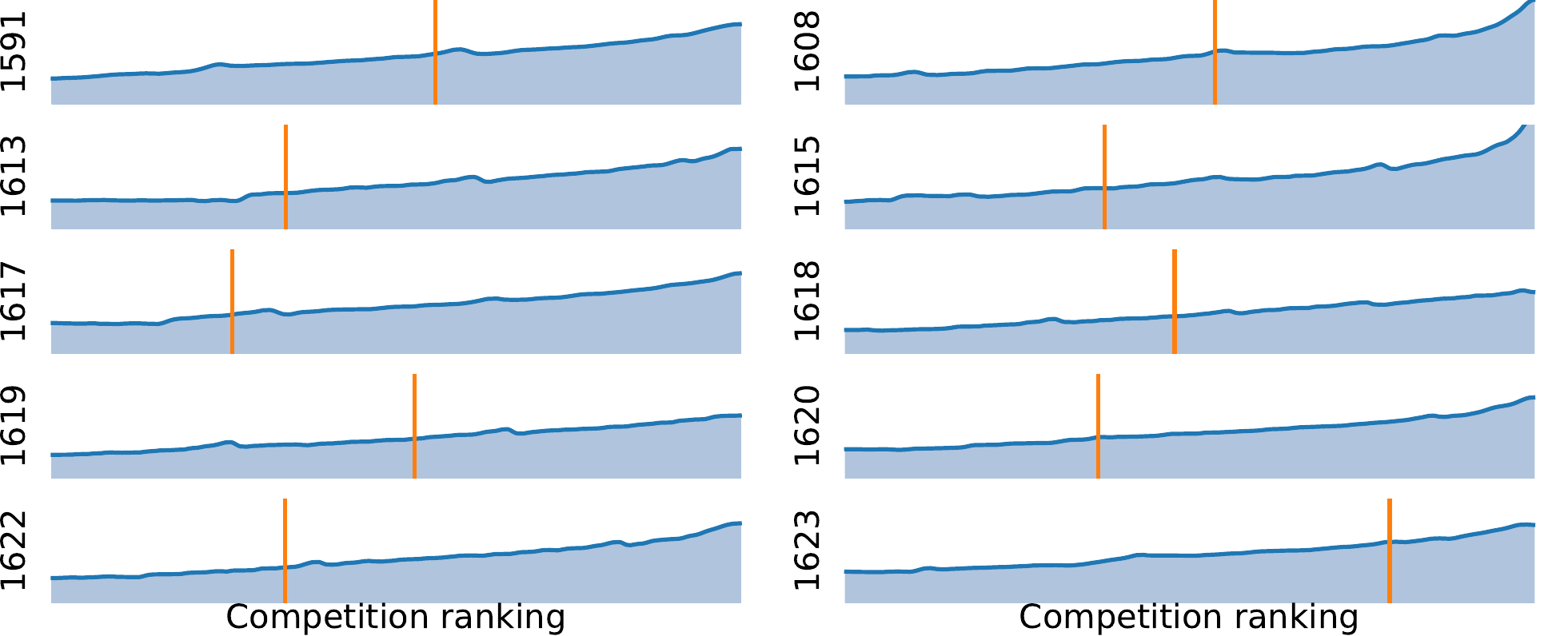}} &
        \includegraphics[width=0.35\textwidth]{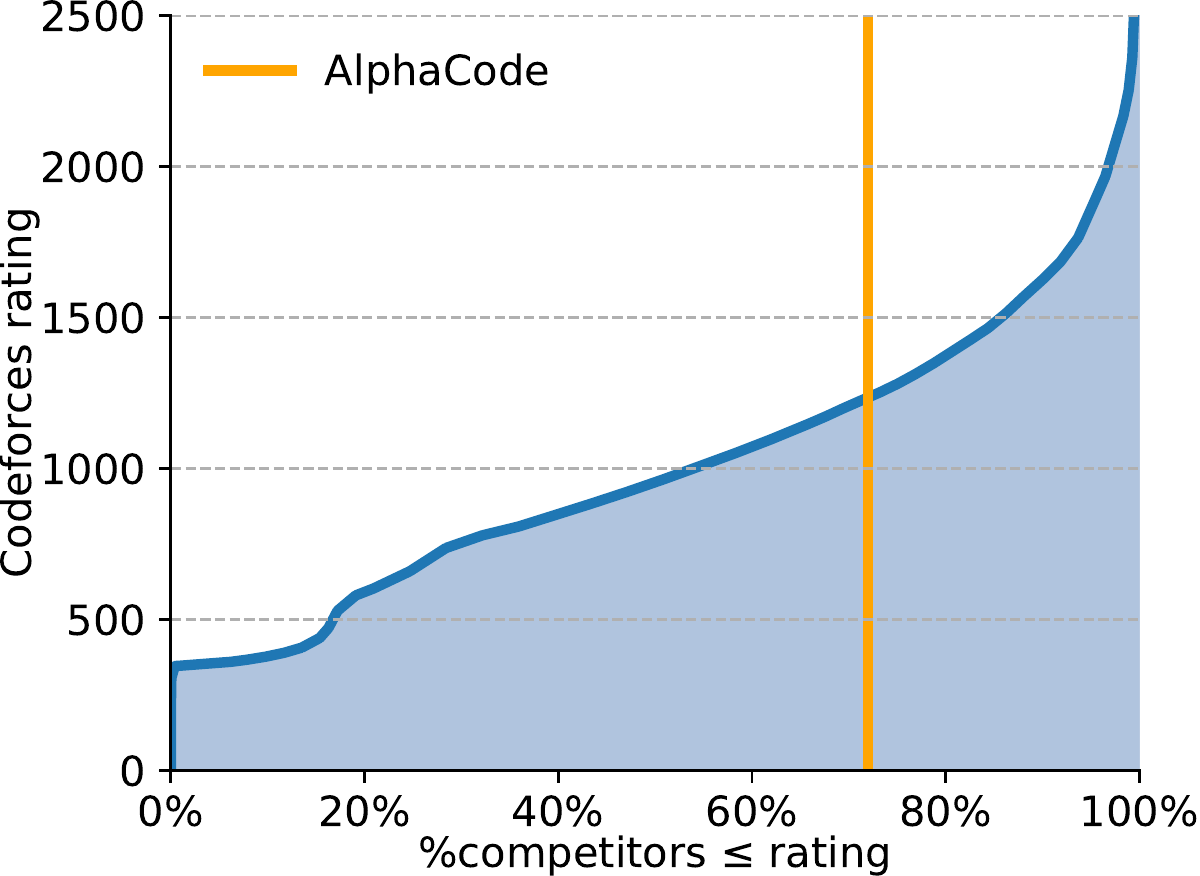} \\
        (a) \OurApproachShort{}'s ranking in 10 contests &
        (b) \OurApproachShort{}'s estimated rating%
    \end{tabular}
    \caption{\textbf{\OurApproachShort{}'s ranking on 10 simulated Codeforces contests and estimated rating (right is better)}. \OurApproachShort{} ranked in the top \HeadlineRanking{}\% among contest participants averaged over 10 contests, and achieved an estimated average rating of \HeadlineElo{}. (a) shows the rating of participants (y-axis) and their rankings in each contest (x-axis), as well as \OurApproachShort{}'s ranking for each of the 10 contests. (b) shows the estimated rating of \OurApproachShort{} among users who have participated in at least 1 contest in the last 6 months.  \OurApproachShort{}'s estimated rating of \HeadlineElo{} is greater than 72\% of these users.
    }
    \label{fig:codeforces-elo-results}
\end{figure}

To further validate our results, we evaluated \OurApproachShort{} on simulated programming competitions hosted on the popular Codeforces platform\footnote{\url{https://codeforces.com/}} (\secref{sec:codeforces-results}). In the evaluation of 10 recent contests with over 5,000 participants each, %
\OurApproachShort{} achieved an average ranking within the top \HeadlineRanking{}\%.
Based on these results, we estimate that our system has achieved a Codeforces rating\footnote{The rating system is similar to the classic Elo score and is primarily explained in three blog posts: \href{https://codeforces.com/blog/entry/102}{1}, \href{https://codeforces.com/blog/entry/20762}{2}, and \href{https://codeforces.com/blog/entry/77890}{3}} of \HeadlineElo{} which is within the top \HeadlineEloRanking{}\%\footnote{\OurApproachShort{}'s overall rating percentile is better than its per-contest percentile. We hypothesise that higher rated competitors compete more regularly than lower rated competitors, and therefore the group ranking above \OurApproachShort{} in contests is relatively more stable than the group ranking below.
} of users who have participated in a contest in the last 6 months (\figref{fig:codeforces-elo-results}) \citep{codeforces-rating}. These evaluations only include users who have tried such competitions, which is a self-selected subset of all programmers. This is the first time that a computer system has achieved such a competitive level in programming competitions.

We also performed a detailed analysis of our system (\secref{sec:limitations}), showing that \OurApproachShort{} does not duplicate sections of code from the training dataset to solve problems, but instead relies heavily on the natural language problem descriptions to create original solutions.
We further examine the types of problems the model can and cannot solve, and discuss how the validation loss is a poor proxy for the solve rate.

\begin{figure}[t]
\footnotesize
\begin{center}
\fbox{
\begin{minipage}[t]{.64\textwidth}
\textbf{Backspace}\\
You are given two strings $s$ and $t$, both consisting of lowercase English letters. You are going to type the string $s$ character by character, from the first character to the last one.\\
\\
When typing a character, instead of pressing the button corresponding to it, you can press the ``Backspace'' button. It deletes the last character you have typed among those that aren't deleted yet (or does nothing if there are no characters in the current string). For example, if $s$ is ``abcbd'' and you press Backspace instead of typing the first and the fourth characters, you will get the string ``bd'' (the first press of Backspace deletes no character, and the second press deletes the character 'c'). Another example, if $s$ is ``abcaa'' and you press Backspace instead of the last two letters, then the resulting text is ``a''.\\
\\
Your task is to determine whether you can obtain the string $t$, if you type the string $s$ and press ``Backspace'' instead of typing several (maybe zero) characters of $s$.\\
\\
\textbf{Input}\\
The first line contains a single integer $q$ $(1 \leq q \leq 10^5)$ the number of test cases.\\
The first line of each test case contains the string $s$ $(1 \leq |s| \leq 10^5)$. Each character of $s$ is a lowercase English letter.\\
The second line of each test case contains the string $t$ $(1 \leq |t| \leq 10^5)$. Each character of $t$ is a lowercase English letter.\\
It is guaranteed that the total number of characters in the strings over all test cases does not exceed $2 \cdot 10^5$.\\
\\
\textbf{Output}\\
For each test case, print ``YES'' if you can obtain the string $t$ by typing the string $s$ and replacing some characters with presses of ``Backspace'' button, or ``NO'' if you cannot.\\
You may print each letter in any case (YES, yes, Yes will all be recognized as positive answer, NO, no and nO will all be recognized as negative answer).

\end{minipage}
}
\hspace{0.1cm}
\begin{minipage}[t]{.3\textwidth}
\footnotesize
\textbf{Example Input}
\begin{lstlisting}
4
ababa
ba
ababa
bb
aaa
aaaa
aababa
ababa
\end{lstlisting}
\textbf{Example Output}
\begin{lstlisting}
YES
NO
NO
YES
\end{lstlisting}
\textbf{Explanation}\\
In order to obtain ``ba'' from ``ababa'', you may press Backspace instead of typing the first and the fourth characters.\\
\\
There's no way to obtain ``bb'' while typing ``ababa''.\\
\\
There's no way to obtain ``aaaa'' while typing ``aaa''.\\
\\
In order to obtain ``ababa'' while typing ``aababa'', you have to press Backspace instead of typing the first character, then type all the remaining characters.
\end{minipage}
\end{center}
\vspace{-0.5cm}
\caption{\textbf{Competitive programming problem statement.} Problem statement of \textbf{\href{https://codeforces.com/problemset/problem/1553/D}{Backspace}}, a Codeforces problem~\citep{codeforces2020}. This is a problem of medium difficulty, with a rating of 1500. The right side shows the public example test case included in the problem description. Hidden tests used to evaluate submissions are shown in \figref{fig:test_cases}. A solution produced by \OurApproachShort{} is shown in \figref{fig:problem_solution}. The entire statement is given to \OurApproachShort{}, and examples of the exact formatting of problem descriptions seen by the model are provided in \appendixref{sec:complete-examples}. %
}
\label{fig:problem_statment}
\end{figure}
\section{Problem setup}
\label{sec:problem-setup}

\subsection{Competitive programming}

Programming competitions first began in the 1970s and have since grown in popularity to include hundreds of thousands of participants %
worldwide. %
The annual International Collegiate Programming Contest attracts almost 60,000 students from over 3,000 universities \citep{icpc-stats}, and companies including Google~\citep{gcj} and Facebook~\citep{facebookhackercup} hold regular competitions. The popular Codeforces platform, used throughout this paper, has more than 500,000 active users and holds weekly competitions with tens of thousands of participants~\citep{codeforces2020}. 

\begin{figure}[t] 

  \begin{minipage}[c]{0.46\textwidth}
    \begin{center}
    \includegraphics[trim=0 450 340 5,clip,width=\textwidth]{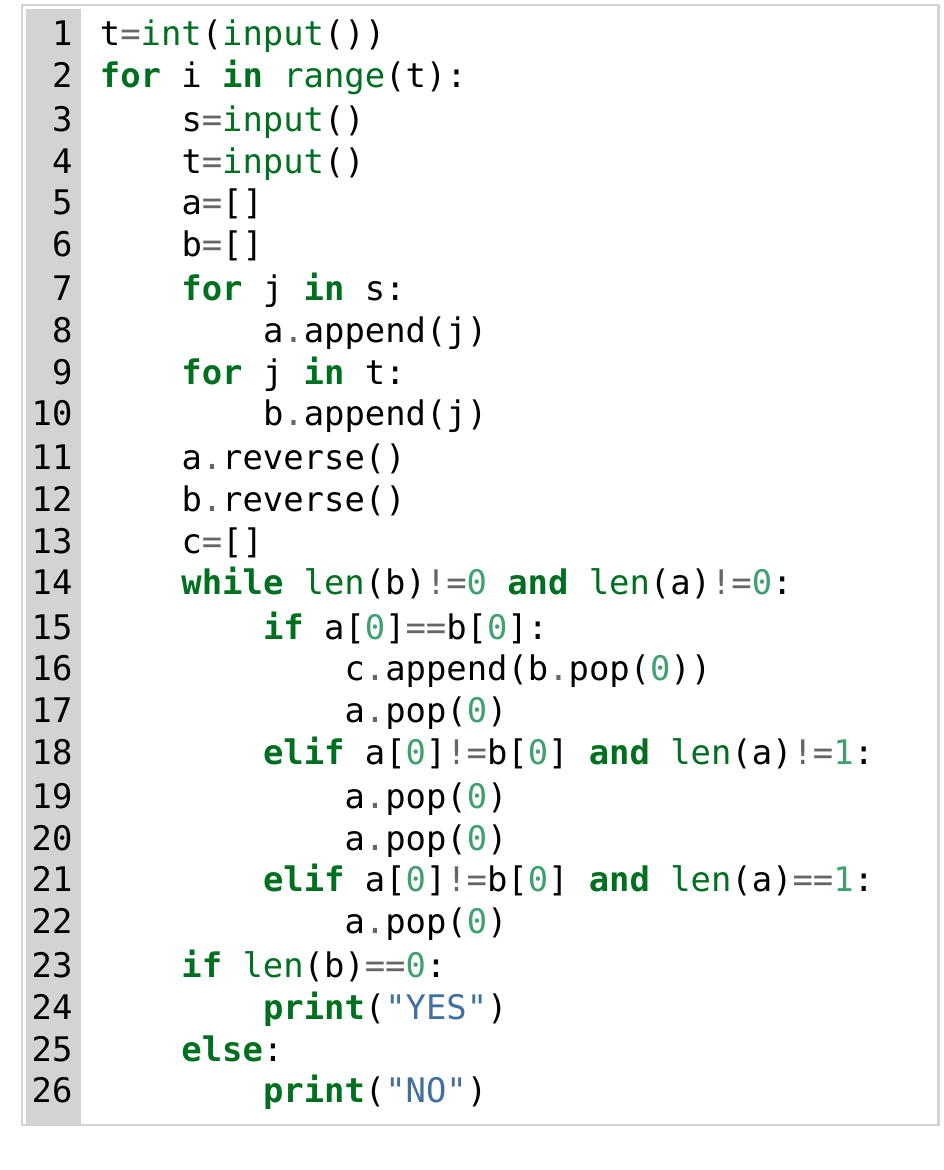}
    \end{center}
  \end{minipage}\hfill
  \begin{minipage}[c]{0.5\textwidth}
    \caption[Test]{\textbf{Solution to \figref{fig:problem_statment} generated by \OurApproachShort{}.} %
    The model successfully extracted the information necessary to solve the problem from the natural language description: %
    \begin{enumerate}
    \item The problem is to figure out if it is possible to convert one phrase to another by pressing backspace instead of typing some letters. So first we read the two phrases (lines 3-4).
    \item If the letters at the end of both phrases don't match, the last letter must be deleted. If they do match we can move onto the second last letter and repeat (11-18).
    \item Backspace deletes two letters. The letter you press backspace instead of, and the letter before it (19-20).
    \item If we matched every letter, it is possible to obtain string $t$ from $s$ (23-26).
    \end{enumerate}
    }
    \label{fig:problem_solution}
  \end{minipage}
  \vspace{-0.5cm}
\end{figure}

The exact format of a programming competition varies between contests, but in general individuals or teams of competitors are given between 5 and 10 problem descriptions (\figref{fig:problem_statment}), and approximately 3 hours to write programs (\figref{fig:problem_solution}) to correctly solve as many problems as possible. The program submissions are sent to a server which automatically evaluates them on an exhaustive set of hidden tests (\figref{fig:test_cases}). Competitors are told whether or not their submission passed all tests, though not necessarily the exact cause of a failure. There are penalties based on the number of incorrect submissions per problem and the amount of time it took to solve each problem~\citep{icpc-rules}. Submissions can be written in a variety of programming languages, among which \CC{} and Python are currently the most popular. Problems are often given ratings to indicate difficulty, and more difficult problems are worth more points.

There are three steps involved in solving a problem. First, participants must read and understand a natural language description spanning multiple paragraphs that contains: narrative background typically unrelated to the problem, a description of the desired solution that the competitors need to understand and parse carefully, a specification of the input and output format, and one or more example input/output pairs (that we call ``example tests''). %

The next step is to create an efficient algorithm that solves the problem.  Going from ``what the problem is'' to ``how to solve the problem'' is a great leap that requires understanding and reasoning about the problem, as well as a deep comprehension of a wide range of algorithms and data structures. This leap is a significant difference from previous works, which tend to explicitly specify what to implement. The algorithm must also be efficient enough to execute in time for the input sizes and time limits specified by the problem,\footnote{The time limit for the problem in Figure \ref{fig:problem_statment} is 2 seconds, using at most 256 MB of memory.} which often eliminates easier, naive attempts. %

Finally, the algorithm must be implemented. Implementation efficiency matters given execution time constraints (harder problems can sometimes only be solved in faster languages such as \CC{}), subtle edge cases can be difficult to account for, and the solution itself can be over a hundred lines of precise code. Participants are given small example test cases to run against, and often debug, fix, and rerun their candidate submission many times before attempting an official submission against the hidden tests cases. An example correct solution generated by \OurApproachShort{} for the problem in \figref{fig:problem_statment} is given in \figref{fig:problem_solution}, and extensive results and analysis can be found in \secref{sec:results} and \ref{sec:limitations}.

\subsection{Evaluation\label{sec:evaluation}}

Though running a system against a live programming competition is an unbiased evaluation, it adds a large degree of complexity and is not a stable benchmark. %
To alleviate this issue, we developed a proxy measure suitable for research iteration similar to the development sets present in most supervised learning datasets. Our measure mirrors the fundamental structure of competitions while simplifying incidental details.  The metric we use is ``percentage of problems solved using $n$ submissions from $k$ samples per problem'', denoted as $n@k$.

This metric indicates the percentage of problems a model can solve if for each problem it is allowed first to create $k$ samples, and then to evaluate $n \leq k$ of these samples against the hidden tests. The problem is considered solved if any of these $n$ evaluations passes all tests. The filtering method is up to the system itself, but should only be based on information available to competitors (e.g. the example tests given as part of the problem description, but not the hidden tests). To decrease variance between runs, assuming both $n$ and $k$ are finite, the metrics we report are expectations computed using bootstrapping on a set of samples typically much larger than $k$ (\appendixref{appendix:evaluation-metrics}). Decreasing variance through expectations makes comparisons of improvements more meaningful, as our validation and test sets are relatively small, and there is significant variance when sampling from a single model.

Limiting the amount of submissions to $n$ emulates the penalty for incorrect submissions and prevents systems from exploiting the evaluation metric by evaluating against the hidden tests an unreasonable number of times. Fixing $k$ is important for comparing different evaluations, as we found that performance increases with the number of samples (\secref{sec:results}). Our use of bootstrapping ensures that we can still benefit from the variance reduction obtained from generating a much larger set of $K \gg k$ samples to estimate the $n@k$ metric.

The setting we use to model programming competitions is $10@k$ -- 10 submissions per problem from $k$ samples. %
We also use $pass@k$ (solve rate with $k$ samples), to be consistent with~\cite{chen2021codex}, which assumes all samples can be submitted for evaluation. $pass@k = k@k$, and is an upper bound metric for using $k$ samples.  We show solve rate with respect to different $k$ values as good results at low sample budgets do not necessarily correlate with good performance at high sample budgets.

\section{Datasets}
\label{sec:dataset}
All our models were first pre-trained on a collection of open-source code from GitHub, and subsequently fine-tuned on a dataset we created (\OurDatasetShort{}, released \href{https://github.com/deepmind/code_contests}{here}) of programming competition data.  The pre-training stage helps the model learn good representations of code and generate code fluently, while the fine-tuning stage helps the model adapt to the target competitive programming domain.

\subsection{Pre-training dataset} %

Our pre-training dataset is based on a snapshot of selected public GitHub repositories
taken on 2021/07/14.
We included all code files from several popular languages: \CC{}, C\#, Go, Java, JavaScript, Lua, PHP, Python, Ruby, Rust, Scala, and TypeScript.
Following previous work~\citep{chen2021codex}, we filtered out all files larger than 1MB or with lines longer than 1000 characters, to exclude automatically generated code. We also removed duplicates of the same file, ignoring whitespace in comparisons. After filtering, our final pre-training dataset contains a total of 715.1 GB of code.  The dataset composition across languages can be found in the appendix (\tabref{tab:dataset-github-stats}). %

\subsection{\OurDatasetShort{} fine-tuning dataset} %
\label{sec:fine-tune}

Models pre-trained on GitHub can generate good code and solve simple programming problems, but as shown in \appendixref{sec:competitive-hardness} they can solve very few competitive programming problems. %
Fine-tuning the model on a dedicated competitive programming dataset is critical for performance.  

To facilitate fine-tuning and evaluation, we curated a new dataset of competitive programming problems, named \OurDatasetShort{}.\footnote{The dataset can be found on \href{https://github.com/deepmind/code_contests}{GitHub}.} %
The dataset includes problems, solutions and test cases we scraped from the Codeforces platform, along with existing public competitive programming datasets mixed into our training set.
More concretely, the training dataset combines newly scraped data from Codeforces~\citep{codeforces2020} with existing data from Description2Code~\citep{Caballero_Description2Code_Dataset_2016}, and CodeNet~\citep{puri2021project}.  The validation and test splits of the dataset consist entirely of newly scraped Codeforces problems.  To guard against data leakage, we adopted a strict temporal split: all pre-training and fine-tuning training data appeared online before any validation problems, and all validation problems before test ones. 
Following our GitHub pre-training dataset snapshot date, all training data in \OurDatasetShort{} was publicly released on or before 2021/07/14. Validation problems appeared between 2021/07/15 and 2021/09/20, and the test set contains problems published after 2021/09/21. This temporal split means that only information humans could have seen is available for training the model (see \appendixref{sec:dataset-temporal-split} for more details and analysis).
Some basic statistics of this dataset are shown in \tabref{tab:dataset-desc2code-stats}.

Our scraped data from Codeforces includes full problem descriptions like that shown in \figref{fig:problem_statment}, along with %
metadata for each problem.  The metadata includes \emph{difficulty ratings} and \emph{tags} that indicate which approaches might be required to solve the problem (e.g. ``greedy'' or ``dp'').  Neither the difficulty rating nor the tags are visible at competition time (and so should not be used at test time). Our dataset also contains both correct and incorrect human submissions written in the most popular submission languages: \CC{}, Python, and Java.  Each problem includes all the test cases that are accessible from the platform: example tests in the problem statements and hidden test cases that are made available at the evaluation result pages once a contest is finished.
To improve data quality and consistency, and to avoid duplication issues involved in merging datasets, we cleaned this data using the procedure outlined in \appendixref{sec:appendix-dataset-merging}. 

The correctness of a program is checked by executing it on the test cases and comparing the program output with the expected correct output.  More details about this correctness checking process are documented in %
\appendixref{sec:program-evaluation}.

\begin{table}[t]
    \centering
    \begin{tabular}{l|r|rrr|r@{\hskip 2pt}rr@{\hskip 2pt}rr@{\hskip 2pt}r}
    \toprule
    & & \multicolumn{3}{c|}{Tests per problem} & \multicolumn{6}{c}{Solutions per problem (\% correct)} \\
    Split & Problems & Example & Hidden & Generated & \multicolumn{2}{c}{\CC}& \multicolumn{2}{c}{Python}& \multicolumn{2}{c}{Java}\\
    \midrule
    Train & 13328 & 2.0 & 14.8 & 79.1 & 493.4 &(27\%) & 281.1 &(47\%) & 147.9 &(46\%) \\
    Valid & 117 & 1.5 & 12.9 & 190.0 & 231.6 &(47\%) & 137.2 &(55\%) & 131.1 &(54\%) \\
    Test & 165 & 1.7 & 9.4 & 192.7 & 196.0 &(45\%) & 97.3 &(54\%) & 105.2 &(51\%)\\
    \bottomrule
    \end{tabular}
    \caption{\textbf{Statistics of our \OurDatasetShort{} dataset}. The number of problems in each split, and the per-problem averages for the number of test cases, number of solutions, and percentage of solutions which are correct.}
    \label{tab:dataset-desc2code-stats}
\end{table}

\subsubsection{False positives and additional generated tests}
\label{sec:false-positives}
We want the test cases to be as exhaustive as possible, so that submissions cannot be marked as correct by exploiting a lack of test coverage. Unfortunately, high-quality test cases are not readily available. For example, the Codeforces platform does not display full test cases when they are longer than approximately 400 characters.
Lack of test coverage leads to 
``false positives'' where incorrect submissions are marked as correct, and ``slow positives'' where correct but algorithmically inefficient solutions that do not fulfill time and memory constraints are marked correct (e.g. a solution that is of the wrong complexity class). These false positives do not effect the evaluation on Codeforces described \secref{sec:codeforces-results}.

Notably, both issues are common in prior datasets and the program synthesis literature, as input/output examples are an under-specification of program behavior~\citep{gulwani2017program}.  \tabref{tab:dataset-test-cases-false-positive} shows the estimated false positive rate of our dataset compared to APPS \citep{hendrycks2021measuring} and HumanEval \citep{chen2021codex}, which both have many false positives.
A high average number of tests per problem does not necessarily indicate exhaustive tests, because some problems may have far fewer tests per problem than average, and some tests may examine similar cases. %

\begin{table}[t]
    \centering
    \begin{tabular}{lccc}
    \toprule
        Dataset & Tests / problem & False Positive (FP) Rate & FP or Slow Rate \\
        \midrule
        APPS & 20.99 & 60\% & 70\%\\
        HumanEval & 7.77 & 30\% & N/A \\
        \OurDatasetShort{} raw & 12.4 & 62\% & 88\% \\
        \OurDatasetShort{} & \textbf{203.7} %
        & \textbf{4\%} & \textbf{46\%} \\
        \bottomrule
    \end{tabular}
    \caption{\textbf{Dataset false positive rates}. The bottom row is the dataset we used, while ``\OurDatasetShort{} raw'' does not use generated tests and does not filter out problems with insufficient tests. %
    Validation splits were used for \OurDatasetShort{} and APPS. We randomly selected 50 problems our 1B parameter model solved (from 10,000 samples per problem for APPS, 200 for HumanEval, and 1,000,000 for \OurDatasetShort{}), and manually examined one solution for each problem to check whether they are false positives or slow solutions.
    HumanEval does not have timing constraints for most problems, so there is no slow rate. %
    }
    \label{tab:dataset-test-cases-false-positive}
\end{table} 

We reduced the false positive rates of our dataset by generating additional test cases, created by mutating existing test inputs. Possible mutations are applying bit flips to binary inputs, randomly incrementing or decrementing integers, and swapping and changing characters in strings. Mutated inputs are verified by running 30 correct solutions on them, and checking that all solutions produce the same output. This process was run on each problem for a maximum of 10 CPU hours or 200 generated tests. Because of complex input formats, we failed to generate the full set of 200 tests for 6.3\% of problems.

Lastly, we filtered out problems in the validation and test splits with insufficient test coverage, keeping only problems with at least 5 hidden or generated test cases that result in at least 2 different outputs. This ensures a model cannot trivially solve problems by always outputting a constant, such as \emph{YES} or \emph{NO}. As seen in \tabref{tab:dataset-test-cases-false-positive}, generated tests and filtering reduced our false positive rates from 62\% to 4\%. \OurDatasetShort{} has significantly better false positive rates than prior work even though we drew fewer samples for both APPS and HumanEval, and the problems in those datasets are relatively less complex (both of which tend to lower the false positive rates). However, there is still a significant number of problems where slow but semantically correct solutions are accepted by the tests.

\begin{figure}[t]
\centering
\includegraphics[width=0.8\textwidth]{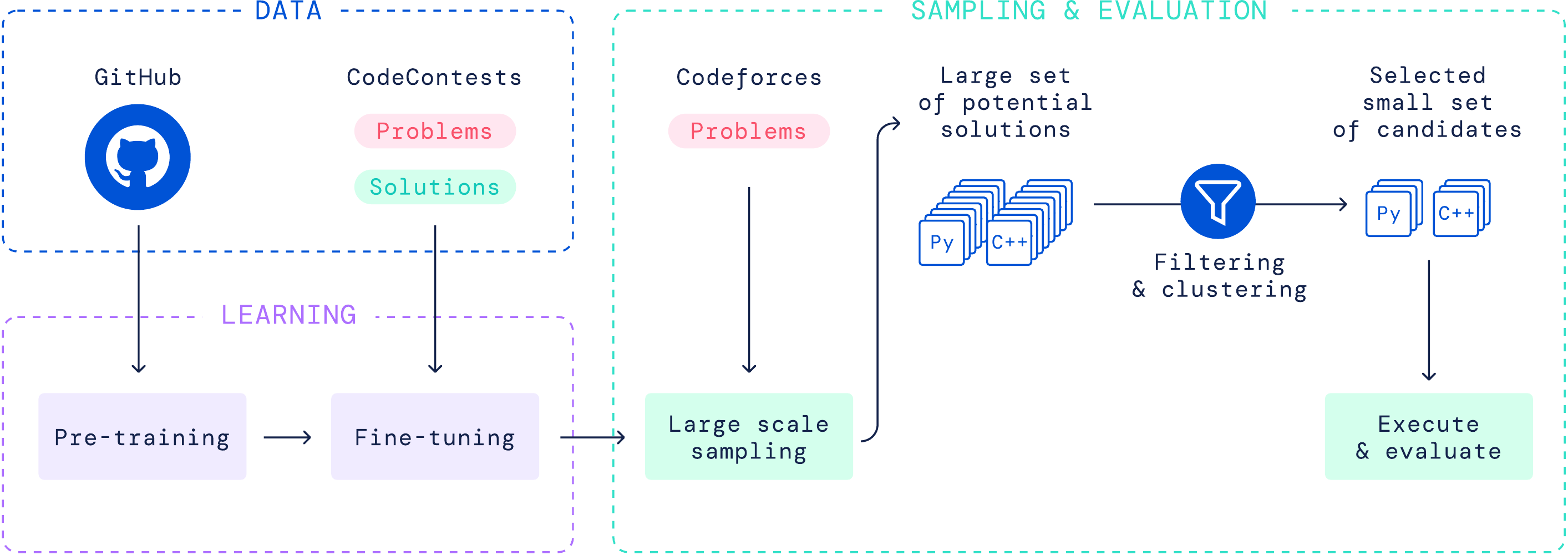}
\caption[]{
\label{fig:method}
\textbf{Overview of \OurApproachShort. %
}}
\end{figure}

\section{Approach}\label{sec:approach}

Generating code that solves a specific task requires searching in a huge structured space of programs with a very sparse reward signal. To make matters worse, for many domains including competitive programming, there is a limited number of examples of such tasks and solutions to learn from. Finally, as we restrict the amount of submissions per problem our model can do,
each submission must be used wisely.

Our system, \OurApproachShort{}, is meant to address all these challenges. A high-level view of our approach can be seen in Figure~\ref{fig:method}. The main process is to:
\begin{enumerate}
    \item Pre-train a transformer-based language model on GitHub code with standard language modelling objectives. This model can reasonably represent the space of human coding, which greatly reduces the problem search space.
    \item Fine-tune the model on our dataset of competitive programming data, using GOLD~\citep{pang2020text} with tempering~\citep{dabre2020tempering} as the training objective. This further reduces the search space, and compensates for the small amount of competitive programming data by leveraging pre-training.
    \item Generate a very large number of samples from our models for each problem. %
    \item Filter the samples to obtain a small set of candidate submissions (at most 10), to be evaluated on the hidden test cases, by using the example tests and clustering to pick samples based on program behaviour. %
\end{enumerate}

Among these, the large-scale sampling followed by filtering is unique to our setup, and we found that this process greatly improves problem solve rate. Therefore many of our design decisions were made to facilitate efficient and effective sampling.

\subsection{Model architecture}

The competitive programming code generation problem can be viewed as a sequence-to-sequence %
\citep{sutskever2014sequence}
translation task: given a problem description $X$ in natural language (e.g.~\figref{fig:problem_statment}), %
produce a corresponding solution $Y$ in a programming language (e.g.~\figref{fig:problem_solution}). %
This naturally motivates the choice of an encoder-decoder transformer architecture \citep{vaswani2017attention} for \OurApproachShort{}, which models $p(Y|X)$. The architecture takes as input to the encoder the problem description $X$ as a flat sequence of characters (including metadata, tokenized), and samples $Y$ autoregressively from the decoder one token at a time until an end of code token is produced, at which point the code can be compiled and run (see~\appendixref{sec:complete-examples} for example $X,Y$ pairs, and \url{https://alphacode.deepmind.com/} for an interactive model visualisation).

Compared to decoder-only architectures commonly used for language modeling and generation, an encoder-decoder architecture allows a bidirectional description representation (tokens at the beginning of the description can attend to tokens at the end) %
and the extra flexibility to untie the encoder structure from the decoder. %
Because problem descriptions are on average twice as long as their corresponding human solutions, we use an asymmetric architecture with 1536 tokens for the encoder but only 768 tokens for the decoder. We further found that using a shallow encoder and a deep decoder significantly improves the efficiency of training without hurting problem solve rate. The exact architectures for our models are listed in \tabref{tab:approach-model-architectures}. The 9B and 41B models were trained using model parallelism, with 1 key and value head per shard. We built our model using JAX \citep{jax2018github} and Haiku \citep{haiku2020github}, and trained them on TPUv4 accelerators using bfloat16 precision.

\begin{table}[t]
    \centering
    \begin{tabular}{lrrrrrrrrr}
    \toprule
                & & & \multicolumn{2}{c}{Heads} & \multicolumn{2}{c}{Blocks} & \multicolumn{3}{c}{Training} \\
        Name & $n_{params}$ & $d_{model}$ & Query & KV & Enc & Dec & Batch & Steps & Tokens \\
        \midrule
        \OurApproachShort{} 300M & 284M  & 768  &  6 &  1 & 4 & 24 & 256  &  600k &  354B \\
        \OurApproachShort{} 1B   & 1.1B  & 1408 & 11 &  1 & 5 & 30 & 256  & 1000k &  590B \\
        \OurApproachShort{} 3B   & 2.8B  & 2048 & 16 &  1 & 6 & 36 & 512  &  700k &  826B \\
        \OurApproachShort{} 9B   & 8.7B  & 3072 & 24 &  4 & 8 & 48 & 1024 &  530k & 1250B \\
        \OurApproachShort{} 41B  & 41.1B & 6144 & 48 & 16 & 8 & 56 & 2048 &  205k &  967B \\
        \bottomrule
    \end{tabular}
    \caption{\textbf{Architecture configuration of our models at different parameter scales}. This table lists the total number of parameters in the model $n_{params}$, the hidden dimension of the transformer blocks $d_{model}$, the number of query and key-value heads, the number of transformer blocks in the encoder and decoder, the training batch size, the number of gradient update steps, and the number of total training tokens. The head size is always 128, with a feed-forward fan-out ratio of 6. %
    }
    \label{tab:approach-model-architectures}
\end{table}

To reduce the cost of sampling from our models, we take advantage of multi-query attention \citep{shazeer2019fast}. Using a full set of query heads but sharing key and value heads per attention block significantly reduces memory usage and cache update costs, which are the main bottleneck during sampling. This memory reduction also allows larger batch sizes for sampling, further increasing efficiency.

For tokenization we used a SentencePiece tokenizer \citep{kudo2018sentencepiece} with a vocabulary size of 8,000 tokens trained on a mix of GitHub and \OurDatasetShort{} data. The training mix ensures it can effectively tokenize programs from a range of languages, as well as the natural language descriptions of problems. The encoder and decoder in our models use the same tokenizer.

\subsection{Pre-training}

We pre-trained our models on the GitHub dataset described in \secref{sec:dataset}, with a standard cross-entropy next-token prediction loss for the decoder and a masked language modeling loss \citep{devlin2018bert} for the encoder. The masked language modeling loss was essential for improving the representation learning of the encoder.
We split GitHub files by uniformly sampling pivot locations, using content before the pivot as input to the encoder, and content after for the decoder.

Our base 1B parameter model was trained for $10^6$ steps with a batch size of 256. Following \cite{kaplan2020scaling}, we adjusted the amount of training for other model sizes such that larger models are trained more and smaller models are trained less to optimize the use of compute. However, due to resource limitations and to make optimal use of compute, the training of our largest 41B model was stopped early, and therefore this model was relatively undertrained compared to models at other scales
(\tabref{tab:approach-model-architectures}).

We trained all models using the AdamW variant~\citep{loshchilov2017decoupled} of the Adam optimiser~\citep{kingma2014adam} with $\beta_1=0.9$, $\beta_2=0.999$ for \{300M, 1B, 3B\} models, and $\beta_2=0.95$ for \{9B, 41B\} models. We used a weight decay of $0.1$ to enhance regularization. We trained the models with an initial learning rate of $10^{-4}$, which was then cosine decayed to $10^{-5}$ at the end of pre-training. We linearly warmed-up the learning rate from $10^{-9}$ to $10^{-4}$ over the first $1,000$ training steps, and clipped the global gradient norm to stay below $1.0$.

\subsection{Fine-tuning}
\label{sec:finetuning}

We fine-tuned our model on our \OurDatasetShort{} dataset. During fine-tuning, we used the natural language problem description for the encoder and the program solution for the decoder. Similar to pre-training, we used both the standard next-token prediction and masked language modeling losses. We also adopted additional conditioning and modifications that we found improved the overall solve rate: tempering, value conditioning and prediction, and GOLD described below, as well as metadata conditioning described in \appendixref{sec:metadata-conditioning}. We set the initial learning rate as $10^{-5}$, and cosine decayed it to $10^{-6}$ at the end of fine-tuning. We used the same linear warm-up stage for the learning rate over the first $1,000$ training steps.

\textbf{Tempering.~~}
Tempering, introduced by \citet{dabre2020tempering}, is a regularization technique that makes the token probability distribution artificially smoother or sharper at training time by dividing the output logits of a model by a scalar temperature $T$ before the softmax layer. We observed that when using $T=0.2 < 1$, tempering helps avoid overfitting to our fine-tuning dataset by making the training distribution sharper, and consequently the inference distribution smoother. Notably, this is the opposite of the suggestion of \citet{dabre2020tempering} to use $T > 1$ to make a sharper inference distribution. At sampling time, we divided the logits by another temperature $T'$ tuned on the validation set ($T' = 0.12$ for models trained with tempering only; $T'=0.25$ for models trained with tempering and GOLD).

\begin{figure}[t]
    \centering
    \begin{lstlisting}[language={}]
RATING: 1200
TAGS: dp,implementation
LANGUAGE IS python3
CORRECT SOLUTION
Polycarp must pay exactly n burles at the checkout ... (rest of the description)
\end{lstlisting}
    \caption{\textbf{Example format of the additional metadata information}. This is added to the top of problem descriptions. Metadata and problem descriptions are handled identically. See \appendixref{sec:complete-examples} for a full example of what is used in the decoder. The problem in this example can be found \href{https://codeforces.com/problemset/problem/1551/A}{here}.}
    \label{fig:metadata-format}
\end{figure}

\textbf{Value conditioning \& prediction.~~}
\OurDatasetShort{} contains both correct and incorrect problem submissions. We used value conditioning and prediction to discriminate between these two types of submissions, providing an additional training signal and allowing use of data which could otherwise mislead the model. Similar approaches were used in, e.g., \cite{vinyals2019grandmaster}.
In value conditioning, we inserted whether or not a submission was correct into the problem description so that the model can condition on this information, as shown in \figref{fig:metadata-format}. At sampling time, the model was always conditioned on the sample being correct. In value prediction, we added an auxiliary value prediction task during training such that the last layer token representations before projecting to logits are also used in a small Transformer to classify whether the submission is correct. Value prediction was not used during sampling.  %

\textbf{GOLD \citep{pang2020text}.~~}
Solving competitive programming problems from descriptions is inherently a one-of-many task~\citep{nandwani2021neural}: each unique problem allows many distinct solutions that depend on algorithm choice, implementation, etc. \OurDatasetShort{} contains several orders of magnitude more solutions than descriptions (\tabref{tab:dataset-desc2code-stats}).
Standard maximum likelihood objectives minimise loss by putting some weight on each solution in the training set (like recall), whereas our metric measures whether a model can find a single correct solution in the submission attempt budget (like precision).
To resolve this discrepancy, we adopted a variation of GOLD~\citep{pang2020text}, an offline RL algorithm which allows the model to both learn from tokens it already assigns high likelihood to, and to ignore tokens that are not in its distribution (allowing it to concentrate on precision). To combine GOLD and tempering, we introduce a short training phase between pretraining and finetuning. Full details of GOLD and this combination are in \appendixref{appendix:gold}.

\subsection{Large scale sampling}
\label{sec:large-scale-sampling}

Sampling from transformer models can be easily parallelized, which allowed us to scale to millions of samples per problem -- a critical driving force for performance improvement. To ensure sufficient diversity in such a large number of samples, we take a single trained model and: (i) generate half of the samples in Python and half in \CC{}, (ii) randomize the problem tags and ratings in the natural language prompt (see \figref{fig:metadata-format} for an example and \appendixref{sec:metadata-conditioning} for more details), and
(iii) use a relatively high sampling temperature. The single model, via the additional metadata we condition upon, can generate solutions with different languages, tags, and ratings. To make the most effective use of our samples we then apply filtering (\secref{sec:filtering}) and clustering (\secref{sec:clustering}) to obtain a small number of candidate submissions.

For problem tags and ratings conditioning, we picked random tags from the most popular 50 for the model to condition on, and sampled ratings uniformly in the range of 800 to 3500 as these metadata are not visible for new unseen problems in a competition. We found that conditioning on random tags and ratings can improve performance, potentially by increasing diversity of the samples.

The optimal sampling temperature depends on the total number of samples (in general the more samples, the higher the optimal temperature).  However different temperatures in a wide range do not significantly change the solve rates (\figref{fig:different-sampling-settings}).
We therefore use a fixed sampling temperature $T'=0.25$ in all experiments that use tempering and GOLD, $T'=0.12$ when using tempering only, and tune the sampling temperature separately otherwise.

We also experimented with top-$k$ \citep{fan2018topksampling} and nucleus sampling~\citep{holtzman2019nucleus}. As seen in \figref{fig:different-sampling-settings}, despite running exhaustive hyperparameter sweeps we did not observe significant performance improvements with these methods. We therefore use regular sampling with temperature in our experiments. A few complete examples of model prompts and samples are provided in~\appendixref{sec:complete-examples}.

\subsection{Filtering}
\label{sec:filtering}
To accurately represent competitive programming contests and penalties, our formulation limits us to just 10 submissions per problem no matter how many samples we draw. One powerful tool for selecting these submissions is filtering samples to only those that pass the example tests given in the problem statement. %
Filtering removes approximately 99\% of model samples, although the exact amount depends on the problem and model, and filtering can still leave tens of thousands of candidate samples for many problems. Finding solutions that pass example tests is itself a difficult problem, and on approximately 10\% of problems our models cannot find a single such program. Indeed this easier version of our setting is a classic program synthesis formulation, where the task is specified by a list of given input/output pairs \citep{gulwani2017program}. 

\subsection{Clustering}
\label{sec:clustering}
Filtering using example tests can still leave thousands of candidate programs per problem. Randomly picking from this pool wastes the limited submission budget on programs that are syntactically different but semantically equivalent.  Semantically equivalent programs could be detected if we had additional test inputs, by executing all remaining programs on these inputs and grouping programs that produce the same outputs together into clusters. We could then avoid repeatedly picking from the same clusters.

We trained a separate test input generation model, using the same architecture as our main models, and initialised from the same GitHub pre-trained checkpoint.  This model was trained to predict test inputs from problem descriptions, using example, hidden, and generated test inputs as training data.
After training, this model was used to create new test inputs for unseen problems. Note that although these created test inputs are not guaranteed to be valid, especially when problems have complex constraints, imperfect and even invalid test inputs can still be useful for grouping sampled programs.

This learned test input generation model is different from the mutation-based test generation process used in \secref{sec:false-positives} to augment our dataset. The latter requires correct solutions (which are not available at test time) to filter out bad test cases.

After clustering on program behaviour we found that selecting one solution from each cluster from largest to smallest performed best, perhaps because there are many ways solutions can be incorrect while correct solutions tend to behave the same and therefore are grouped into larger clusters. If the candidate solutions for a problem form less than 10 clusters (or more in the case of more than 10 submissions), after reaching the smallest cluster, we repeat from the first cluster skipping samples that have already been submitted.

\section{Results}\label{sec:results}

In this section we present experimental results that give insights into our model performance, and evidence that guided our design decisions.  We highlight the results obtained by evaluating on the Codeforces platform (\secref{sec:codeforces-results}) and on \OurDatasetShort{} (\secref{sec:codecompete-eval}), present a detailed study of model performance on our dataset in \secref{sec:codecompete-results}, and conclude by comparing to published models in the literature on the public APPS \citep{hendrycks2021measuring} benchmark of programming problems in \secref{sec:public-benchmark-results}.  To ensure that our baseline models are comparable to past work we also compare our decoder-only baseline directly to~\cite{chen2021codex} on the HumanEval benchmark in \appendixref{sec:appendix-humaneval}.

\subsection{Codeforces competitions evaluation}\label{sec:codeforces-results}

Evaluating on programming competitions checks program correctness more thoroughly, compared to evaluating on our dataset which has known weaknesses including false positives, accepting algorithmically inefficient solutions, and handling problems with multiple acceptable outputs. Additionally, evaluating in the real setting allows us to benchmark against the best performers on this task: human competitors. 

We evaluated our best system on all Codeforces competitions from 2021/12/01 to 2021/12/28 with more than 5,000 participants per contest, a total of 10 competitions. The system was an ensemble of 41B and 9B models with clustering, which performed best on our validation set but turned out to be slightly worse than using the 41B model alone with clustering (see \appendixref{app:ensemble} for more on ensembling). For each contest, we simulated running \OurApproachShort{} live, generating samples for each problem, filtering with example tests,\footnote{For problems permitting multiple correct outputs, we change the example test outputs to be the most canonical, which gives our approach a slight advantage in the evaluation. See \appendixref{sec:codeforces-eval} for more details.} and then clustering to get candidate submissions.  We submitted these selected candidates to the Codeforces platform,\footnote{Submitted programs can be found on our 3 accounts on Codeforces: \href{https://codeforces.com/submissions/SelectorUnlimited}{SelectorUnlimited},  \href{https://codeforces.com/submissions/WaggleCollide}{WaggleCollide}, and  \href{https://codeforces.com/submissions/AngularNumeric}{AngularNumeric}. Attention visualizations for these problems can be found \href{https://alphacode.deepmind.com/}{here}.} and computed \OurApproachShort{}'s placement in each contest. After the first run, we repeated this procedure two more times to measure variance and performance with more than 10 submissions.  Sources of variance include problem distribution, model training, sampling, filtering, and clustering. %
See \appendixref{sec:codeforces-eval} for the exact evaluation procedure, and \tabref{tab:codeforces-result} and \tabref{tab:codeforces-contest-result} for full results.

\begin{table}[t]
    \centering

\begingroup
\setlength{\tabcolsep}{3pt}
{\small
\begin{tabular}{r|cccccccccc|c}
\toprule
{Contest ID} & {1591} & {1608} & {1613} & {1615} & {1617} & {1618} & {1619} & {1620} & {1622} & {1623} & {Average} \\
\midrule
Best & {\cellcolor[HTML]{F2F6EC}} \color[HTML]{000000} 43.5\% & {\cellcolor[HTML]{F2F6EC}} \color[HTML]{000000} 43.6\% & {\cellcolor[HTML]{FAEDF3}} \color[HTML]{000000} 59.8\% & {\cellcolor[HTML]{FAECF3}} \color[HTML]{000000} 60.5\% & {\cellcolor[HTML]{FBE7F2}} \color[HTML]{000000} 65.1\% & {\cellcolor[HTML]{E9F5D8}} \color[HTML]{000000} 32.2\% & {\cellcolor[HTML]{F5F7F2}} \color[HTML]{000000} 47.1\% & {\cellcolor[HTML]{F8F3F6}} \color[HTML]{000000} 54.0\% & {\cellcolor[HTML]{F9EFF4}} \color[HTML]{000000} 57.5\% & {\cellcolor[HTML]{D6EEB6}} \color[HTML]{000000} 20.6\% & {\cellcolor[HTML]{F6F7F5}} \color[HTML]{000000} \textbf{48.4\%} \\
Estimated & {\cellcolor[HTML]{F3F6ED}} \color[HTML]{000000} 44.3\% & {\cellcolor[HTML]{F4F7F0}} \color[HTML]{000000} 46.3\% & {\cellcolor[HTML]{FBE6F1}} \color[HTML]{000000} 66.1\% & {\cellcolor[HTML]{FAEAF2}} \color[HTML]{000000} 62.4\% & {\cellcolor[HTML]{FCDDED}} \color[HTML]{000000} 73.9\% & {\cellcolor[HTML]{F8F5F6}} \color[HTML]{000000} 52.2\% & {\cellcolor[HTML]{F5F7F2}} \color[HTML]{000000} 47.3\% & {\cellcolor[HTML]{FBE9F2}} \color[HTML]{000000} 63.3\% & {\cellcolor[HTML]{FBE6F1}} \color[HTML]{000000} 66.2\% & {\cellcolor[HTML]{D8EFB9}} \color[HTML]{000000} 20.9\% & {\cellcolor[HTML]{F8F3F6}} \color[HTML]{000000} \textbf{54.3\%} \\
Worst & {\cellcolor[HTML]{FCDBED}} \color[HTML]{000000} 74.5\% & {\cellcolor[HTML]{EFB0D6}} \color[HTML]{000000} 95.7\% & {\cellcolor[HTML]{FBD9EC}} \color[HTML]{000000} 75.0\% & {\cellcolor[HTML]{F3BCDD}} \color[HTML]{000000} 90.4\% & {\cellcolor[HTML]{F7CCE5}} \color[HTML]{000000} 82.3\% & {\cellcolor[HTML]{F8F3F6}} \color[HTML]{000000} 53.5\% & {\cellcolor[HTML]{F4C1DF}} \color[HTML]{000000} 88.1\% & {\cellcolor[HTML]{FBD9EC}} \color[HTML]{000000} 75.1\% & {\cellcolor[HTML]{F8CEE6}} \color[HTML]{000000} 81.6\% & {\cellcolor[HTML]{F9F1F5}} \color[HTML]{000000} 55.3\% & {\cellcolor[HTML]{FAD6EA}} \color[HTML]{000000} \textbf{77.2\%} \\
\bottomrule
\end{tabular}}
    \caption{\textbf{Estimated
    percent ranking of our system in 10 Codeforces competitions (lower is better)}.  For each contest, we show ranking using simulated time and incorrect submission penalties (Estimated), as well as the best and worst possible rankings using minimum and maximum possible time penalties as estimates, averaged over 3 evaluations. Percents are how many users performed better than \OurApproachShort{}.  Our system achieved an overall ranking of top \HeadlineRanking{}\% averaged across the 10 contests.  %
    }
    \label{tab:codeforces-results}
\endgroup
\end{table}

\tabref{tab:codeforces-results} shows evaluation results across the 10 competitions.  For each competition, we show the estimated percentile ranking using a simulated penalty, and upper and lower bounds assuming zero and maximum submission time penalties. The bounds represent how ranking depends on the number of accelerators used to draw samples during competition. For the second and third runs, \tabref{tab:codeforces-contest-result} shows the estimated percentile when not limiting to 10 submissions per problem (still taking into account penalties for incorrect submission), which although not human-like does follow competition rules.
We found that the model still continued to solve problems when given more attempts, though at a decreased rate. The model tends to solve the easier problems in competitions, but it does manage to solve harder problems including one rated 1800.

Overall our system achieved an average ranking of top \HeadlineRanking{}\% limiting to 10 submissions per problem, with an actual average of \AverageSubmissionsTenAttempt{} submissions for each problem solved.\footnote{Our estimated performance is closer to its upper bound than its lower bound, because human solutions (and our solutions) are typically submitted early in the contest, especially for easier problems.}  When allowed more than 10 submissions per problem (the second and third evaluation), \OurApproachShort{} achieved a ranking of top \RankingUnlimitedAttempt{}\%, with an actual average of \AverageSubmissionsUnlimitedAttempt{} submissions for each problem solved. Our 10 submissions per problem result corresponds to an estimated Codeforces rating of \HeadlineElo{}, which is within the top \HeadlineEloRanking{}\% of users who have participated in a contest in the last 6 months (a small and selected subset of all programmers). To the best of our knowledge, this is the first time that a computer system has been competitive with human participants in programming competitions.

\subsection{\OurDatasetShort{} evaluation}\label{sec:codecompete-eval}

As well as the Codeforces evaluation, we evaluated our model on the validation and test sets of \OurDatasetShort{}. The test set is a superset of the competitions used in \secref{sec:codeforces-results}.\footnote{Except one problem that does not have 5 test cases and is therefore not included in our test set.} The metrics on our dataset are lower variance and easier to measure, since they do not involve submitting to an external site. For \OurDatasetShort{} (both here and in \secref{sec:codecompete-results}), we focus on the two main metrics discussed in~\secref{sec:evaluation}:
\begin{itemize}
    \item ${\mathbf{pass@k}}$:  The percentage of problems solved when we take $k$ samples from the model for each problem and submit all of them for evaluation on the hidden tests.  If any solution in the specified sample budget solves a problem, the problem is counted as solved.  Therefore this metric measures mostly the search aspect of the sampling process, and is used in \secref{sec:codecompete-results}. %
    \item $\mathbf{10@k}$:  The percentage of problems solved when we take $k$ samples from the model for each problem but can only submit $10$ of them for evaluation on the hidden tests.  This measures factors including the filtering process and how models behave at a very large number of samples.
\end{itemize}

\begin{table}[t]
\begingroup
\setlength{\tabcolsep}{4pt}
    \centering
    \begin{tabular}{l|cccc|ccc}
    \toprule
        \multirow{2}{*}{Approach} & \multicolumn{4}{c|}{Validation Set} &
        \multicolumn{3}{c}{Test Set} \\
        & 10@1k & 10@10k & 10@100k & 10@1M
        & 10@1k & 10@10k & 10@100k \\
    \midrule
        9B & 16.9\% & 22.6\% & 27.1\% & 30.1\% & 14.3\% & 21.5\% & 25.8\% \\
        41B & 16.9\% & 23.9\% & 28.2\% & 31.8\% & 15.6\% & 23.2\% & 27.7\% \\
        41B + clustering & 21.0\% & 26.2\% & 31.8\% & 34.2\% & 16.4\% & 25.4\% & 29.6\% \\
    \bottomrule
    \end{tabular}
    \caption{\textbf{Solve rates of our best systems on the validation set and test set
    }.}
    \label{tab:valid-test-comparison}
\endgroup
\end{table}

The results are shown in \tabref{tab:valid-test-comparison}. With up to a million samples per problem, we can solve 34.2\% of problems in our validation set; and with one hundred thousand samples, we solve 31.8\% of problems in our validation set, and 29.6\% of problems in our test set. Because of the temporal split, no problem in either set was seen by our model during training. Given the difficulty of these problems (since they are problems given to the self-selected group of those who try competitive programming), this is a substantial proportion of the dataset.

Differences in solve rates between the validation and test sets are caused by variation in problem distributions (as the test set and validation set were collected in temporally disjoint periods), as well as some overfitting. However, the difference in performance between the two sets remains limited. The 41B consistently outperforms the 9B model, and clustering consistently provides an improvement.

\subsection{\OurDatasetShort{} ablations \& results \label{sec:codecompete-results}}

This section contains results that support our design decisions described in \secref{sec:approach}. All results are on the \OurDatasetShort{} validation set, with models fine-tuned on the \OurDatasetShort{} training set and not using clustering unless otherwise noted.

\subsubsection{Solve rates scale with respect to parameter count, compute, number of samples, and dataset size} \label{sec:large-scale-sampling-and-models}

As would be expected, scaling up the number of model parameters or the size of the dataset greatly improves model performance (see \figref{fig:dataset-scaling} for scaling with dataset size). However, even when only $10$ samples can be submitted, scaling up the total number of samples leads to massive improvements in model solve rate.

\begin{figure}[t]
    \centering
    \begin{tabular}{cc}
        \includegraphics[width=0.48\textwidth]{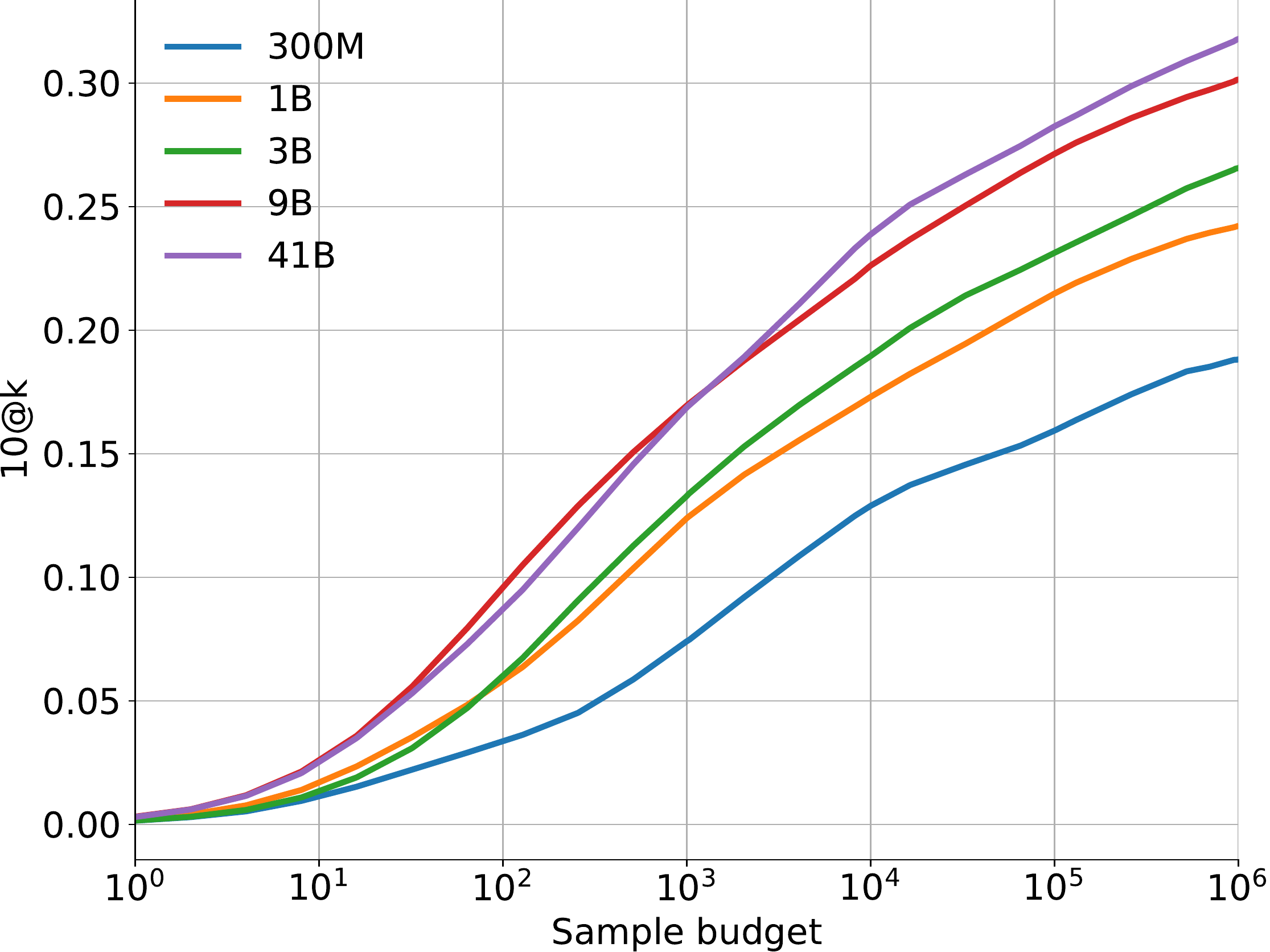} &
        \includegraphics[width=0.48\textwidth]{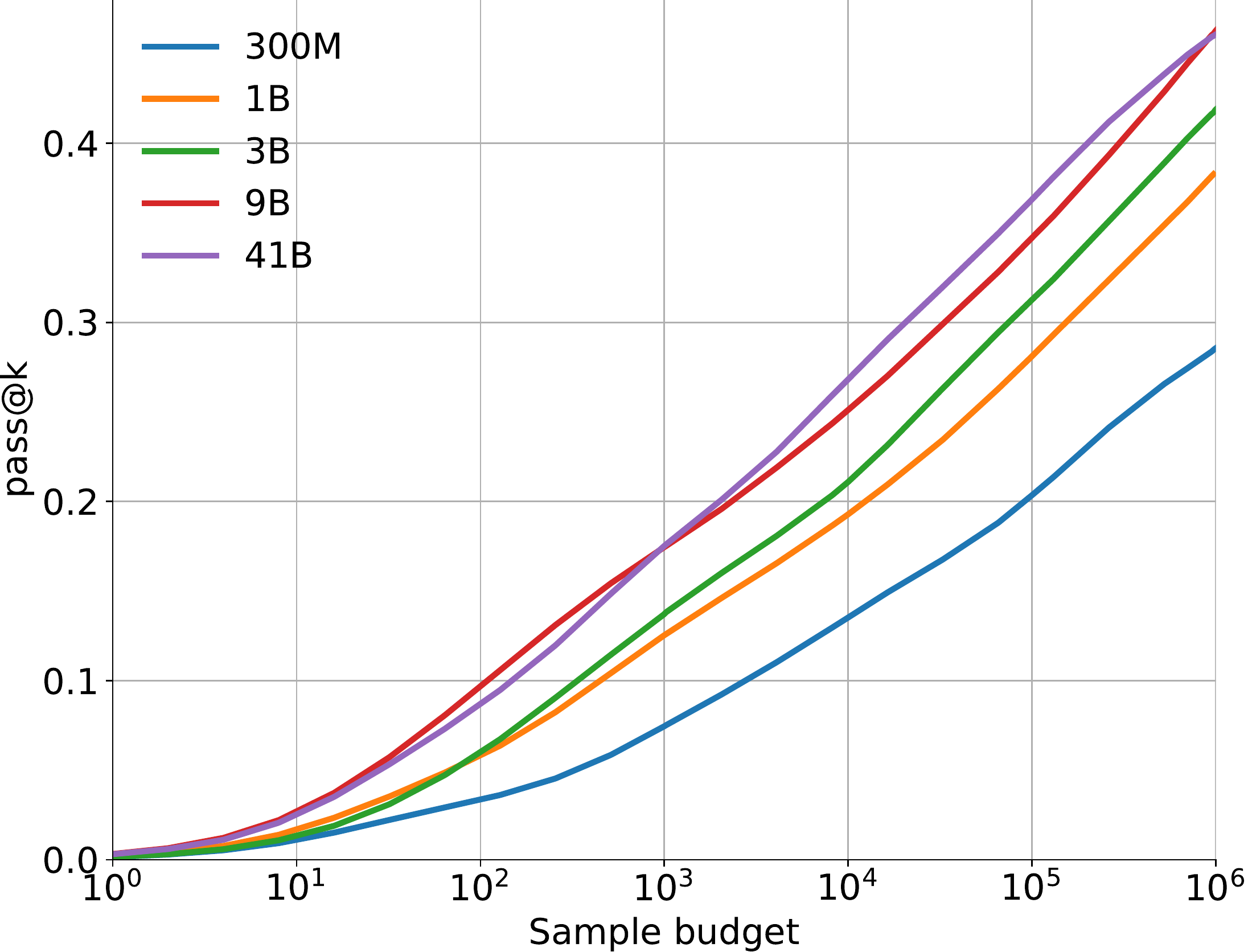} \\
        (a) 10 attempts per problem & (b) Unlimited attempts per problem
    \end{tabular}
    \caption{\textbf{Solve rate scaling vs. number of samples}. The solve rate scales approximately log-linearly with the number of samples, although this tapers off slightly in the $\mathbf{10@k}$ setting. The better, larger-parameter models have higher scaling slopes in this log-linear plot.  %
    }
    \label{fig:sample-scaling-curves}
\end{figure}

\begin{figure}[t]
    \centering
    \begin{tabular}{cc}
        \includegraphics[width=0.48\textwidth]{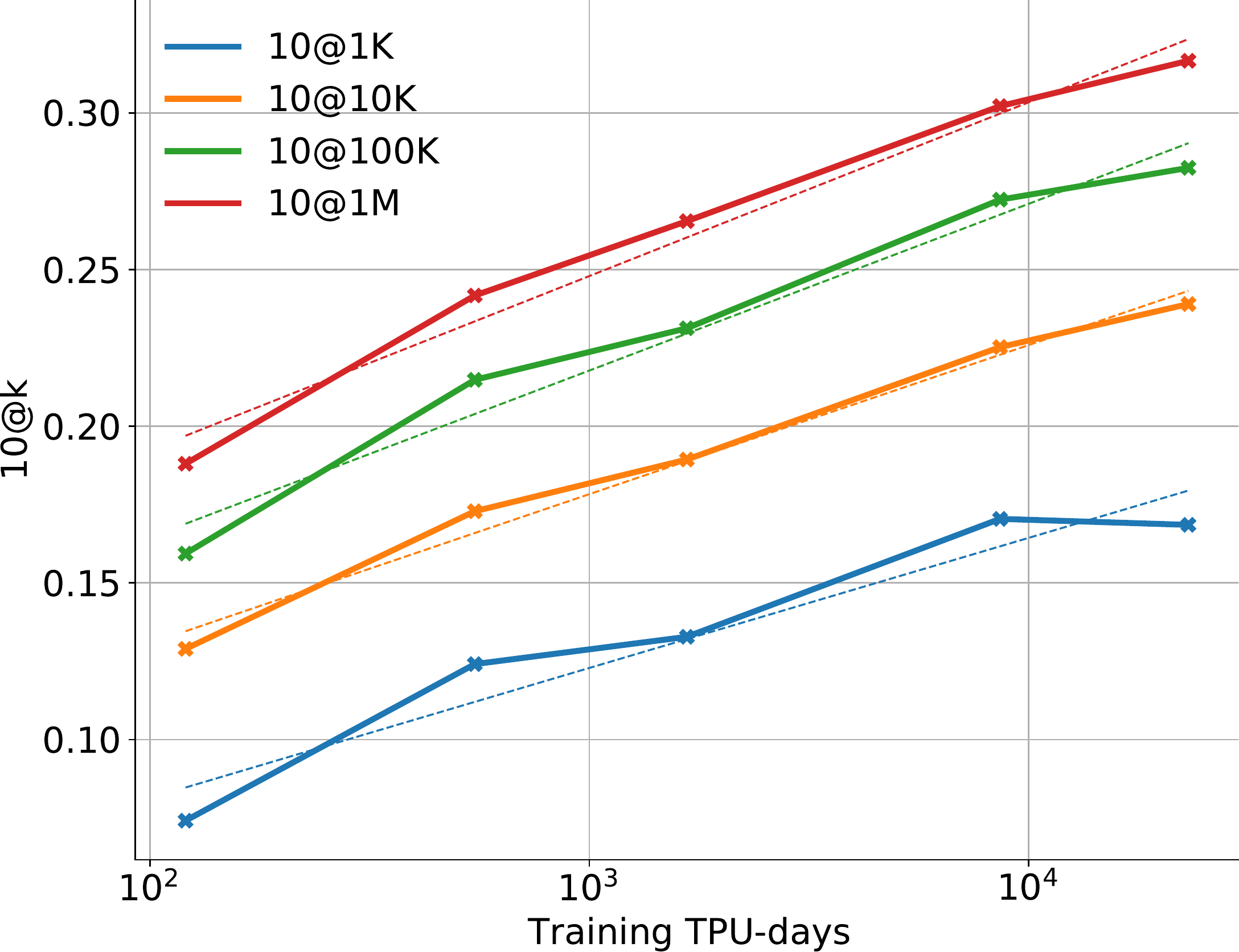} &
        \includegraphics[width=0.48\textwidth]{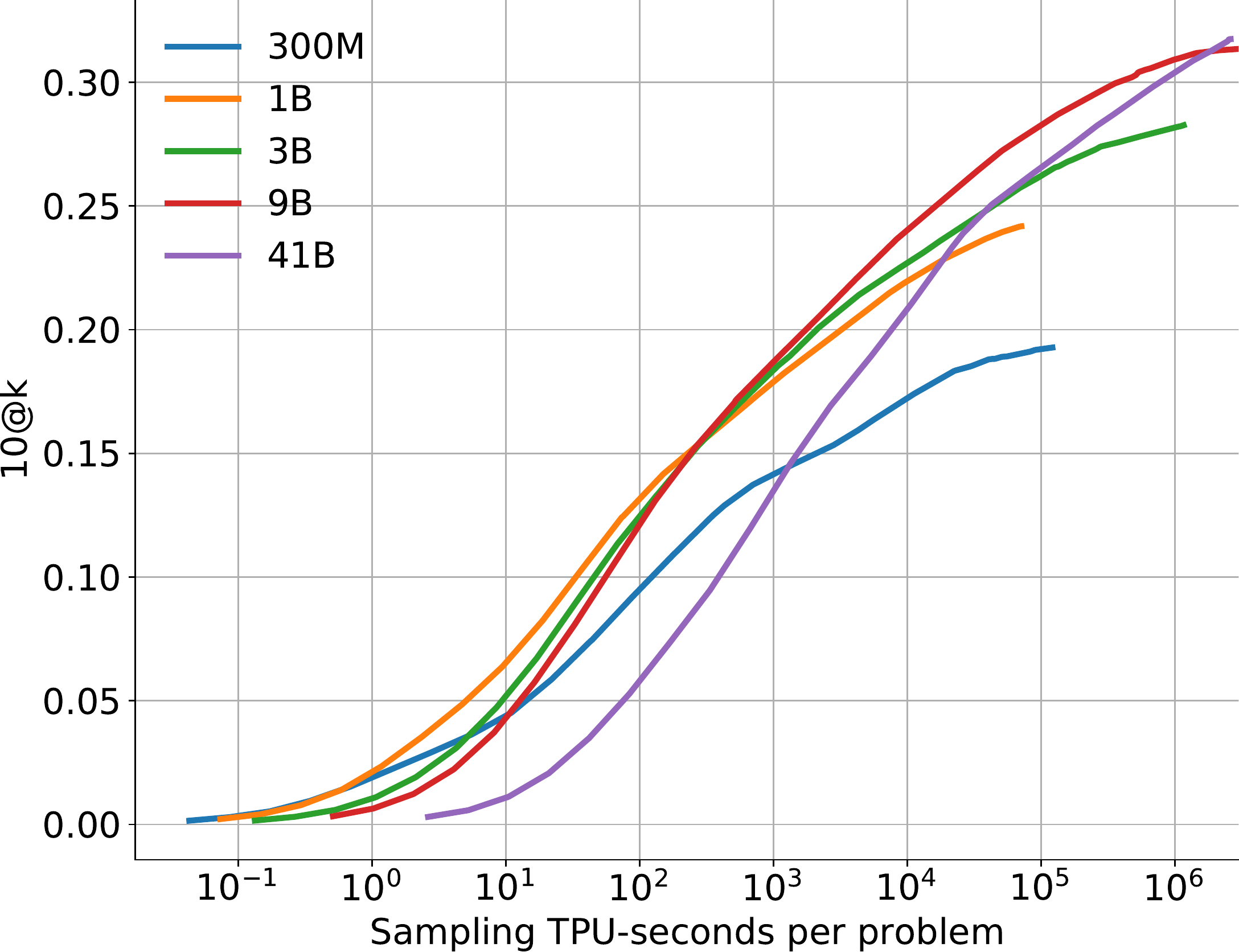} \\
        (a) Solve Rate vs. Training Compute & (b) Solve Rate vs. Sampling Compute
    \end{tabular}
    \caption{\textbf{Solve rate scaling vs. Compute}. The solve rate scales approximately log-linearly with the training compute when we choose model sizes close to optimal for each compute allocation.  Similarly, as we increase the amount of compute we use for sampling the optimal model size increases. %
    }
    \label{fig:sample-scaling-curves-2}
\end{figure}

\figref{fig:sample-scaling-curves} shows how the model performance scales on the $10@k$ and $pass@k$ metrics with more samples, i.e.~as we increase $k$. The difference between the two metrics highlights the importance of selecting which samples to submit.  
\figref{fig:sample-scaling-curves-2} shows how performance scales with the amount of compute used for training and for sampling.
These scaling curves highlight a few interesting facts about this problem domain and our models:

\textbf{Solve rates scale log-linearly with more samples.~~}
Both the 10@k and pass@k solve rates scale %
approximately log-linearly with $k$, with the 10@k curve bending down slightly at high sample budgets.  %
The fact that sampling significantly more than 10 still improves the 10@k solve rate shows how important it is to sufficiently explore the search space before committing to the final 10 submissions per problem. 
However, %
improving solve rate requires exponentially increasing amounts of samples and the costs quickly become prohibitive.

\textbf{Better models have higher slopes in the scaling curve.~~}
Another observation from \figref{fig:sample-scaling-curves} is that larger models tend to have better model quality, reflected as better solve rate with the same number of samples and higher slope in this log-linear scaling curve.  Because of log-linear scaling, a better model with a higher slope can reach the same solve rate with exponentially fewer samples than worse models. This points to improving model quality as an effective way to counter the exponential explosion of sample budget required to reach a higher solve rate.

\textbf{Solve rates scale log-linearly with more compute.~~} As shown in \figref{fig:sample-scaling-curves-2}(a), the solve rate also scales approximately log-linearly with more training compute.  Each point on the curves corresponds to one model size. \figref{fig:sample-scaling-curves-2}(b) shows how solve rate scales with sampling compute, and highlights that larger models take more compute to draw each sample, but they eventually outperform smaller models even with the same sampling compute as the better quality of samples from the larger models become the dominant factor for performance.  These results present an interesting trade-off between how much of the available compute should be used to train a model compared to sampling from it. Both ways of leveraging more compute demonstrate log-linear scaling.

\subsubsection{Architecture changes to improve sampling speed\label{sec:architecture}}
Because drawing more samples is important for improving performance, architecture changes that increase sampling speed would also increase the overall solve rate within a certain compute budget. Therefore, we made two architecture decisions: using (1) an encoder-decoder architecture with asymmetric encoder and decoder structures and (2) the multi-query attention setup from \citet{shazeer2019fast} which uses one shared attention head for keys and one for values each block.

\begin{table}[t]
    \centering
    \begin{small}
    \begin{tabular}{crrrrccccc}
    \toprule
        & \multicolumn{2}{c}{Blocks} & \multicolumn{2}{c}{Seq. length} & Hidden & Fan-Out & & Samples / & \\
    Model & Enc. & Dec. & Enc. & Dec. & Size & Ratio & Params & TPU sec & 10@10K \\
    \midrule
        \OurApproachShort{} model & 5 & 30 & 1536 & 768  & 1408 & 6   & 1.15B & 4.74 & 17.3\% \\
        Decoder-only              & - & 40 & -    & 2304 & 1408 & 6   & 1.17B & 1.23 & 18.5\% \\
        Std MH attention          & 5 & 30 & 1536 & 768  & 1408 & 4.3 & 1.16B & 0.37 & 17.0\% \\
        \bottomrule
    \end{tabular}
    \end{small}
    \caption{\textbf{Architecture comparison}. Architecture changes increase sampling speed without significantly impacting the solve rate. %
    }
    \label{tab:architecture-ablations}
\end{table}

To investigate the effects of these decisions, we compared our base 1B parameter model against the two alternatives that remove each of the changes.  We pre-trained and fine-tuned the standard multi-head attention model in exactly the same way as our base 1B model. %
The decoder-only model was trained with the same amount of compute.  However, due to the significantly longer decoder sequence length (2304 tokens), with the same amount of training compute it consumes 
50\% more loss tokens than training the encoder-decoder models. %

\tabref{tab:architecture-ablations} shows that our encoder-decoder model with multi-query attention significantly improves the sampling speed while keeping the sample quality at the same level as the more expensive alternatives. %

\subsubsection{Choice of the pre-training dataset}

\tabref{tab:pretraining-dataset} compares our base 1B model trained on our full GitHub dataset with equivalent models that are pretrained on (1) the Python-only portion of GitHub, (2) the MassiveText generic text dataset \citep{rae2021scaling} which also includes a portion of GitHub or (3) not pre-trained at all.  The pre-trained models are then fine-tuned and sampled in exactly the same way, except that the model pre-trained on Python-only data is also fine-tuned on Python-only data and only samples Python solutions. %

As \tabref{tab:pretraining-dataset} shows, pre-training on the full GitHub dataset with all languages leads to significantly better results than pre-training either on Python alone, or on the MassiveText dataset that mostly consists of natural language text.  Any pre-training significantly improves the results over training from scratch on \OurDatasetShort{}.

\begin{table}[th]
    \centering
    \begin{tabular}{ccccc}
    \toprule
    \multirow{2}{*}{Pre-training dataset} & \multicolumn{3}{c}{Solve rate} \\
     & 10@1K & 10@10K & 10@100K \\
    \midrule
        
         No pre-training & 4.5\% & 7.0\% & 10.5\% \\
         GitHub (Python only) & 5.8\% & 11.1\% & 15.5\% \\
         MassiveText & 9.7\% & 16.1\% & 21.2\% \\
         GitHub (all languages) & 12.4\% & 17.3\% & 21.5\% \\
        \bottomrule
    \end{tabular}
    \caption{\textbf{Model solve rate with different pre-training settings and datasets}.
    }
    \label{tab:pretraining-dataset}
\end{table}

\subsubsection{Model enhancements}

As discussed in \secref{sec:approach}, we adopted training and model enhancements which significantly improved the solve rate relative to the standard encoder-decoder transformer setup.  \tabref{tab:fine-tuning-ablations} presents the results of a build-up ablation of the enhancements we added to \OurApproachShort{}, starting from the base setting with no enhancements (beyond the multi-query attention change discussed in \secref{sec:architecture}). We added one new setting at a time, with the final setting that corresponds to \OurApproachShort{} reported at the bottom of the table. Each additional setting improves performance and combining the 5 enhancements together increases the 10@100k solve rate from 15.2\% to 24.1\%, although the contribution depends on the number of samples.

\begin{table}[th]
    \centering
    \begin{small}
    \begin{tabular}{lcccc}
    \toprule
    & \multicolumn{4}{c}{Solve rate} \\
    Fine-tuning setting &  \multicolumn{1}{c}{10@1K} &  \multicolumn{1}{c}{10@10K} &  \multicolumn{1}{c}{10@100K} & \multicolumn{1}{c}{10@1M} \\
    \midrule
    No Enhancements &     6.7\% (6.5-6.8) &   10.4\% (9.6-11.0) &  15.2\% (14.3-15.9) &  19.6\% (18.2-20.4) \\
            \ + MLM &     6.6\% (6.2-7.0) &  12.5\% (12.1-12.7) &  17.0\% (16.5-17.2) &  20.7\% (19.1-21.3) \\
    \   + Tempering &     7.7\% (7.2-8.5) &  13.3\% (12.5-13.8) &  18.7\% (18.0-19.2) &  21.9\% (20.7-22.6) \\
    \ + Tags and Ratings &     6.8\% (6.4-7.0) &  13.7\% (12.8-14.9) &  19.3\% (18.1-20.0) &  22.4\% (21.3-23.0) \\
        \ + Value &   10.6\% (9.8-11.1) &  16.6\% (16.4-16.9) &  20.2\% (19.6-20.7) &  23.2\% (21.7-23.9) \\
            \ + GOLD &  12.4\% (12.0-13.0) &  17.3\% (16.9-17.6) &  21.5\% (20.5-22.2) &  24.2\% (23.1-24.4) \\
       \ + Clustering &  12.2\% (10.8-13.4) &  18.0\% (17.3-18.8) &  24.1\% (23.2-25.0) &  28.4\% (27.5-29.3) \\
    \bottomrule
    \end{tabular}
    \end{small}
    \caption{\textbf{Build-up ablation for model enhancements}. Effect of each additional model enhancement building up from {\em No enhancements} which is a plain fine-tuned 1B encoder-decoder model trained with the standard next token prediction loss. Numbers in parentheses represent 95\% confidence intervals.  For each setting we fine-tuned and sampled from at least 3 different models from the same pre-trained checkpoint, and computed means and confidence intervals using a combination of subsampling and bootstrapping as discussed in~\appendixref{appendix:evaluation-metrics}.%
    \label{tab:fine-tuning-ablations}}
\end{table}

\subsubsection{Filtering \& clustering}

To solve problems within a realistic evaluation budget, we rely on filtering and clustering to select a small number of samples to evaluate from the large amount of model samples we generate.

\textbf{Filtering using example tests.~~}
\tabref{tab:filtering-with-public-tests} shows the percentage of model samples that pass example tests and the percentage of problems where at least one sample passes example tests.  Note that these percentages are calculated based on the full set of samples, without first filtering out programs that have syntax errors (see \secref{sec:solution-characteristics} for more on syntactic correctness of the samples). Overall less than 1\% of samples from our models pass example tests, though the percentage varies greatly across problems, which means that filtering using example tests removes more than 99\% of the model samples.  On problems where our models do find a correct solution, the fraction of samples that pass example tests roughly doubles but still remains at a low level.  The non-uniform distribution of $p_\text{pass example tests}$ across problems is highlighted more in \appendixref{appendix:filtering-clustering}.

Another observation from \tabref{tab:filtering-with-public-tests} is that larger and better quality models produce samples more likely to pass example tests, and pass example tests for significantly more problems.  With $10^6$ samples, our largest 41B models can generate solutions that pass example tests for over 90\% of problems, a remarkable success as finding programs that satisfy I/O example constraints remains a very challenging problem.

\begin{table}[t]
    \centering
    \begin{tabular}{cccc}
    \toprule
    & \% Problems with $\ge 1$ & Average $p_\text{pass example tests}$ & Average $p_\text{pass example tests}$ \\
        Model & samples pass example tests & on all problems & on solved problems   \\
        \midrule
         300M & 82.05\% & 0.39\% & 1.18\% \\
         1B & 87.18\% & 0.59\% & 1.40\% \\
         3B & 87.18\% & 0.49\% & 0.98\% \\
         9B & 89.74\% & 0.76\% & 1.52\% \\
         41B & 92.31\% & 0.73\% & 1.47\% \\
        \bottomrule
    \end{tabular}
    \caption{\textbf{Example test statistics}. Example tests help us filter out more than 99\% of model samples, and as models get better with larger scales, they are more likely to find samples that pass example tests for more problems.  One million samples were drawn per problem from each model.
    }
    \label{tab:filtering-with-public-tests}
\end{table}

\textbf{Clustering.~~}
A solution has to pass hidden tests in addition to example tests, so we must further select correct samples from those that pass all public tests. Filtering 99\% of a million samples still leaves thousands of samples per problem to select from. We cluster the remaining samples based on their behaviour on generated test inputs, to make the most of the evaluation budget. 

\begin{figure}
    \centering
    \includegraphics[width=0.5\textwidth]{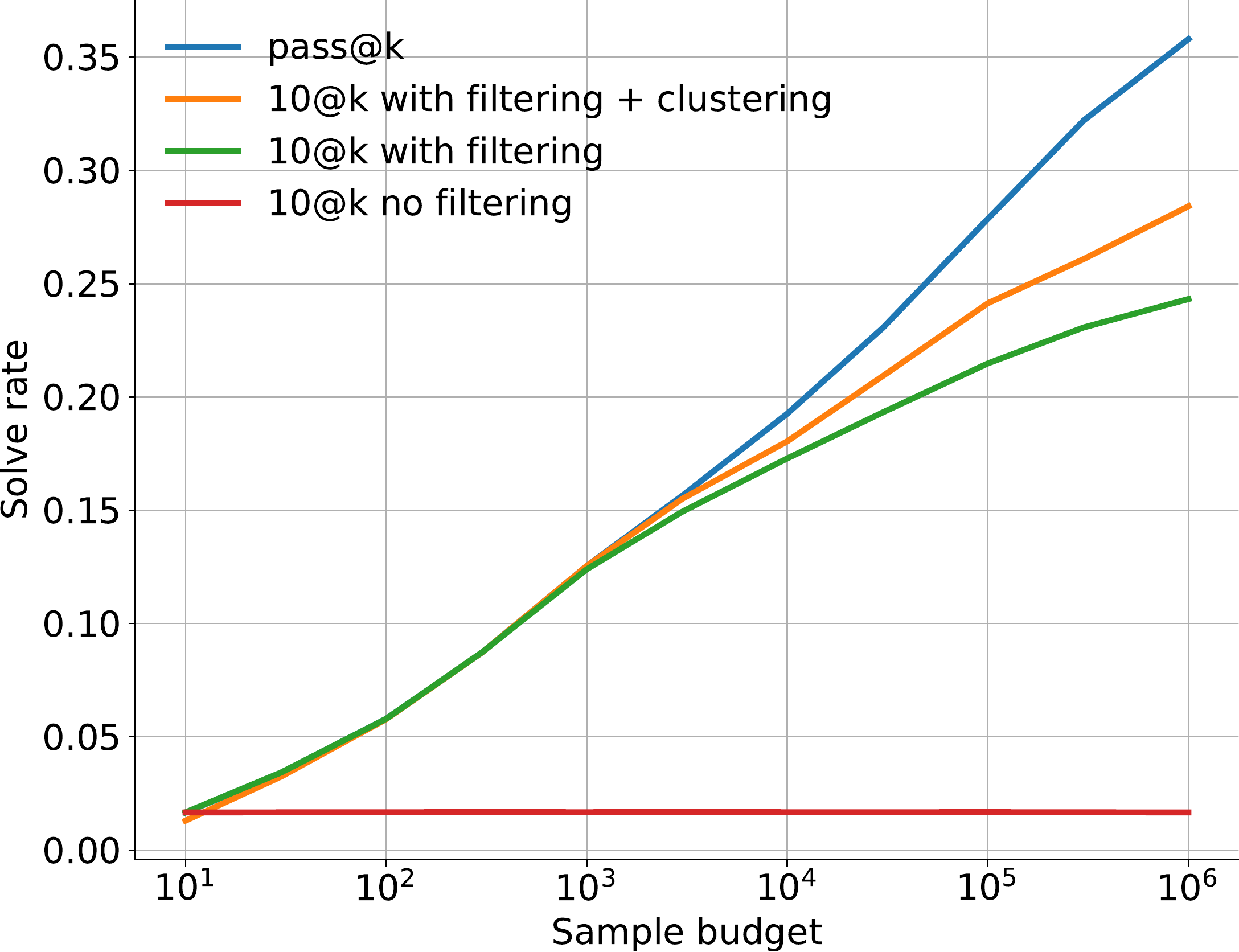}
    \caption{\textbf{Comparison of different sample selection methods}. We show random selection (``10@k no filtering''), filtering using example tests (``10@k with filtering''), clustering after filtering (``10@k with filtering + clustering''), and perfect sample selection (``pass@k''). %
    }
    \label{fig:compare-filtering-clustering}
\end{figure}

\figref{fig:compare-filtering-clustering} shows a comparison between (i) randomly picking model samples without filtering, (ii) filtering and then randomly selecting from the filtered samples, (iii) filtering and then using clustering to select samples, and (iv) allowing unlimited evaluation attempts, which gives us the upper bound performance attainable with a perfect sample selection method.  Filtering and clustering clearly enable scaling, as otherwise the solve rate remains flat. However there is still a large gap between them and the theoretical upper bound.

\subsection{Results on APPS}\label{sec:public-benchmark-results}

In addition to evaluating on Codeforces competitions and \OurDatasetShort{}, we performed evaluations on the previously published APPS benchmark to directly compare to previous work. The APPS dataset \citep{hendrycks2021measuring} contains a total of 10,000 programming problems divided equally between training and test sets. Because of missing information in the dataset, we could not apply our full method. We therefore followed the settings we used for pre-training on GitHub, and fine-tuned our pre-trained models on the APPS training set without using clustering, tags, ratings, value conditioning, or prediction, and with sampling temperature 0.25 and nucleus sampling. Other settings were the same as our main models.

\tabref{tab:apps} compares our model with existing large language models fine-tuned on this dataset as reported by \cite{hendrycks2021measuring}, as well as the 1-shot performance of the Codex model reported by \cite{chen2021codex}.  A small 1B parameter model already outperforms the GPT-NEO baseline on all difficulty levels, and outperforms Codex 12B on the interview and competition difficulty levels. We highlight that \OurApproachShort{} still improves when increasing the number of samples per problem, showing support for our claim of the importance of large scale sampling. Differences in performance between APPS results and \OurDatasetShort{} could be attributed to dataset quality (e.g. the high APPS false positive rate shown in \secref{sec:false-positives}), dataset size, missing components of \OurApproachShort{}, and tuning for the problem distribution.

\begin{table}[t]
\begin{center}
\begin{tabularx}{\textwidth}{l c r c c c}
\toprule
& Filtered From ($k$) & Attempts ($n$) & Introductory & Interview & Competition \\
& & & n@k & n@k & n@k\\
\midrule
GPT-Neo 2.7B & N/A & 1 & 3.90\% & 0.57\% & 0.00\% \\
GPT-Neo 2.7B & N/A & 5 & 5.50\% & 0.80\% & 0.00\% \\
\midrule
Codex 12B & N/A & 1 & 4.14\% & 0.14\% & 0.02\% \\
Codex 12B & N/A & 5 & 9.65\% & 0.51\%  & 0.09\% \\
Codex 12B & N/A & 1000 & 25.02\% & 3.70\%  &  3.23\% \\
\midrule
Codex 12B & 1000 & 1 & 22.78\% & 2.64\% & 3.04\% \\
Codex 12B & 1000 & 5 & 24.52\% & 3.23\% & 3.08\% \\
\midrule
\OurApproachShort{} 1B & N/A & 1000 & 17.67\% & 5.24\% & 7.06\% \\
\midrule
\OurApproachShort{} 1B & 1000 & 5 & 14.36\% & 5.63\% & 4.58\% \\
\OurApproachShort{} 1B & 10000 & 5 & 18.18\% & 8.21\% & 6.65\% \\
\OurApproachShort{} 1B & 50000 & 5 & 20.36\% & 9.66\% & 7.75\% \\
\bottomrule
\end{tabularx}
\end{center}
\vspace{-0.5cm}
\caption{
\label{tab:apps}
\textbf{n@k results on APPS}. If there is no filtering, then $n=k$ and the metric is pass@k. Finetuned GPT-Neo numbers reported from  \cite{hendrycks2021measuring}, Codex numbers from \cite{chen2021codex}.  %
We used a time limit of 3 seconds per test to match Codex 12B, and report average numbers over 3 different fine-tuning runs for \OurApproachShort{}. Note that this does not include all components described in \secref{sec:approach}, and does not use the \OurDatasetShort{} dataset.
}
\end{table}

\section{\OurApproachShort{}'s capabilities \& limitations}\label{sec:limitations}

We performed a detailed analysis of the capabilities and limitations of our models. In particular, we find that our models are not simply copying from the training set (\secref{sec:lcs}) and our models are sensitive to various changes in the problem descriptions and metadata used for conditioning (\secref{sec:sensitivity-to-rewordings} and \ref{sec:sensitivity-to-provided-metadata}), both of which indicate that we are not solving problems by exploiting obvious weaknesses in the task structure.  

We also analyze the characteristics of the solutions the model finds, for syntactic correctness, dead code, and the types of problems it can solve (\secref{sec:solution-characteristics}). We further show that using validation loss as a proxy for model performance has several issues (\secref{sec:validation-solve-rate}).  More analysis of our model and approach are included in \appendixref{appendix:capabilities-and-limitations}, and an attention visualization %
as well as example problems and solutions generated by the model can be found at \url{https://alphacode.deepmind.com/}.  All analysis results are reported without clustering unless otherwise noted. 

\subsection{Copying from training data}
\label{sec:lcs}
A commonly raised concern for large language models trained on large amounts of data is that they may solve downstream problems by simply memorising the training set (e.g. \cite{ziegler2022research,carlini2021extracting}). For competitive programming to be a good test of problem-solving ability we expect that models need to come up with novel solutions to solve new problems. 

\begin{figure}[t]
\begin{center}

\includegraphics[width=0.9\textwidth]{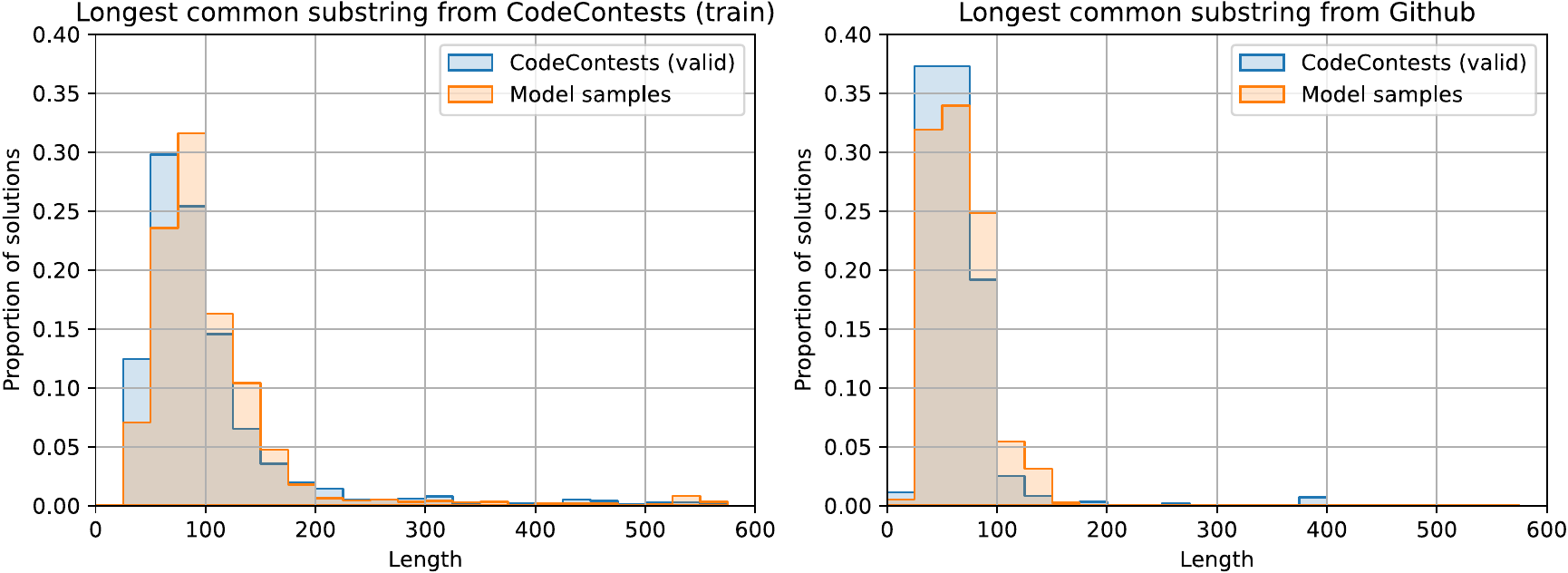} 
\end{center}
\vspace{-0.8cm}
\caption[]{
\label{fig:lcs_hists}
\textbf{Length of longest common substrings between solutions and training data.} The human solution distribution is in blue and the model is in orange, and the results are compared against the GitHub and \OurDatasetShort{} datasets. Model and human solutions have similar distributions. On \OurDatasetShort{}, approximately 3\% of human solutions and less than 1\% of model solutions had a common substring of length greater than 600. %
}
\end{figure}

Based on the results in \appendixref{sec:dataset-temporal-split}, simply copying full solutions from the training set is not sufficient to solve any problems in the unseen validation set. %
However, it might be possible to solve problems by duplicating large or critical parts of previous solutions, if problems are sufficiently similar to previous ones. To investigate this, we found the longest common substrings between correct validation problem solutions generated by the model and the entire training dataset (GitHub + \OurDatasetShort{}, ignoring whitespace), and compared the distribution of the lengths of these matches to human solutions. \figref{fig:lcs_hists} contains these results, using 50 \CC{} and 50 Python solutions from a selection of 16 validation problems that had that many solutions. 

The figure shows that model and human solutions share substrings with the training data at similar rates, although the average longest common substring is slightly higher for model solutions, and human solutions have a heavier tail.
The common substrings between model solutions and training data mostly contained boilerplate code for reading and parsing the input data format, rather than key logic for solving problems
(for example, a FastIO Python class has length 805 and appears in 1.66\% of all human Python solutions). AlphaCode thus does not seem to solve problems by copying long blocks of code.

\begin{figure}[tp]
\begin{center}

\begin{tabular}{ll}
    Model-generated solution & Source document of LCS \\
    \\
    \includegraphics[width=0.42\textwidth]{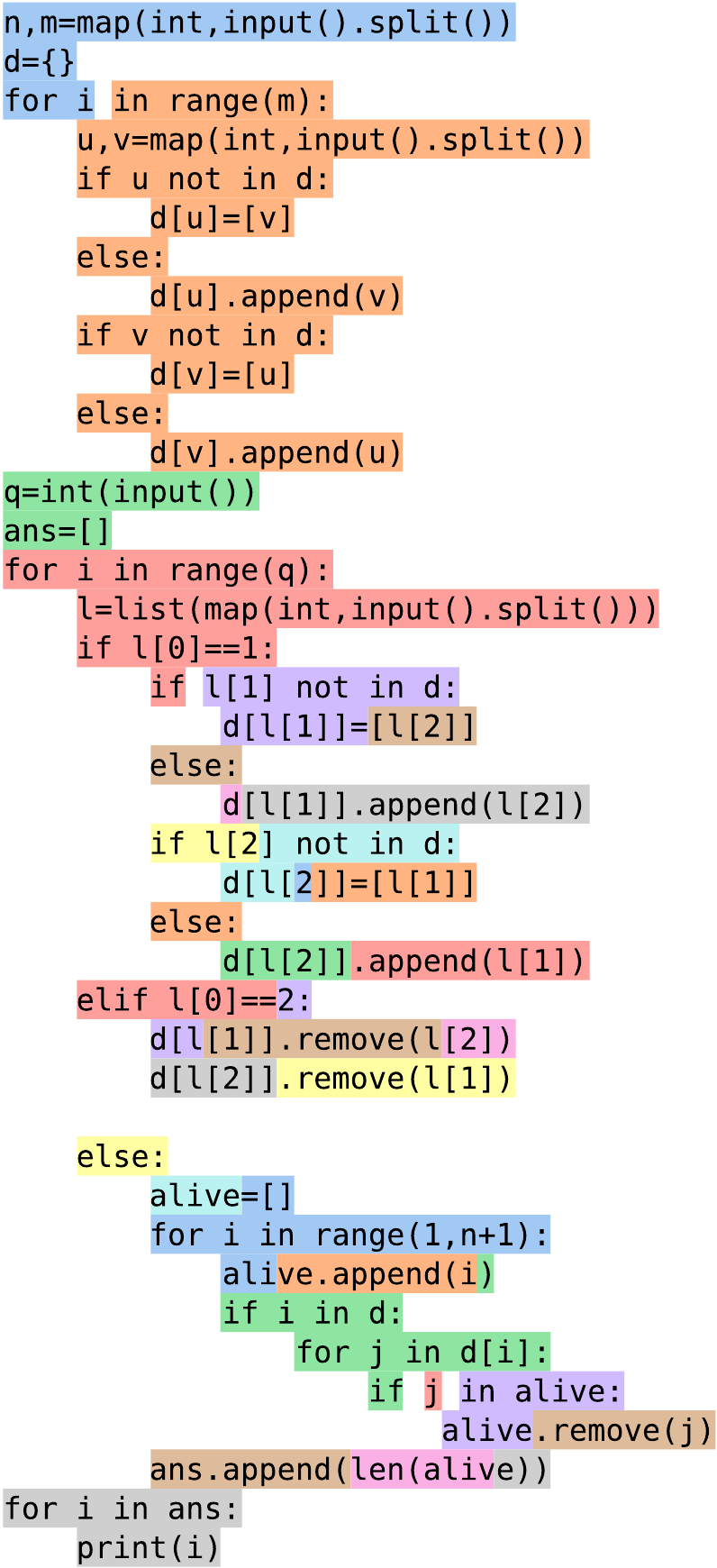} &
    \includegraphics[width=0.45\textwidth]{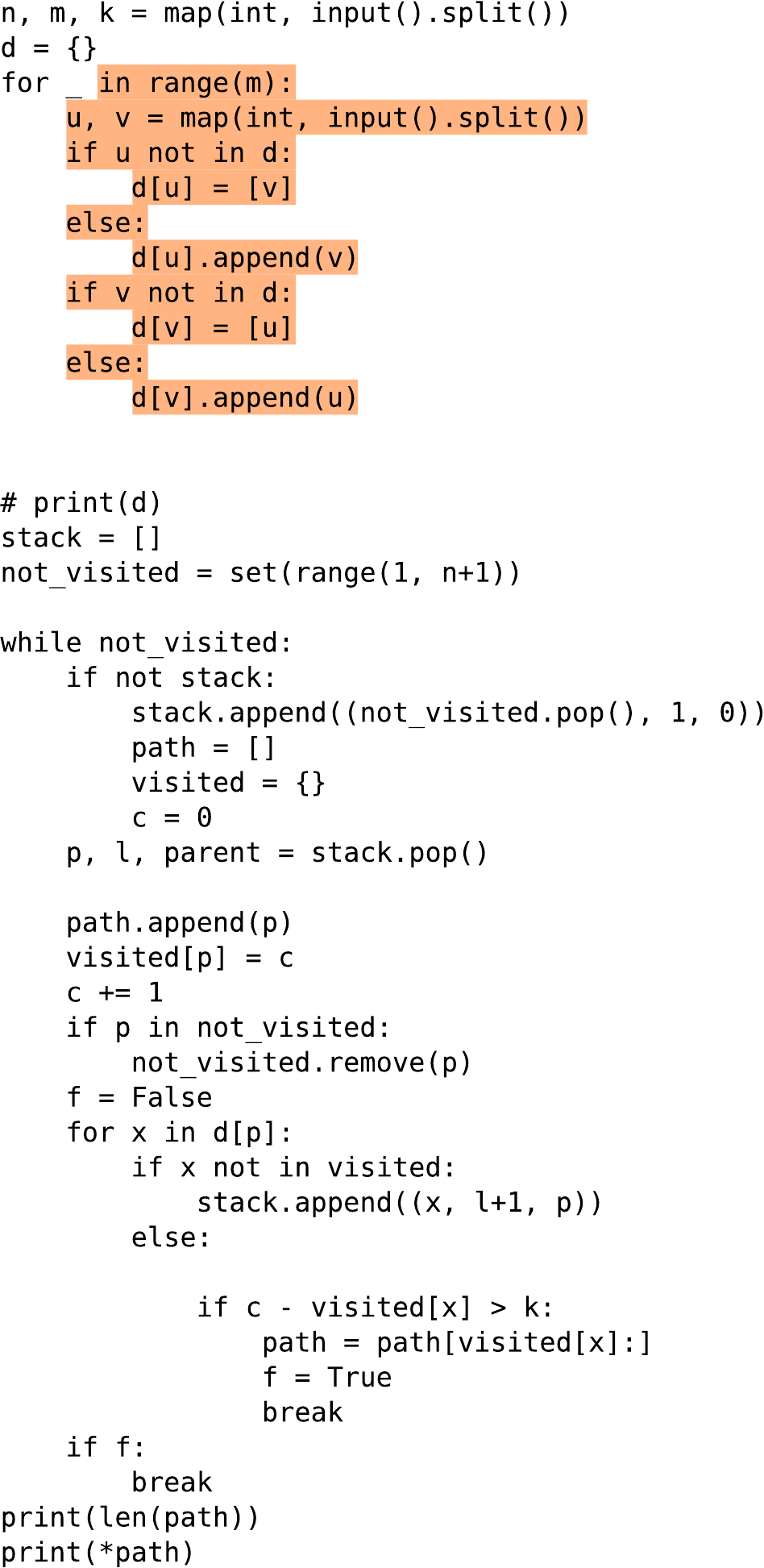}
\end{tabular}

\caption{\textbf{Left}: Example decomposition of model solution into substrings from the fine-tuning dataset. Each new color denotes a new substring. \textbf{Right:} Document that shares the longest substring (183 chars). This solution doesn't duplicate core logic. However, it does share snippets that parse input and construct graph adjacency lists. Document sourced from \href{https://codeforces.com/}{Codeforces}. 
}
\label{fig:lcs_decomposition}
\end{center}
\end{figure}

To investigate beyond just the longest common substrings between the solution and the training data, we performed a more detailed qualitative analysis of 50
model-generated solutions. We took each solution and iteratively removed the longest common substring from it, creating a partitioning of each solution consisting of substrings from the finetuning training data. \figref{fig:lcs_decomposition} shows an example.
We found no evidence that our model copies core logic from the training data. Further examples are provided in \appendixref{appendix:verbatim-copying}.

\subsection{Model solution characteristics}\label{sec:solution-characteristics}

We measured the proportion of samples from the model that are syntactically correct (i.e.~compile for \CC{}, and do not generate a SyntaxError for Python) for each language and model size. 
As shown in \tabref{tab:syntax}, our models tend to produce mostly syntactically correct programs for Python, and \CC{} syntax is harder to master than Python. %

\begin{figure}[t]
\includegraphics[width=\textwidth]{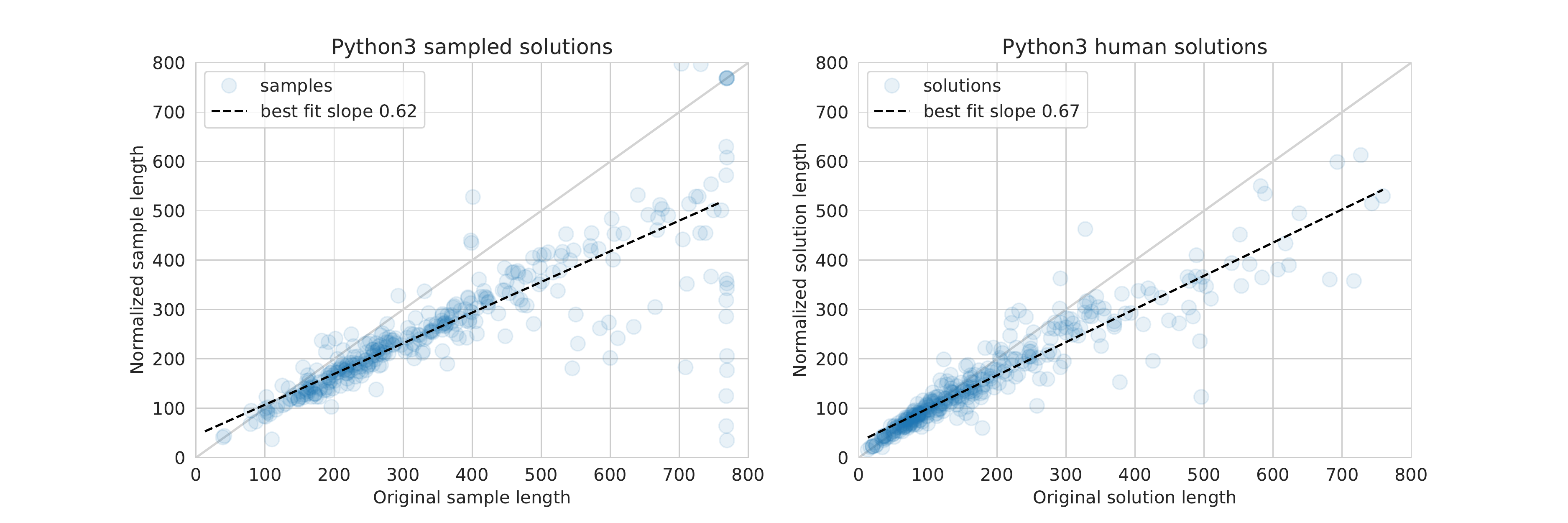} 
\caption[]{
\label{fig:deadcode}
\textbf{Sample and human solution lengths in tokens before and after dead code removal}. Note that our samples contain a maximum of 768 tokens, and standard formatting can sometimes make the code longer.
}
\end{figure}

We further analysed the amount of dead code (i.e.~lines of code that have no impact on program behaviour) in our solutions. Such code is present in our training data; for example competitors will sometimes copy-paste unused imports or functions. High amounts of dead code could indicate the model has a poor understanding of what it is generating.
\figref{fig:deadcode} shows the results of applying a standard code formatter and Python's Abstract Syntax Tree (ast) module on correct Python human and model solutions to remove unused imports, functions and classes. \OurApproachShort{} generates approximately the same amount of dead code as humans.

\begin{table}[tb]
\begingroup
\setlength{\tabcolsep}{2.5pt}
    \centering
    \begin{tabular}{ccccccccccc}
\toprule
 \multirow{2}{*}{Model} &
 \multirow{2}{*}{Greedy} &
 \multirow{2}{*}{Math} &
 \multirow{2}{*}{DP} &
 Constructive &
 Brute &
 Data &
 Implem- &
 \multirow{2}{*}{Graphs} &
 \multirow{2}{*}{Bitmasks} &
 \multirow{2}{*}{Sortings} \\
 & & & & Algorithms & Force & Structures & entation & & & \\
\midrule
 
 300M & 13.1\% & 19.3\% & 4.5\% & 7.5\% & 9.8\% & 8.8\% & 5.0\% & 0.2\% & 22.2\% & 16.9\% \\
 1B & 19.7\% & 22.7\% & 4.5\% & 9.1\% & 12.0\% & 10.5\% & 14.1\% & 5.9\% & 26.8\% & 21.5\% \\
 3B & 19.9\% & 22.7\% & 4.9\% & 11.2\% & 13.2\% & 11.9\% & 13.4\% & 8.8\% & 25.4\% & 23.8\% \\
 9B & 23.7\% & 29.4\% & 7.1\% & 13.8\% & 19.5\% & 16.9\% & 16.4\% & 16.6\% & 27.4\% & 27.8\% \\
 41B & 25.0\% & 28.2\% & 8.8\% & 14.9\% & 20.4\% & 15.7\% & 16.5\% & 13.6\% & 33.8\% & 25.5\% \\
\bottomrule
\end{tabular}
    \caption{\textbf{Solve rate (10@10k) for the 10 most common problem types at different model sizes.} The tags are ordered by popularity, with ``Greedy'' as the most popular.}
    \label{tab:solve-rate-across-tags}
\endgroup
\end{table}

\tabref{tab:solve-rate-across-tags} shows our models' solve rates across different problem types specified using tags for each problem.  Notably the solve rates across all tags overall go up as models improve with larger scales.  Our models are relatively better at problems that deal with bitmasks, sorting, maths, and greedy algorithms, but notably worse at dynamic programming (DP) and constructive algorithms.

\appendixref{appendix:solve-rate-for-different-ratings} contains the solve rate of our models in different problem difficulty rating buckets.  Unsurprisingly our models have significantly higher overall solve rates for easier problems, and lower solve rates for harder problems. %

\subsection{Sensitivity to problem descriptions}\label{sec:sensitivity-to-rewordings}
We performed a detailed analysis of our models' sensitivity to the problem description in order to measure the importance of the description to the model performance. Overall, \OurApproachShort{} seems to respond appropriately to important changes in the problem description and makes use of extra information available in it. This indicates that, for example, the system does not ignore most of the description and brute force every possible solution that fits the category of problem (e.g. algorithms related to prime numbers if the word "prime" is mentioned in the description).

\begin{table}[t]
\begingroup
\setlength{\tabcolsep}{4pt}
    \centering
    \begin{tabular}{lllr}
    \toprule
    & & \multicolumn{2}{r}{\% correct} \\
    
    \midrule
    \multirow{2}{*}{\textbf{(a)} Description simplification}
    &\multirow{2}{*}{\tabref{tab:rewording-solves}} & Original & 3.0\% \\
    & & Simplified & 15.7\% \\
    
    \midrule
    
    \multirow{6}{*}{\textbf{(b)} Description rewording}
    & \multirow{6}{*}{\tabref{tab:rewording-simplified}} & Original & 17.1\% \\
    & & Opposite & 0.1\% \\
    & & Related & 3.2\% \\
    & & Underspecified & 0.03\% \\
    & & Verbose & 19.4\% \\
    & & Algorithm described in words only & 19.7\% \\
    
    \midrule
    \midrule
    & & \multicolumn{2}{r}{10@1024} \\

    \midrule
    
    & & Original & 13.5\% \\
    
    \midrule
    
    \multirow{2}{*}{\textbf{(c)} Variable renaming}
    & \multirow{2}{*}{\figref{fig:variable-renaming-scaled}} & $\leq 6$ variables consistently renamed & 12.1\% \\
    & & $\leq 6$ variables inconsistently renamed & 10.1\% \\
    
    \midrule
    
    \textbf{(d)} Type information & \tabref{tab:type-removal} & No type information & 13.3\% \\
    
    \midrule
    
    \textbf{(e)} Typos & \figref{fig:word-level-changes} & 30 typos & 11.3\% \\
    
    \midrule
    
    \multirow{3}{*}{\textbf{(f)} Missing description sections}
    & \multirow{3}{*}{\tabref{tab:prompt_ablation}} & Description + Specification & 10.4\% \\
    & & Description + IO & 4.8\% \\
    & & Specification + IO & 6.9\% \\
    
    \midrule
    
    \textbf{(g)} Synonym substitution &  \figref{fig:word-level-changes} & 7 synonyms substituted & 12.5\% \\
    
    \midrule
    
    \multirow{2}{*}{\textbf{(h)} Word permutation and deletion}
    & \multirow{2}{*}{\figref{fig:word-level-changes}} & Distance 7 permuted words & 8.0\% \\
    & & 0.40 probability of deletion & 6.7\% \\
    \bottomrule
    \end{tabular}
    \vspace{-0.2cm}
    \caption{\textbf{Summary of solve rates in response to description changes}. Solve rate deteriorates consistently when removing or obscuring information in the description, demonstrating that the model relies on this information. All solve rates are for the 1B model; the model was not retrained for description changes.  Some of the studies were done on the full validation set, while others were done on selected subsets. %
    \textbf{(a)} and \textbf{(b)} use the percentage of correct samples on selected problems, and \textbf{(c-h)} use 10@1024 solve rates.}
    \label{tab:rewordings-summary}
\endgroup
\end{table}

Full details of this study are in \appendixref{sec:sensitivity-to-rewordings-appendix}, and a summary of results is shown in \tabref{tab:rewordings-summary}. \textbf{(a)} shows that when given a simplified description of the problem (not available in real evaluations), the model solves it at a much higher rate. \textbf{(b)} shows that the solve rate goes down dramatically when given related but different problems, and is not very affected by different ways of describing the same the problem. \textbf{(c, d, e, g, h)} show that the model is largely unaffected by changes that do not seem significant (like replacing words with synonyms or removing some type details), but responds more to larger changes (like deleting words or making the problem ill-posed). Notably, in \appendixref{sec:appendix-var-relations}, we see that the model deteriorates relatively more in response to lower quality descriptions as model quality increases, indicating that better models are more capable of paying attention to subtle but important description changes. Finally, \textbf{(f)} shows that the model relies on different sections of the description, and particularly on the specification. This makes sense because the specification describes how to read the input, and otherwise the model would have to guess the input format.

\subsection{Sensitivity to provided metadata}\label{sec:sensitivity-to-provided-metadata}
As described in \secref{sec:large-scale-sampling}, at sampling time we provide randomised metadata to \OurApproachShort{} to increase sample diversity. This includes tags (e.g. whether the problem is of type ``binary search'' or ``brute force''), ratings (how difficult the problem is), programming language, and whether or not the solution being generated is correct.

There are two natural questions to explore: does the model respond appropriately to variations in this conditioning metadata, and what are the best settings to generate this metadata with at test time when they are not available. We find that the model is indeed conditioned on this metadata; providing different tags changes what algorithms the model generates. We also find that we should sample randomly for tags and ratings, and condition on the solution being \texttt{CORRECT}.  Here we present an analysis of tag conditioning, and additional results on ratings and correctness conditioning are included in \appendixref{sec:sensitivity-to-metadata-appendix}.

We examined the model's tag conditioning behaviour on an example problem in our validation set: 
Codeforces problem 1549A, \emph{Gregor and Cryptography} (\figref{fig:varname_replacement_example}).  In this problem, we are given a prime number, $P$, and need to find two integers, $a$ and $b$, such that $P \;\mathrm{mod}\; a = P \;\mathrm{mod}\; b$ and $2 \leq a < b \leq P$. It's tempting to solve this via brute-force, but there is a simple number theory solution: $P$ must be odd, so $P \;\mathrm{mod}\; 2$ and  $P \;\mathrm{mod}\; (P-1)$ both equal 1.

\begin{figure}[tp]
\begin{center}
\begin{minipage}[t]{.5\textwidth}
\small
Code generated with tag ``brute force'':
\begin{lstlisting}[language=Python]
t = int(input())

for _ in range(t):
    p = int(input())
    
    for a in range(2, p):
        b = p - a + 1
        if p %
            print(a, b)
            break
\end{lstlisting}
\end{minipage}
\begin{minipage}[t]{.4\textwidth}
\small
Code generated with tag ``number theory'':
\begin{lstlisting}[language=Python]
t = int(input())
while t:
    p = int(input())
    print('2 %
    t-=1
\end{lstlisting}
\end{minipage}
\caption{\textbf{Sensitivity to tags}. The first successful samples when attempting \emph{Gregor and Cryptography} with different tags provided as part of the prompt.}
\label{fig:cryptography-samples}
\end{center}
\end{figure}

We sampled from our model, first with the tag ``brute force'', and then with the tag ``number theory''. These tags changed the sample distribution as demonstrated by the first successful samples in the two sampling runs, shown in \figref{fig:cryptography-samples}. The ``brute force'' approach is just that -- although in reality it is guaranteed to break out of its loop on the first iteration -- whereas the ``number theory'' approach simply outputs the answer with no loop structure.
This pattern continued in the following 2048 samples. The model solved the problem three times more often with the ``number theory'' tag (29 instead of 9 solutions), and output a perfect loop-free solution (other than reading the input)  four times more often (12 instead of 3 solutions).

\begin{table}[tp]
    \centering
    \begin{tabular}{lccc}
    \toprule
    Tag sampling scheme & 300M & 1B & 3B \\
    \midrule
    Random per sample (default) & $8.2\%$  & $13.5\%$ & $14.9\%$\\
    True problem tags & $8.1\%$ & $13.3\%$ & $13.9\%$ \\
    Random per problem & $6.0\%$  & $11.9\%$ & $13.3\%$\\
    \bottomrule
    \end{tabular}
    \caption{
    \label{tab:tags-sampling-scheme}
    \textbf{Solve rate (10@1024) according to different methods of sampling tags at test time}.}
\end{table}

There are two possible ways that tag conditioning can improve the model solve rate: random tags may increases sample diversity or sampling correct tags may provides a useful hint. To distinguish between them, we compared solve rates when
sampling a set of random tags for each sample (default), when providing the true problem tags, and when sampling a set of random tags for each problem (and reusing it for all samples for the same problem). The results are shown in \tabref{tab:tags-sampling-scheme}. Providing the true tags for a problem is impossible at test time, but is better than providing a fixed set of random tags. However, the best results come from sampling with random tags per sample, showing that the extra diversity of samples is important to increasing the solve rate. %

\subsection{Loss is a poor proxy for solve rate}\label{sec:validation-solve-rate}

\begin{figure}[h]
    \centering
    \includegraphics[width=0.5\textwidth]{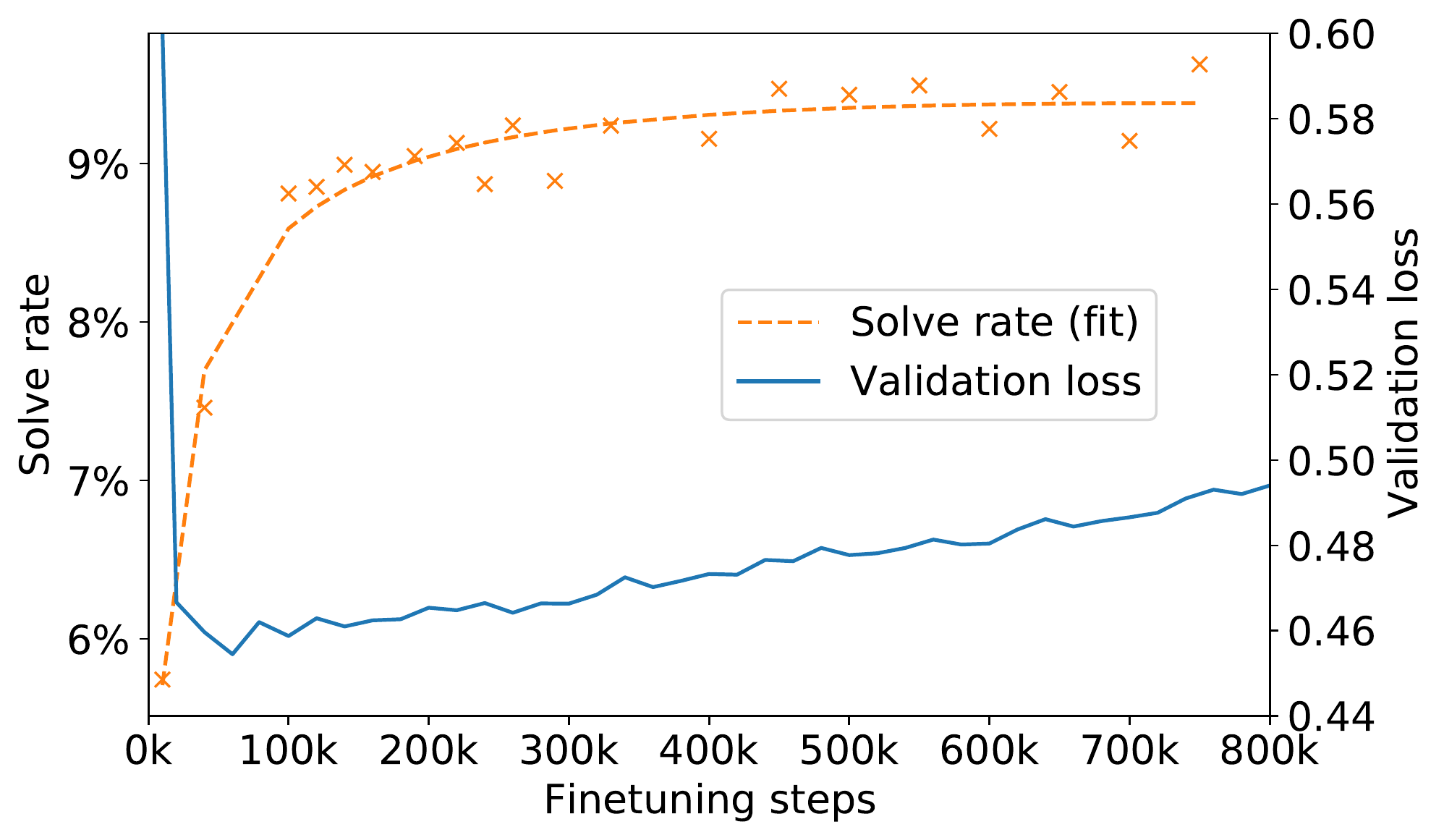}
    \caption{\label{fig:valid_loss_vs_solve_rate} \textbf{Validation loss and solve rate (10@1024) by finetuning steps}. The validation loss starts to increase early in finetuning, indicating overfitting, while the solve rate keeps improving.}
\end{figure}

When finetuning \OurApproachShort{}~models we observed that the validation language modelling loss starts increasing after around 50k steps for an early training run of the 1B model, while the training loss still decreases. This normally indicates overfitting.
However, contrary to the validation loss, our target metric solve rate continues to improve well past 50k steps as shown in \figref{fig:valid_loss_vs_solve_rate}.

As discussed in~\secref{sec:finetuning}, solving a problem is a one-of-many task, i.e.~as long as one of many samples solves a problem, the problem is considered solved.  Our finetuning dataset \OurDatasetShort{} contains many solutions per problem.  Our main solve rate metric, 10@k, also uses $k$ samples rather than a single sample. %
We hypothesize that the model reallocates probability mass from some atypical solutions towards more typical solutions, leading to a worse validation loss overall, but a higher probability of producing more typical solutions and therefore a better solve rate.

Although solve rate is the ultimate metric, its high variance and computational cost make it difficult to use for decisions like the number of training steps. Improving validation loss to better correlate with performance could guide the decision making process better. 
We leave a full investigation of the relationship between validation loss and solve rate to future work.

\section{Related work}

\subsection{Program synthesis}

Program synthesis consists of automatically generating a program that satisfies a task specification. Possible ways of expressing the task include natural language descriptions, a set of input/output examples, or a series of constraints.
As a research topic, program synthesis has a long history.  Most classic approaches formulate the problem as searching for programs in a search space defined by the underlying programming language, where the programs must satisfy all the constraints which define the task.  A notable example is the deductive synthesis approach \citep{green1969application,manna1971toward}, that transforms the task specification into constraints, uses a theorem prover to find a proof that satisfies all the constraints, and extracts the program from the proof. 
Later on, input/output-based task specifications became more popular, with notable examples like FlashFill \citep{gulwani2011automating}. Finally, sketch-based approaches \citep{solar2008program} that synthesize programs from a provided skeleton of the target program greatly reduce the search space. \cite{gulwani2017program} provides an excellent survey of these program synthesis approaches.

In recent years, deep learning has emerged as a useful tool for program synthesis. 
\cite{yin2017syntactic} used recurrent networks with attention to map text to abstract syntax trees and then code. \cite{ling-etal-2016-latent} used similar models, with pointer networks, %
to generate complex class structures from mixed natural language and structured specifications of Hearthstone cards.
Learned models can now be used to guide program search \citep{balog2016deepcoder}, generate program sketches \citep{murali2017neural,guo2021learning}, convert pseudocode to code \citep{kulal2019spoc}, directly generate a target program \citep{devlin2017robustfill}, or even generate programmatic policies in reinforcement learning settings \citep{trivedi2021learning}.

Automatic code completion is also relevant to our work and has become an integral part of most code editors and integrated development environments (IDEs). While typing, these tools suggest possible continuations, greatly improving programming productivity. The earliest code completion systems were purely syntax-based. \citet{hindle2012naturalness} provided empirical evidence that code can be modeled by statistical $n$-gram language models, and capitalised on this to develop a simple code
completion engine for Java.  More recent intelligent code completion systems can learn from history \citep{robbes2008program} and large amounts of existing code data \citep{svyatkovskiy2020intellicode,aye2021learning}.

However, until very recently most code completion systems only generated suggestions for at most a single line. Similar trends are present throughout program synthesis: either restricting to short programs in narrowly defined domain-specific languages, or short code snippets of general-purpose programming languages.  Scaling up, which increases both the depth and width of the search problem, has proven to be a difficult challenge.

\subsection{Transformers for program synthesis}
Recently, the successes of large transformers in natural language modelling~\citep{brown2020language} have created a surge of interest in using transformer models for code retrieval, translation and generation \citep{feng2020codebert,clement2020pymt5,chen2021codex}, making significant progress on program synthesis challenges. Trained on huge datasets covering a wide spectrum of text on the Internet, these models are capable of generating text with unprecedented fluency. The most relevant work to ours is the recent Codex system \citep{chen2021codex}, a GPT language model \citep{radford2019language} trained on public code from GitHub. This model demonstrated impressive performance, achieving a high success rate at correctly completing hand-specified Python functions given the function signature and docstring, especially after fine-tuning on a similar dataset.  Codex was used to build interactive program synthesis systems that are capable of solving university-level linear algebra and probability and statistics questions in \citep{tang2021solving,drori2021solving}, and further used to create an advanced autocomplete system in \href{https://copilot.github.com/}{GitHub Copilot}.
A similar model to Codex was trained by \cite{austin2021program}, who also show that fine-tuning on a portion of a programming task dataset can improve the success rate on similar tasks.

However, the programming tasks these works address are simple compared to the full scope of competitive programming problems, where both the task specification and the solutions are more involved.  For example, in our dataset the median problem description length is 1,628 characters and the median solution length is 606 characters, while the HumanEval benchmark \citep{chen2021codex} has a median description length of 396 characters and solution length of 148.5 characters, about 4 times shorter for both. HumanEval problems also tend to include instructions about exactly what to implement, as opposed to competitive programming problems which pose a problem with no suggested implementation. Finally, it is not clear how unseen tasks are.
Though they are hand-written instead of copied from an existing source, tasks like sorting arrays or checking if a number is prime have solutions that can be copied from the training dataset. Our work uses transformers but pushes model performance a significant step forward, from generating function completions to creating full solutions to held-out competitive programming problems. 

\subsection{Scaling sampling}
Similar to our sampling and filtering, though on a much smaller scale, \cite{chen2021codex}, \cite{austin2021program}, and \cite{cobbe2021training} found that repeated sampling on the same problem significantly increases the probability of finding a correct solution. \cite{cobbe2021training} further introduced a way of selecting a small number of final submissions from multiple samples by majority voting. They also demonstrate verifiers (value functions) used to judge the correctness of model samples as a way of reranking samples.

\subsection{Evaluation metrics}
Evaluation metrics have evolved as models themselves have improved. Early work evaluated performance by measuring how well the generated code matched the ground truth reference code at the token level, syntax tree level, or full program level \citep{ren2020codebleu}. These metrics determine whether code matches rather than whether a program is correct (syntactically different programs can be functionally identical), so as models have improved to be capable of fully solving programming problems, evaluation processes that execute programs and measure functional correctness~\citep{kulal2019spoc,chen2021codex,hendrycks2021measuring} have become more popular. However both \citet{chen2021codex} and \citet{hendrycks2021measuring} are somewhat limited as benchmarks because we have found that it is common for incorrect programs to be marked correct due to limited test coverage (\tabref{tab:dataset-test-cases-false-positive}).

\subsection{Competitive programming}
Lastly, progress in developing models that can solve competitive programming problems would not be possible without competitive programming datasets for training and evaluation. \cite{Caballero_Description2Code_Dataset_2016} released a dataset of a few thousand competitive programming problems, and corresponding Python and \CC{} solutions gathered from popular competitive programming platforms.  \cite{zavershynskyi2018naps} introduced another dataset of competitive programming problems and solutions, though they converted solutions into an intermediate programming language which makes using pre-trained models difficult.  \cite{puri2021project} released a dataset of a large number of solutions in a wide variety of programming languages, with correct and incorrect solutions, and rich meta-data. 
Finally \cite{hendrycks2021measuring} introduced the APPS dataset, a collection of 10,000 coding competition problems, and were the first to evaluate large transformer language models on competitive programming. The authors found that the overall solve rate on interview or competition level problems using large language models remained close to 0\%. However, their evaluation format is not representative of competitive programming and, as noted above, this solve rate is an upper bound because of false positives due to a lack of test coverage (\tabref{tab:dataset-test-cases-false-positive}). Our dataset is built upon \cite{Caballero_Description2Code_Dataset_2016} and \cite{puri2021project}, with our own additional scraping from the Codeforces platform.

\section{Broader impact}
Good code generation models have the potential to have a positive, transformative impact on society, with a wide range of applications including computer science education, developer tooling, and making programming more accessible. However, like most technologies, these models might enable applications with societal harms which we need to guard against, and desire to have a positive impact is not itself a mitigation against harm.

\subsection{Applications}
Although there are few direct applications of this work outside of competitive programming, improving human readable code generation opens the door to many future applications with large real-world impact. All these applications require varying amounts of future work.

Automating code generation could make existing programmers more productive, with potential applications ranging from suggesting extended code completions to optimizing blocks of code. Further in the future, advanced code generation models could let developers operate at a higher level of abstraction that elides details, much in the same way that modern software engineers typically no longer write in assembly.

Derived tools could make programming more accessible or help educate new programmers. Models could suggest alternative, more efficient or idiomatic, ways of implementing programs, which would allow one to improve their coding style. A code-to-documentation tool~\citep{feng2020codebert}, would make it easier to understand what a complex section of code does. More extreme systems that operate entirely in natural language, like that proposed in Codex~\citep{chen2021codex}, could make it so that no knowledge of coding is required to create software.

However, code generation tools could also be used by bad actors. Better tools may make it easier to create new versions of malware, thus helping them avoid detection by security software (which often rely on databases of file fingerprints). Tools that improve the productivity of developers would also improve the productivity of developers writing malicious code~\citep{weid2021ethical, chen2021codex}. Alternatively, competitive programming code generation models in particular could give users unfair advantages in programming competitions or technical interviews.

\subsection{Potential risks and benefits}

\textbf{Interpretability.~~}
One major advantage of code generation models is that code itself is relatively interpretable.
Understanding the behavior of neural networks is challenging, but the code that code generation models output is human-readable and can be analysed by traditional methods (and is therefore easier to trust). Proving a sorting algorithm is correct is usually easier than proving a network will sort numbers correctly in all cases.

Interpretability makes code generation safer for real-world environments and for fairer machine learning. We can examine code written by a human-readable code generation system for bias, and understand the decisions it makes.

\textbf{Generalisation.~~}
Code %
generated by code generation models tends to generalise; passing a sufficient number of tests makes code more likely to also pass even out-of-distribution tests. This level of generalisation over the full domain of inputs and outputs is often hard to obtain with neural networks, and could make code generation models more reliable in real-world out-of-distribution applications.

\textbf{Bias, fairness, and representation.~~}
Similar to natural language models~\citep{brown2020language}, code generation models are prone to reproducing the deficiencies and biases of their training data. When trained on diverse corpora of human data, these models can reinforce and perpetuate societal stereotypes, leading to disproportionate impact on marginalized communities. For example, programs can contain culture and location-specific assumptions about names~\citep{falsehoodsnames}, addresses~\citep{falsehoodsaddressess}, or time~\citep{falsehoodstime}, excluding underrepresented users.

Bias can also lead to low quality code that perpetuates bugs or the use of outdated APIs,\footnote{This issue also exists for human programmers who copy old code.} resulting in performance and security issues. This could decrease uptake of new libraries or programming languages.

\textbf{Security.~~}
As mentioned above, code generation can have security risks and benefits. Models can generate code with exploitable weaknesses, either unintentional vulnerabilities from outdated code or intentional ones injected by malicious actors into the training set~\citep{pearce2021cybersecurity}. Further, code generation could enable both threat actors and threat defenders, increasing productivity and enabling new techniques. For example, polymorphic malware changes its implementation to hide from detection~\citep{chen2021codex}.

\textbf{Environmental impact.~~}
Like large-scale language models, training transformer-based code generation models takes a significant amount of compute. Further, because we found that large-scale sampling is critical to improving performance, relatively more compute is spent executing our model compared to traditional language models. Both sampling and training from our model required hundreds of petaFLOPS days.\footnote{Experiments were run in Google datacenters, which purchase renewable energy equal to the amount consumed~\citep{ursrenewable}.}

However, one comparative advantage of code generation models is that once a program is synthesized it can generally be executed cheaply by any computer, unlike neural network models that typically need to be run on accelerators. Therefore code generation models %
can potentially be scaled to many applications more easily.

\textbf{Intellectual property.~~}
There are intellectual property concerns with large training corpora used to train code generation models. Whether training on publicly available data is fair use is an open question for some systems~\citep{copilotcopyright}, although this is less relevant to \OurApproachShort{} which filters its dataset based on licenses. There still remains the decision of how to credit and use people's code, even with permissive licenses.

\textbf{Automation.~~}
As programming becomes more accessible and productive, and code generation can automate some simple tasks, it's possible that there could be increased supply and decreased demand for programmers. This is partially mitigated because writing code is only one portion of the job, and previous instances of partially automating programming (e.g. compilers and IDEs) have only moved programmers to higher levels of abstraction and opened up the field to more people.

\textbf{Advanced AI risks.~~}
Longer term, code generation could lead to advanced AI risks. Coding capabilities could lead to systems that can recursively write and improve themselves, rapidly leading to more and more advanced systems.

\section{Conclusion}
In this work, we present \OurApproachShort{}, a system applied to code generation for competitive programming that can generate novel solutions to unseen programming problems. Evaluated on Codeforces, \OurApproachShort{} performs roughly at the level of the median competitor. We find that 
massively scaling up sampling and then filtering and clustering samples to a small set, together with new sampling-efficient transformer architectures to support large-scale sampling, are essential to achieving good performance.  Our clean dataset and robust evaluation procedure also contributed significantly to guiding our research progress.
We also show through detailed analysis that there is no evidence that \OurApproachShort{} copies important parts of previous solutions or exploits weaknesses in the problem structure. This indicates that our model indeed is able to solve problems it has never seen before, even though those problems require significant reasoning. Finally, we present the results of various model probings, and discuss broader impact and hazards of such code generation models.

\section*{Acknowledgements}
We would like to thank
Trevor Cai, Jack Rae, Sebastian Borgeaud, Mia Glaese, Roman Ring, Laurent Sifre, Jordan Hoffman, John Aslanides, Jean-Baptiste Lespiau, Arthur Mensch, Erich Elsen, George van den Driessche, and Geoffrey Irving for developing tools we use to train large language models, and for lending their expertise in model training; 
Kirsty Anderson, Claudia Pope, and Rachel Foley for project management in early stages;
Yee Whye Teh, Chris Dyer, David Silver, Amin Barekatain, Anton Zhernov, Matt Overlan, and Petar Veličković for research advice and assistance;
Karen Simonyan, Chris Dyer, and Dani Yogatama for reviewing the paper; 
Lorrayne Bennett, Kareem Ayoub, and Jeff Stanway for logistically making the project possible;
Sumanth Dathathri for analysing our model;
Ethan Caballero for giving us permission to use Description2Code data;
Rosemary Ke for helping connect us with Ethan;
Pablo Heiber for helping connect us with Codeforces;
Petr Mitrichev for helping connect us with Codeforces, and lending competitive programming expertise when writing the paper;
Mike Mirzayanov for allowing us to evaluate on Codeforces;
and everyone at DeepMind for their insight and support. %

\section*{Author Contributions}

\textbf{Agustin Dal Lago} worked on development of the dataset, evaluation, and general infrastructure.

\textbf{Cyprien de Masson d'Autume} worked on model development and analysis.

\textbf{Daniel J. Mankowitz} worked on clustering.

\textbf{David Choi} was the technical lead, developed initial prototypes for solving competitive programming problems, and contributed to aspects including general infrastructure, metrics, large-scale training, model development, sampling, evaluation, datasets, training, and paper writing.

\textbf{Esme Sutherland Robson} worked on project management.

\textbf{Felix Gimeno} worked on model development, metrics, datasets (notably the APPS benchmark), and clustering.

\textbf{Igor Babuschkin}\footnote{Work conducted at DeepMind, now at OpenAI.} worked on initial prototypes of code generation models and contributed to infrastructure tools.

\textbf{James Keeling} worked on code execution and evaluation, sampling infrastructure and scaling, clustering, and paper writing.

\textbf{James Molloy} worked on improving the efficiency of our models on accelerators.

\textbf{Julian Schrittwieser} worked on datasets, evaluation, model development and training losses, tokenization, visualisations, and paper writing.

\textbf{Junyoung Chung} worked on initial prototypes of code generation models, model development and tuning, training losses, training pipeline, model performance, datasets, sampling, large-scale models, running most final experiments (notably the main experiments), and paper writing.

\textbf{Nate Kushman} worked on initial sample scaling efforts, metrics, evaluation, model development and tuning, training losses, datasets (notably the HumanEval benchmark), large-scale models, analysing scaling behavior, running final experiments, and paper writing.

\textbf{Oriol Vinyals} was an early advocate for code generation models, supported and advised the project throughout, and was involved in project management and paper writing.

\textbf{Peter Choy} worked on model development, sampling, and copying analysis.

\textbf{Rémi Leblond} worked on model development and tuning, optimisation, improved training losses, model analysis, sampling efficiency and scaling, datasets, and paper writing.

\textbf{Thomas Hubert} worked on model development, infrastructure, and additional training objectives.

\textbf{Tom Eccles} worked on datasets, metrics, evaluation, clustering, general infrastructure, ensembling, metadata conditioning, model analysis, and paper writing.

\textbf{Xinyun Chen}\footnote{Work conducted during a DeepMind internship, UC Berkeley affiliation.} worked on model development.

\textbf{Yujia Li} was the project lead, developed initial prototypes for solving competitive programming problems, and contributed to aspects including the development of datasets, metrics, evaluation, models, training, clustering, infrastructure tools, and paper writing.

\textbf{Alexey Cherepanov, Johannes Welbl, Po-Sen Huang, and Sven Gowal} worked on model analysis.

\textbf{Nando de Freitas, Koray Kavukcuoglu, and Pushmeet Kohli} advised the project, and Nando de Freitas further contributed to paper writing.

\section*{Data availability}
The datasets used in the experiments have been made available for download on \href{https://github.com/deepmind/code_contests}{GitHub}.

\bibliographystyle{abbrvnat}
\setlength{\bibsep}{5pt} 
\setlength{\bibhang}{0pt}
\bibliography{refs}

\newpage

\section{Appendix}
\part{Appendix}
\vspace{-2.3cm}
{
\setlength{\columnseprule}{0.5pt}
\etocsettocstyle{}{}
\etocsetnexttocdepth{subsection}
\localtableofcontents
}
\appendix

\renewcommand{\figurename}{Appendix Figure}
\setcounter{figure}{0}
\renewcommand{\thefigure}{A\arabic{figure}}

\renewcommand{\tablename}{Appendix Table}
\setcounter{table}{0}
\renewcommand{\thetable}{A\arabic{table}}
\section{Problem setup}

\subsection{Hidden tests}

Competitive programming problems typically contains example tests in the problem statement, and also hidden tests not visible to the competitors that are used for evaluation.
\figref{fig:test_cases} contains a hidden test case for the example problem in \figref{fig:problem_statment}.

\begin{figure}[h]
\begin{center}
\begin{minipage}[t]{.48\textwidth}
\footnotesize
\textbf{Input}
\begin{lstlisting}
10
paxghjnihn
hn
hdmevxvn
n
azdfhfxem
xem
eowhldode
dode
wlclsnht
ct
bpflheocamv
v
flejfh
hixqqbnikthccagc
dugt
eebmbpykcsmi
oivgrzwppny
zhfyiuu
ebkqjcbcwviqkojnzyruwygtbvwws
bofzr
\end{lstlisting}
\end{minipage}
\begin{minipage}[t]{.48\textwidth}
\footnotesize
\textbf{Output}
\begin{lstlisting}
YES
YES
YES
YES
YES
YES
NO
NO
NO
NO
\end{lstlisting}
\end{minipage}

Followed by over a hundred more extensive tests that probe various cases.

\caption{\textbf{Hidden test cases used to verify correctness of solutions to the problem from Figure \ref{fig:problem_statment}}. Compared to the example tests seen by participants (and used within \OurApproachShort{}), the held-out hidden test cases used to evaluate correctness are substantially longer and more demanding. Hidden test case sourced from \href{https://codeforces.com/problemset/problem/1553/D}{Codeforces}.
}
\label{fig:test_cases}
\end{center}
\end{figure}

\subsection{Program judging}
\label{sec:program-evaluation}

When submitting to Codeforces, as in \secref{sec:codeforces-results}, we can use the same judging system used in competitions. We try to emulate this system within \OurDatasetShort{}: we check the correctness of a program by executing it on the test cases and comparing the program outputs with the expected correct outputs.
However, judging whether the program outputs are correct can be challenging, and involves more than checking for an exact match against the correct outputs. Each problem can have specific rules including case sensitivity, whitespace, format, and floating point precision.
Further, problems may have multiple correct outputs (e.g. permitting any sequence that follows a constraint), or multiple possible inputs (e.g. an interactive problem where the input depends on what the program outputs). The judging process described below takes place both for final submissions, and for filtering based on example tests.

Because these constraints are difficult to extract from the problem, our judging code takes a permissive view of formatting. Floating point numbers are considered equivalent if their difference is less than $10^{-5}$, string comparison is case insensitive, and whitespace differences are ignored. This does not exactly match formats given by problems, but we found these issues are not too significant. When verifying dataset false positive rates, we did not find any problems that were incorrectly marked correct because of this issue.

We determined which problems have multiple correct outputs heuristically using human solutions; if any test case had at least $5$ distinct outputs from human-written correct solutions, or $2$ outputs produced by multiple human solutions, we assumed it had multiple outputs. 
About 1/4 of our validation set problems are multiple output problems by this criteria. These problems are judged using the same permissive formatting and against a single correct output, where the correct output is chosen to be what the majority of human solutions output.  Because we assume a single correct output, our judging can underestimate the actual model performance. An alternative is to accept any output that a human solution outputs, which decreases the false negative rate in judging, but we found that this leads to significantly increased false positives.

Interactive problems are substantially rarer than multiple output problems, and we do not explicitly handle them, which could lead to both false negatives and false positives. We did not find any interactive problem false positives.

Competitive programming problems also often include time and memory limits, and we use these limits when executing submissions.

\subsection{Evaluation metrics}
\label{appendix:evaluation-metrics}

As described in \secref{sec:evaluation}, we use the $n@k$ solve rate to evaluate model performance, which measures the fraction of problems a model can solve when allowed to generate $k$ samples but only submit $n$ of them for evaluation.

There are two sources of variance in the computation of this metric:
\begin{itemize}
    \item If we train the same model architecture with the same data but different random seeds, we will end up with different models which will produce different samples.
    \item If we take $k$ samples from the same trained model multiple times, we will end up with a different set of samples each time.
\end{itemize}

We use the technique discussed below to reduce the second source of variance in all of our reported results except for clustering results (which are discussed below). Reducing the first source of variance is more challenging, however, and given the computational cost of training our models it is not practical to pre-train and fine-tune multiple models for all results in the paper.  Nonetheless, given the importance of the ablation results in \tabref{tab:fine-tuning-ablations}, for only this table, we fine-tuned at least 3 models for each setting from the same pre-trained checkpoint, and reported average $n@k$ results over these models.

To compute $n@k$ from a single model we:
\begin{itemize}
    \item Draw a set of $K\ge k$ samples from the model.
    \item Draw $S$ sub-samples of size $k$ from the full set of $K$ samples without replacement.
    \item Calculate solve rates with $n$ submissions for each sub-sample.
    \item Report the average solve rate across all of these sub-samples.
\end{itemize}

This estimation process is outlined in Algorithm \ref{alg:n_at_k_with_filtering}.  To create a scaling curve, we use this procedure to calculate $n@k$ for different values of $k$, using the same set of $K$ samples.

To generate the confidence intervals for the ablation results in \tabref{tab:fine-tuning-ablations}, we use bootstrap re-sampling. Specifically, we:
\begin{itemize}
    \item Re-sample with replacement $m$ models from the $m$ models we have trained.
    \item For each model, re-sample with replacement $k$ samples from the $K$ we have taken from that model.
    \item Compute $n@k$ with the re-sampled models and samples using the process described above.
\end{itemize}
We perform this re-sampling and estimation process many times, and report the 95\% confidence interval as the 2.5th percentile and 97.5th percentile from the resulting set of estimates.

It's difficult to use the sub-sampling process described above for results which include clustering, because computing clusterings adds additional computational complexity. Thus unless otherwise noted, we compute all clustering results as follows.  For each size $k$ and each model trained for the base setting used for clustering, we run clustering on five different subsamples.  The reported means are the average of these 5 data points across all trained models for each size $k$.  The confidence intervals are computed using bootstrap re-sampling similar to the process above, where we first re-sample $m$ models from the $m$ we have available, and then re-sample five clustering runs from the five we have available for each model and each size $k$.  The clustering results in \tabref{tab:fine-tuning-ablations} and \figref{fig:compare-filtering-clustering} were based on 5 models fine-tuned in the "+ GOLD" setting.  The clustering results for the 9B and 41B models are based on single models due to the cost of training multiple models at these sizes.

\begin{algorithm}[tp]
\caption{Algorithm for computing n@k with filtering using example tests.}
\label{alg:n_at_k_with_filtering}
\hspace*{\algorithmicindent} \textbf{Input} $n=$ the number of allowed submissions in $n@k$ \\
\hspace*{\algorithmicindent} \textbf{Input} $k=$ the number of allowed samples in $n@k$ \\
\hspace*{\algorithmicindent} \textbf{Input} $e_p=$ the number of samples which pass the example tests for each problem $p$ \\
\hspace*{\algorithmicindent} \textbf{Input} $s_p=$ the number of samples which solve the problem (pass all tests) for each problem $p$ \\
\hspace*{\algorithmicindent} \textbf{Input} $K=$ the number of samples actually taken per problem \\
\hspace*{\algorithmicindent} \textbf{Hyperparameter} $S=$ the number of subsamples to use for calculation\\
\begin{algorithmic}[1]

\For{each problem $p$ in the problem set}
    \For{each of the $S$ subsamples}
        \State Sample $e_p'\sim$Hypergeometric$(e_p, K-e_p, k)$\Comment{\# samples out of $k$ which pass examples tests.}
        \State $n'\leftarrow \min(e_p', n)$ \Comment{Only submit samples that pass the example tests.}
        \State Sample $s_p'\sim$Hypergeometric$(s_p, e_p - s_p, n')$ \Comment{\# correct solutions out of $n'$ submissions.}
        \State solved$_{p}$ $ = 1$ if $s_p' > 0$ else $0$\Comment{Problem is solved if any submission is correct.}
    \EndFor
    \State Compute $n@k$ for this problem as the average of all solved$_p$.
\EndFor
\State \textbf{return} the average $n@k$ across all problems.
\end{algorithmic}
\end{algorithm}

\section{Datasets}

\subsection{GitHub dataset composition}

\begin{table}[th]
    \centering
    \begin{tabular}{lrrr}
    \toprule
    Language & Files (Millions) & Bytes (GB) & Bytes percentage \\
    \midrule
\CC{} & 21.50 & 290.5 & 40.62\% \\
C\# & 6.73 & 38.4 & 5.37\% \\
Go & 2.19 & 19.8 & 2.77\% \\
Java & 19.35 & 113.8 & 15.91\% \\
JavaScript & 10.55 & 88.0 & 12.31\% \\
Lua & 0.57 & 2.9 & 0.41\% \\
PHP & 11.03 & 64.0 & 8.95\% \\
Python 2 & 1.00 & 10.7 & 1.50\% \\
Python 3 & 6.09 & 43.6 & 6.10\% \\
Ruby & 4.45 & 11.6 & 1.62\% \\
Rust & 0.32 & 2.8 & 0.39\% \\
Scala & 0.83 & 4.1 & 0.57\% \\
TypeScript & 1.69 & 24.9 & 3.48\% \\
\midrule
Total & 86.31 & 715.1 & 100.00\% \\
\bottomrule
    \end{tabular}
    \caption{\textbf{Composition of our GitHub pre-training dataset.} Python 2 and 3 are distinguished by whether the code can be successfully parsed using Python 3's parser.}
    \label{tab:dataset-github-stats}
\end{table}

Our GitHub pre-training dataset contains a total of 715GB of data across a range of different programming languages.  The exact composition is listed in \tabref{tab:dataset-github-stats}.

\subsection{Dataset cleaning}
\label{sec:appendix-dataset-merging}
To avoid data quality and duplication issues involved in combining datasets from different sources, and to make our scraped code more consistent, we performed the following data-cleaning steps:
\begin{enumerate}
    \item Removed problems that are duplicates of each other, ignoring whitespace. Submissions for duplicate problems were merged.
    \item Removed submissions that are duplicates of others, ignoring whitespace.
    \item Cleaned \CC{} submissions to compile with our compiler and sandboxes, for example by adding \texttt{int} in front of \texttt{main()} where it was missing.  We further formatted \CC{} code using \texttt{clang-format}, replaced all the includes with \texttt{\#include <bits/stdc++.h>}, and expanded all other preprocessor directives and \texttt{typedef}s,\footnote{\texttt{\#define}s and \texttt{typedef}s are widely used by competitive programmers to introduce idiosyncratic abbreviations for common constructs, e.g. \texttt{\#define rep(i,n) for(int i=0;i<n;i++)} or \texttt{typedef vector<pair<ll,ll>> vpll;}.} which also made the code shorter.  Note that this cleaning step did not work for a portion of our dataset; when the cleaning failed we simply keep the solution as is.
    \item Executed all Python and \CC{} solutions on all the test cases, and removed in order:
    \begin{enumerate}
        \item all submissions that pass no tests,
        \item all tests that less than 10\% of the remaining submissions produce non-empty outputs on,
        \item all submissions that pass less than 10\% of the remaining tests.
    \end{enumerate}
\end{enumerate}

\subsection{Data leakage and temporal split}\label{sec:competitive-hardness}
\label{sec:dataset-temporal-split}
Transformer language models trained on large datasets of text from the Internet can generate text with impressive fluency. However, due to the scale of the training corpus many downstream evaluation tasks derived from data available on the Internet are subject to training / evaluation data leakage. This constitutes an obvious issue, especially since these models sometimes copy verbatim from the training set \citep{brown2020language}.
While such copying can be beneficial~\citep{borgeaud2021improving}, referring to information that was not available at the time of the competition would constitute cheating. 

In code, data leakage and duplication across training and evaluation sets are particularly common \citep{allamanis2019adverse}, and many participants publish their solutions online after competitions. %
The strict temporal split of our datasets %
can guard against this type of leakage, as it
ensures that our training data only includes information that would be available to a typical human participant in a competition.

We verified the temporal split for both pre-training and fine-tuning by examining the solve rate on \OurDatasetShort{} validation problems. For the fine-tuning dataset, a baseline consisting of evaluating one solution from each training problem on the validation set reached a solve rate of 4.1\% with a random split, but 0\% with a temporal split. When using a random validation split a 1B parameter model pretrained on GitHub had a solve rate of 0.8\% with 13k samples per problem, while the temporal split solve rate remained 0\%. 13k was chosen to match the number of problems (and therefore solutions) used in the baseline. The 0\% solve rate with our temporal split means that models must go beyond simply remembering the training set, and instead they need to create novel solutions to unseen problems.

\section{Approach and Results}

\subsection{Ensembling}
\label{app:ensemble}

Ensembling is another approach we tried to more effectively search in the program space.  To ensemble a set of models, we pool their samples together before running filtering and clustering.  Since different models can have different strengths and weaknesses, ensembling them can increase the diversity of the samples.  However, it is also possible to hurt the performance of good models by ensembling them with significantly worse ones, because this wastes the sample budget.  Finding the right composition for the ensembling therefore is important for performance improvement.

\begin{figure}[tp]
    \centering
    \begin{tabular}{cc}
        \includegraphics[width=0.47\textwidth]{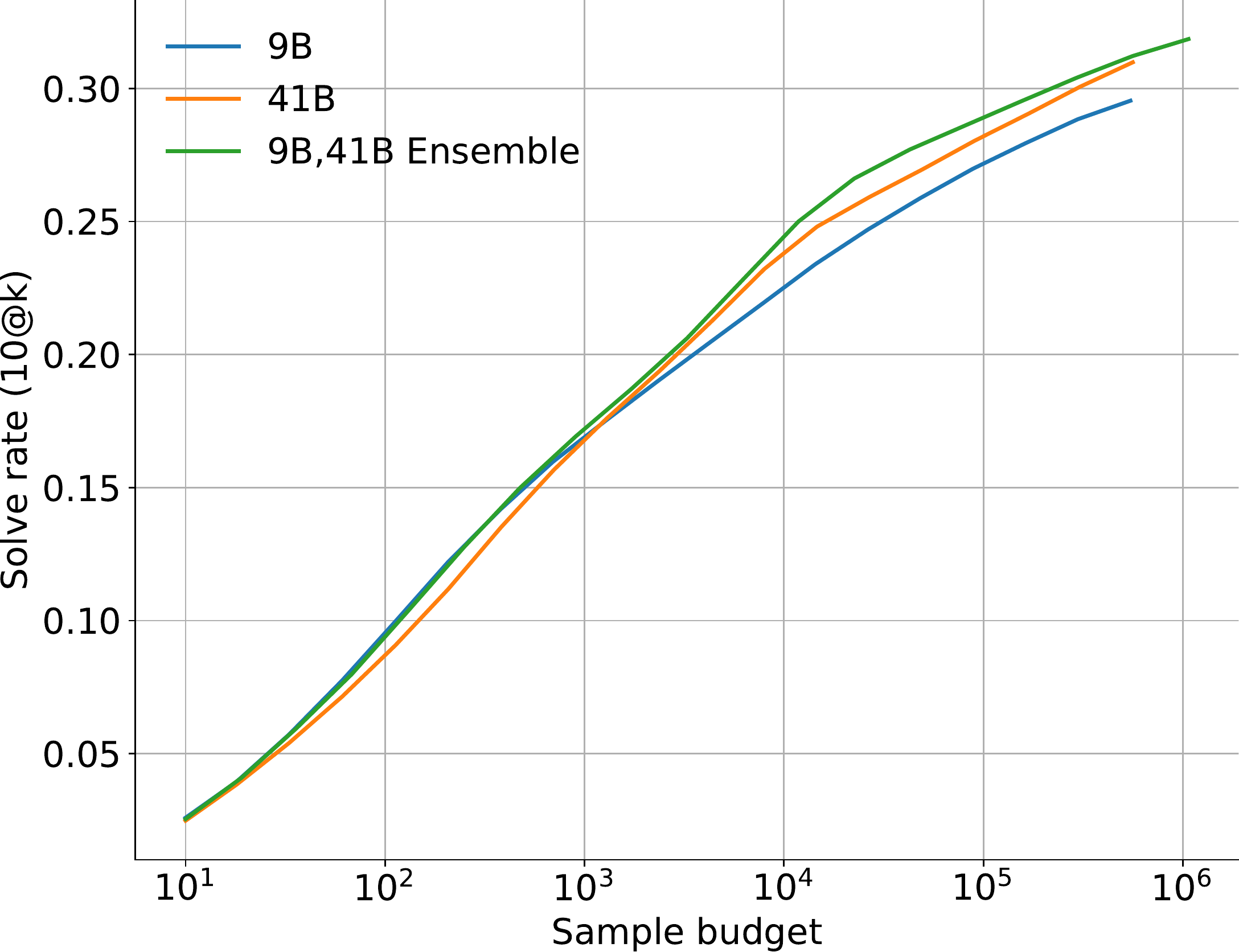} & \includegraphics[width=0.47\textwidth]{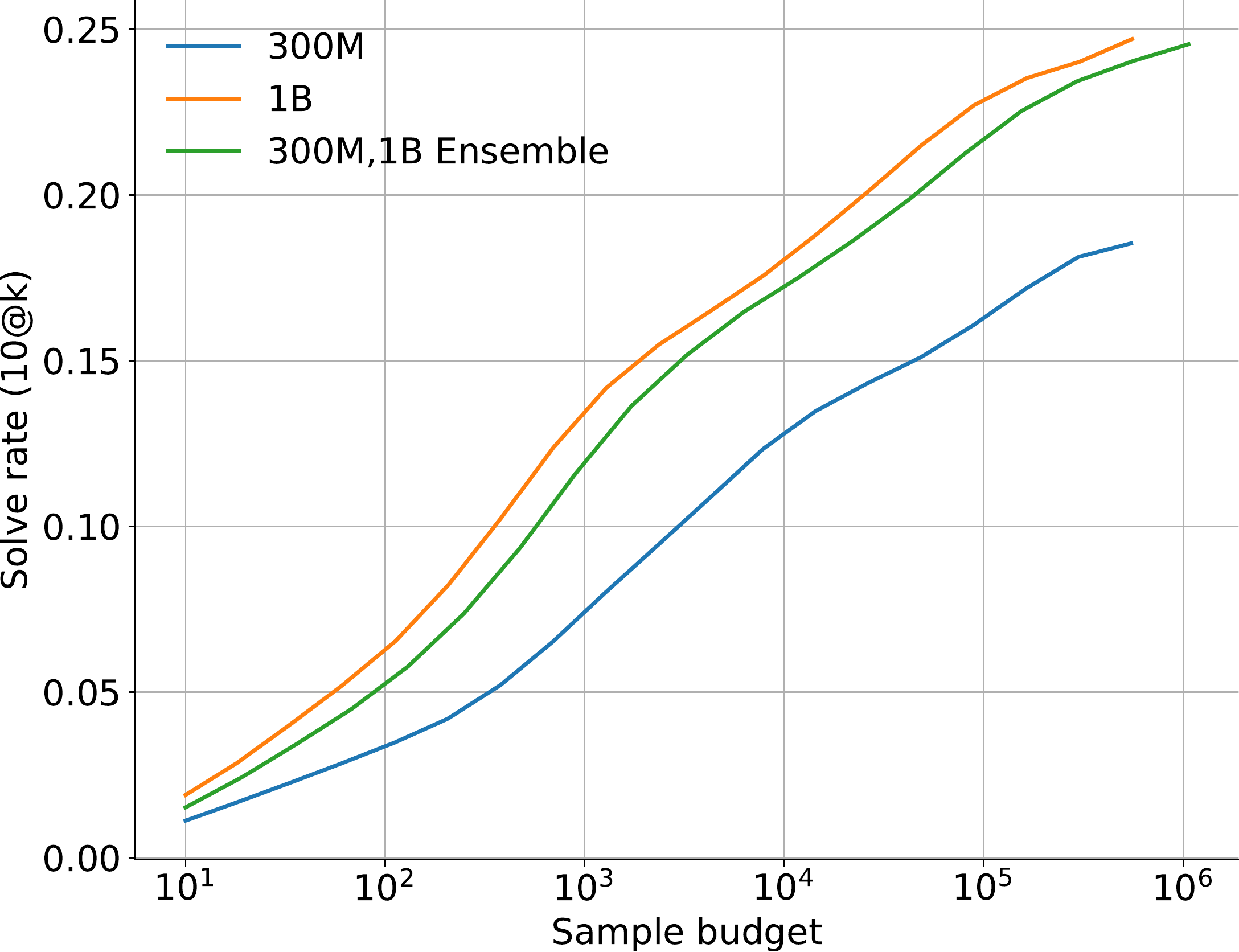} \\
        (a) Ensemble better than all components & (b) Strong components better than ensemble
    \end{tabular}
    \caption{\textbf{Ensemble performance}. Good ensembles can be better than individual components, but bad ensembles can be worse than an individual component.
    }
    \label{fig:ensemble-examples}
\end{figure}

We found that ensembling can indeed increase or decrease performance relative to the individual components in the ensemble.
In particular, when the performance difference of individual runs is large, the ensemble tends to be dominated by the better run and is typically a bit worse than it. %
When the performance difference of individual runs is smaller, the ensemble is more likely to be better, because it covers the search space more effectively than any single component.  Two examples of ensembling are shown in \figref{fig:ensemble-examples}.

The ensemble of our best models at 41B and 9B scales, using equal amounts of samples from each model, outperforms the individual models by a consistent yet small margin. With 1 million samples per problem, the 41B + 9B ensemble reached a solve rate of 32\% on our validation set with 10 submissions, and 35.5\% with clustering.
We therefore used the ensemble of 41B and 9B models for our evaluation on Codeforces described in \secref{sec:codeforces-results}.

However, this ensemble turns out to be slightly worse, or at least not better, than using our 41B model alone on the test set.  Given the small margin of improvement from ensembling, this performance regression is not entirely unexpected.

\begin{table}[th]
\begingroup
\setlength{\tabcolsep}{4pt}
    \centering
    \begin{tabular}{l|cccc|ccc}
    \toprule
        \multirow{2}{*}{Approach} & \multicolumn{4}{c|}{Validation Set} &
        \multicolumn{3}{c}{Test Set} \\
        & 10@1k & 10@10k & 10@100k & 10@1M
        & 10@1k & 10@10k & 10@100k \\
    \midrule
        9B & 16.9\% & 22.6\% & 27.1\% & 30.1\% & 14.3\% & 21.5\% & 25.8\% \\
        41B & 16.9\% & 23.9\% & 28.2\% & 31.8\% & 15.6\% & 23.2\% & 27.7\% \\
        41B + clustering & 21.0\% & 26.2\% & 31.8\% & 34.2\% & 16.4\% & 25.4\% & 29.6\% \\
        \midrule
        41B + 9B & 17.2\% & 24.6\% & 29.0\% & 32.0\% & 15.8\% & 23.0\% & 27.5\% \\
        41B + 9B + clustering & 19.5\% & 26.2\% & 32.9\% & 35.5\% & 15.6\% & 24.3\% & 29.7\% \\
    \bottomrule
    \end{tabular}
    \caption{\textbf{Comparison of ensemble performance on validation set vs. test set.} The ensemble of our best 41B and 9B models performed better than individual runs by a small margin on the validation set, but not better than 41B on the test set.}
    \label{tab:valid-test-comparison-with-ensemble}
\endgroup
\end{table}

\subsection{Metadata conditioning}
\label{sec:metadata-conditioning}

Codeforces problems in our \OurDatasetShort{} dataset contain rich metadata, notably tags and difficulty ratings.  Problems have zero or more tags that suggest what kind of algorithms may be useful for approaching the problem, for example ``divide and conquer'', ``dynamic programming'', and ``data structures''.  Difficulty ratings are values in the range $[800, 3500]$,
where higher ratings correspond to more difficult problems. Tags and ratings are only available after a programming contest has ended. Solutions also contain information about what programming language the solution is.

At sampling time, we do not access the actual ratings and tags as they are not available during competitions.  We found however that solve rates improved by conditioning (i) on ratings sampled uniformly from 800 to 3500 in increments of 100, (ii) on sets of tags sampled uniformly at random from the 50 most popular combinations, and (iii) on language sampled uniformly between \CC{} and Python. We believe that sampling metadata leads to a more diverse set of model samples, a strategy similar to that used by \citet{vinyals2019grandmaster}, and allows our model to take advantage of the relative strengths across the metadata distribution. Because we take a large number of samples, exploring different approaches is more important than maximizing per-sample reward. 

When conditioning on metadata, the chosen metadata values are added as a prefix to the natural language prompt, as shown in \figref{fig:metadata-format} and \figref{sec:complete-examples}.

\subsection{GOLD}\label{appendix:gold}
We want to train a model that is capable of finding any correct solution to a problem (like precision), rather than one that over-focuses on capturing the entire training distribution (like recall). To avoid the tendency of maximum-likelihood objectives to put some weight on every solution, we used a variation of the $\delta$-reward version of GOLD~\citep{pang2020text}, an offline RL algorithm which adds an off-policy importance weight to the standard maximum likelihood objective gradient:
\begin{equation}
    \nabla \mathcal{L_{\text{GOLD}}(\theta)} = - \sum_{s \in \text{Solution tokens}} P_\theta(s) \nabla \log P_\theta(s),
\end{equation}
where $\theta$ are the model parameters, and $\log P_\theta(s)$ is the standard log-likelihood objective for predicting the next token $s$.  The additional $P_\theta(s)$ multiplicative importance weight allows the model to both learn from tokens it already assigns high likelihood to, and to ignore tokens that are not in its distribution. This way, the model can concentrate on precision rather than recall, and increase its chance of getting at least one correct sample.
To mitigate instabilities during training, we replace $P_\theta(s)$ in the importance weight with $\max(P_\theta(s)^\alpha, \beta), \alpha = \frac{1}{2}, \beta = 0.05$. 

Combining GOLD with tempering presents a difficulty of which distribution should be used for the reweighting term. 
The non-tempered distribution becomes smooth as fine-tuning progresses, which means losing the selection benefits of GOLD.
The tempered distribution stays relatively sharp during fine-tuning, but is too sharp at the beginning of fine-tuning (as our pre-trained model is not trained with tempering), leading to overly strong GOLD selection.

To resolve this issue, we add a short training phase between pre-training and fine-tuning, during which we apply tempering but crucially not GOLD. This allows the initial pre-trained distribution to transition to a smoother distribution.
During fine-tuning, we can then use the tempered distribution by dividing the logits by the temperature before computing the loss, so both the log-loss term and the importance weight use the tempered distribution. 

\subsection{Additional results for filtering and clustering}\label{appendix:filtering-clustering}

\begin{figure}[th]
    \centering
    \includegraphics[width=0.6\textwidth]{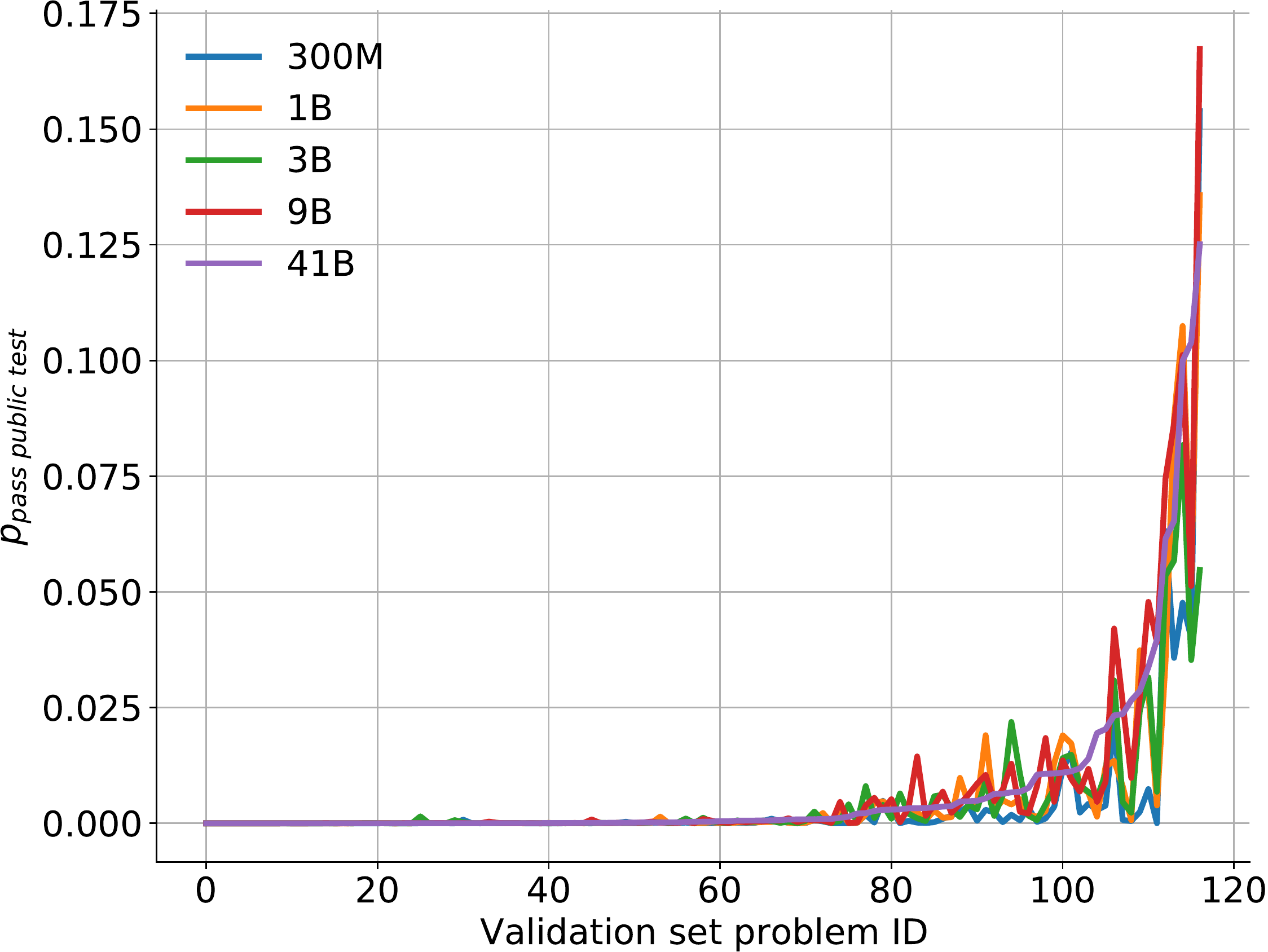}
    \caption{\textbf{Fraction of samples that pass example tests}. $p_\text{pass example test}$ for each problem in the validation set for each model size.  The problems are sorted by the 41B model's $p_\text{pass example test}$.}
    \label{fig:pass-public-test-histogram}
\end{figure}

\textbf{Probability of samples passing example tests varies significantly across problems.~~} \figref{fig:pass-public-test-histogram} shows the distribution of $p_\text{pass example test}$ across problems in our validation set for each model. Notably, this distribution is far from uniform across problems; just over 1/3 of the problems (around the number of problems we solve) have a $p_\text{pass example test}$ significantly higher than zero.

\textbf{Settings for clustering.~~}
Since the test input generation model we trained  is not perfect, it can generate test inputs that are invalid or cannot sufficiently distinguish correct solutions from incorrect samples.  These test inputs can still be used to cluster model samples, but this process may give us ambiguous clusters containing both correct and incorrect samples. 
Additionally, even correct solutions may be put into multiple clusters, as their behaviour on invalid test inputs may still differ.  We tuned two hyperparameters for clustering on the evalidation set: the number of test inputs to use in clustering as well as the maximum number of model samples to consider for clustering. We used 50 test inputs with 8192 model samples for all the clustering results reported in this paper. Performance did not continue to increase with higher numbers of either hyperparameter.

\subsection{HumanEval comparison\label{sec:appendix-humaneval}}

\begin{table}[th]
\begin{center}
\begin{tabular}{c c c c}
\toprule
& \multicolumn{3}{c}{pass@$k$} \\
& $k=1$ & $k=10$ & $k=100$ \\
\midrule
GPT-Neo 125M & 0.75\% & 1.88\% & 2.97\% \\
GPT-Neo 1.3B & 4.79\% & 7.47\% & 16.30\% \\
GPT-Neo 2.7B & 6.41\% & 11.27\% & 21.37\% \\
GPT-J 6B & 11.62\% & 15.74\% & 27.74\% \\
\midrule
TabNine & 2.58\% & 4.35\% & 7.59\% \\
\midrule
Codex-12M & 2.00\% & 3.62\% & 8.58\% \\
Codex-25M & 3.21\% & 7.10\% & 12.89\% \\
Codex-42M & 5.06\% & 8.80\% & 15.55\% \\
Codex-85M & 8.22\% & 12.81\% & 22.40\% \\
Codex-300M & 13.17\% & 20.37\% & 36.27\% \\
Codex-679M & 16.22\% & 25.70\% & 40.95\% \\
\midrule
Pretrained Decoder-only 13M & 1.5\% & 3.6\% & 8.6\% \\
Pretrained Decoder-only 29M & 3.4\% & 5.8\% & 11.2\% \\
Pretrained Decoder-only 55M & 4.2\% & 8.2\% & 16.9\% \\
Pretrained Decoder-only 89M & 4.3\% & 12.2\% & 20.0\% \\
Pretrained Decoder-only 302M & 11.6\% & 18.8\% & 31.8\% \\
Pretrained Decoder-only 685M & 14.2\% & 24.4\% & 38.8\% \\
Pretrained Decoder-only 1.1B & 17.1\% &28.2\% & 45.3\% \\

\bottomrule
\end{tabular}
\end{center}
\vspace{-0.5cm}
\caption{
\label{tab:human-eval}
\textbf{Decoder-only Results on HumanEval}. GPT-Neo, TabNine, and Codex numbers from \cite{chen2021codex}.  Our decoder-only models are pre-trained on the Python-only subset of GitHub and evaluated without any fine-tuning.  The pass@k rates are computed using the algorithm from \cite{chen2021codex} with 1000 samples.  For each row, column pair we only report the best nucleus sampling result from the temperatures 0.0, 0.2, 0.4, 0.6 and 0.8.
}
\end{table}

To ensure that our baseline decoder-only models are as comparable as possible with Codex, we evaluated our models on the HumanEval benchmark from \cite{chen2021codex}.  %
From Table~\ref{tab:human-eval} we can see that our pretrained decoder-only baseline models obtain HumanEval solve rates which are within about 1-3\% of the comparable Codex model for most settings, and significantly better than GPT-Neo and GPT-J at all comparable settings.  The HumanEval results for all of our encoder-decoder models (including the final \OurApproachShort{} model) are significantly worse than the decoder-only models, so we do not report them here. We believe this performance difference stems from the fact that encoder-decoder models are well aligned with the competition programming setting where we have a dataset with clear inputs (programming contest problem descriptions) and outputs (solution code), as well as example tests for effective filtering.  Therefore the encoder-decoder models can both learn effectively and sample efficiently.  However, encoder-decoders are not well aligned with the HumanEval setting where the only training data is GitHub code which cannot easily be split into meaningful inputs and outputs.  As a result, the fact that decoder-only models compute loss on all tokens (rather than only one-third of tokens) enables them to learn more efficiently in the HumanEval setting.

\subsection{APPS dataset settings}

The \OurDatasetShort{} training set has a non-empty intersection with the APPS test set, and therefore \OurDatasetShort{} cannot be used during training when evaluating on the APPS benchmark.

The example tests from problem descriptions were not %
parsed in the APPS dataset, and therefore we parsed them using \figref{fig:apps-num-example-tests}. Our code fails to produce example tests for less than 2\% of APPS test problems, including problems in Russian and a problem without a description (APPS test 4109).

The Codex authors write ``In coding competitions and in the APPS datasets, tasks are provided with 3 input/output examples included in the task description'' \citep{chen2021codex}, but not all problems in APPS have 3 example input/output pairs. Some have 0 (like APPS test 4109 and APPS test 4671), 1 (APPS test 4659) and 2 (APPS test 2844).  We assumed that they only pass 3 example input/output pairs if available, and otherwise parse fewer.

\begin{figure}
\small
\begin{lstlisting}[language=Python]
Copyright 2022 DeepMind Technologies Limited.
SPDX-License-Identifier: Apache-2.0
def _extract_num_public_testcases_from_description(desc: str) -> int:
  """Parses the description to extract the number of public testcases."""
  if '---Sample Input' in desc:
    return desc.count('---Sample Input')
  elif '---Example Input' in desc:
    return desc.count('---Example Input')
  else:
    for section in desc.split('\n-----'):
      if section.startswith('Example'):
        n_tests = max(section.count('\nInput'), section.count('\nSample Input'))
        if n_tests:
          return n_tests
  return 0
\end{lstlisting}
\caption{\textbf{Code used to extract the number of example tests from the descriptions in APPS problems for the filtering system used within \OurApproachShort{}}. %
}
\label{fig:apps-num-example-tests}
\end{figure}

\subsection{Best settings for sampling}

\begin{figure}[p]
    \centering
    \begin{tabular}{cc}
        \includegraphics[width=0.48\textwidth]{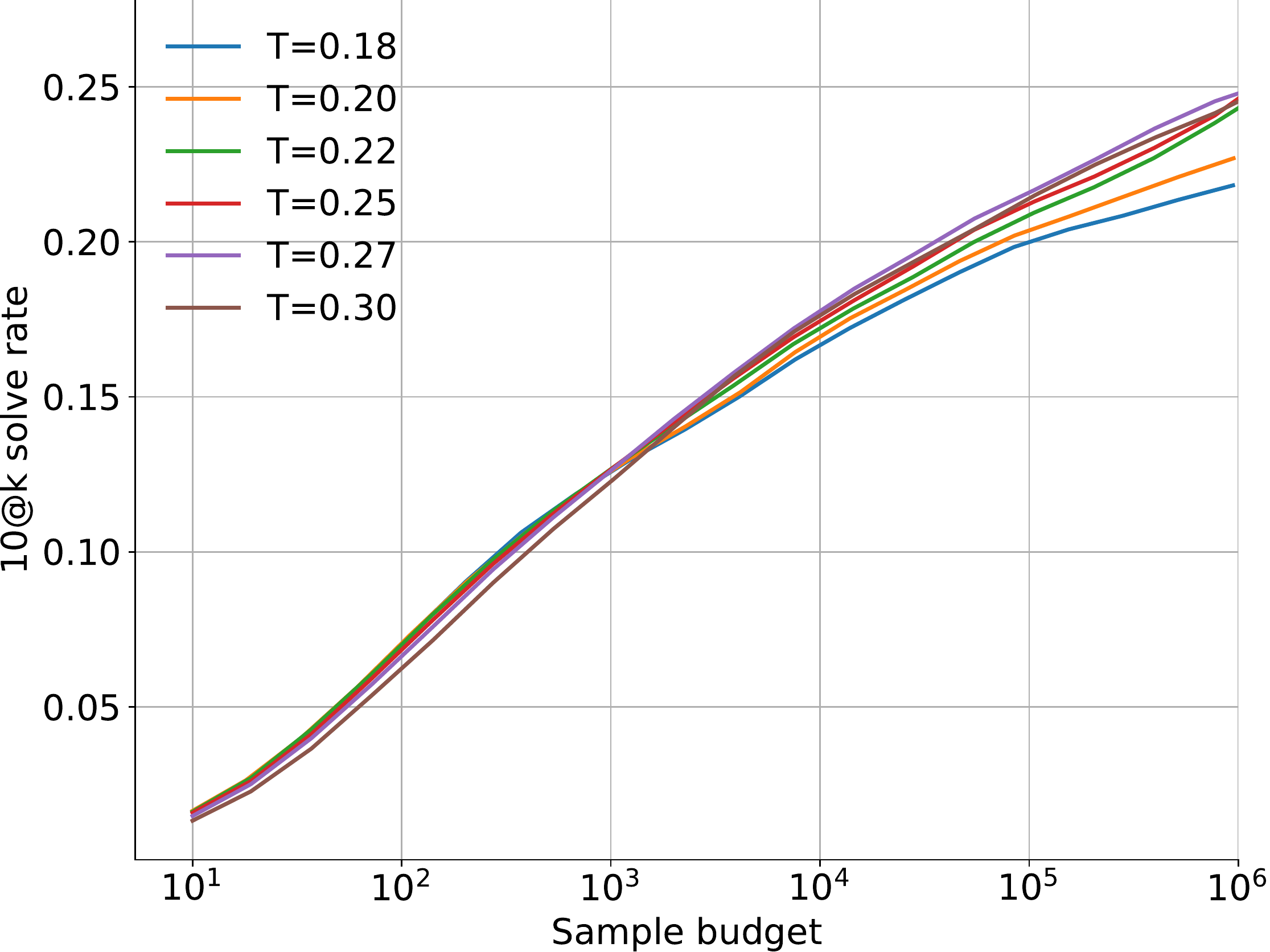} &
        \includegraphics[width=0.48\textwidth]{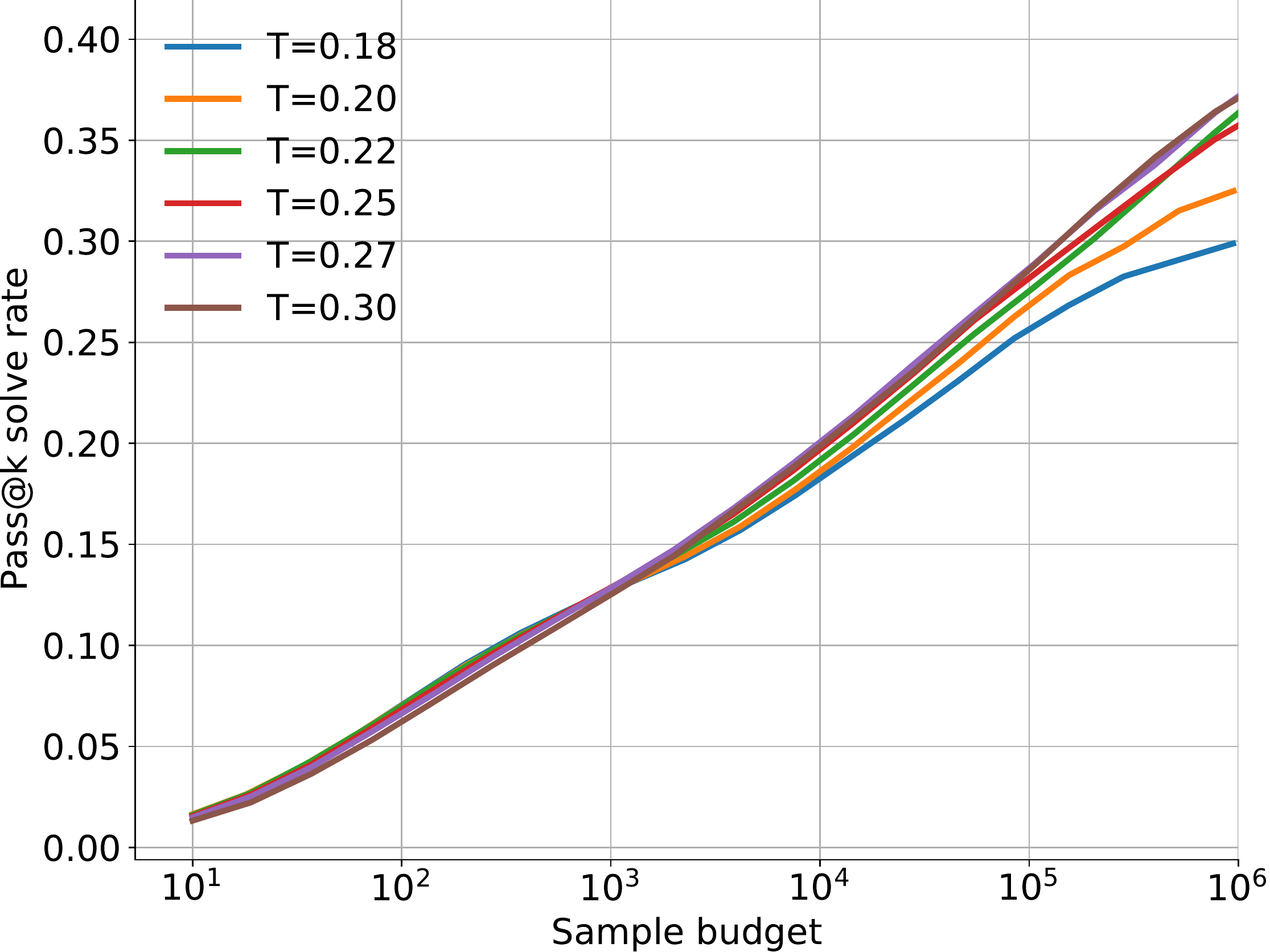}
        \\
        (a) Temperature sensitivity: 10@k & (b) Temperature sensitivity: pass@k \\
        \includegraphics[width=0.48\textwidth]{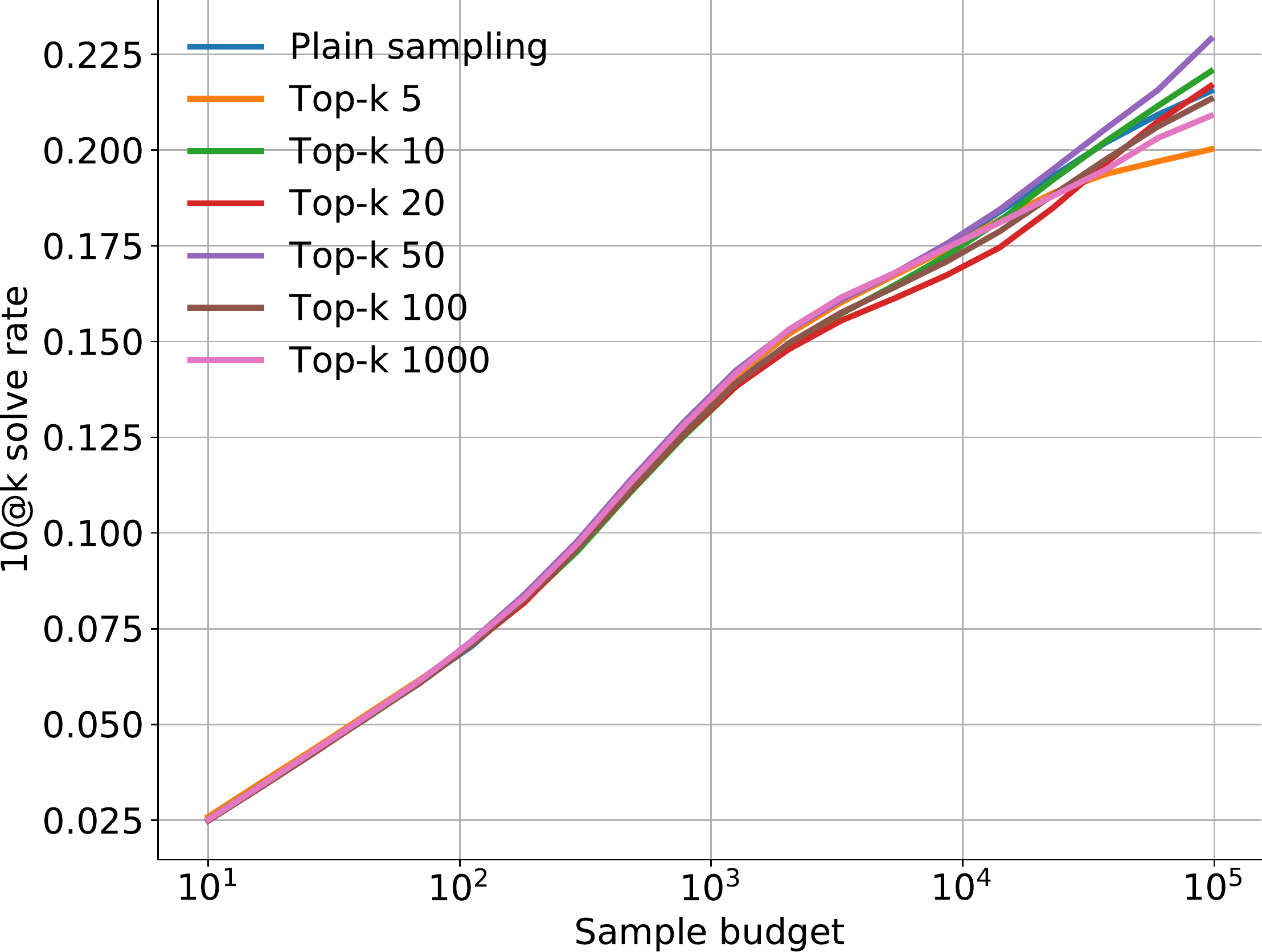} &
        \includegraphics[width=0.48\textwidth]{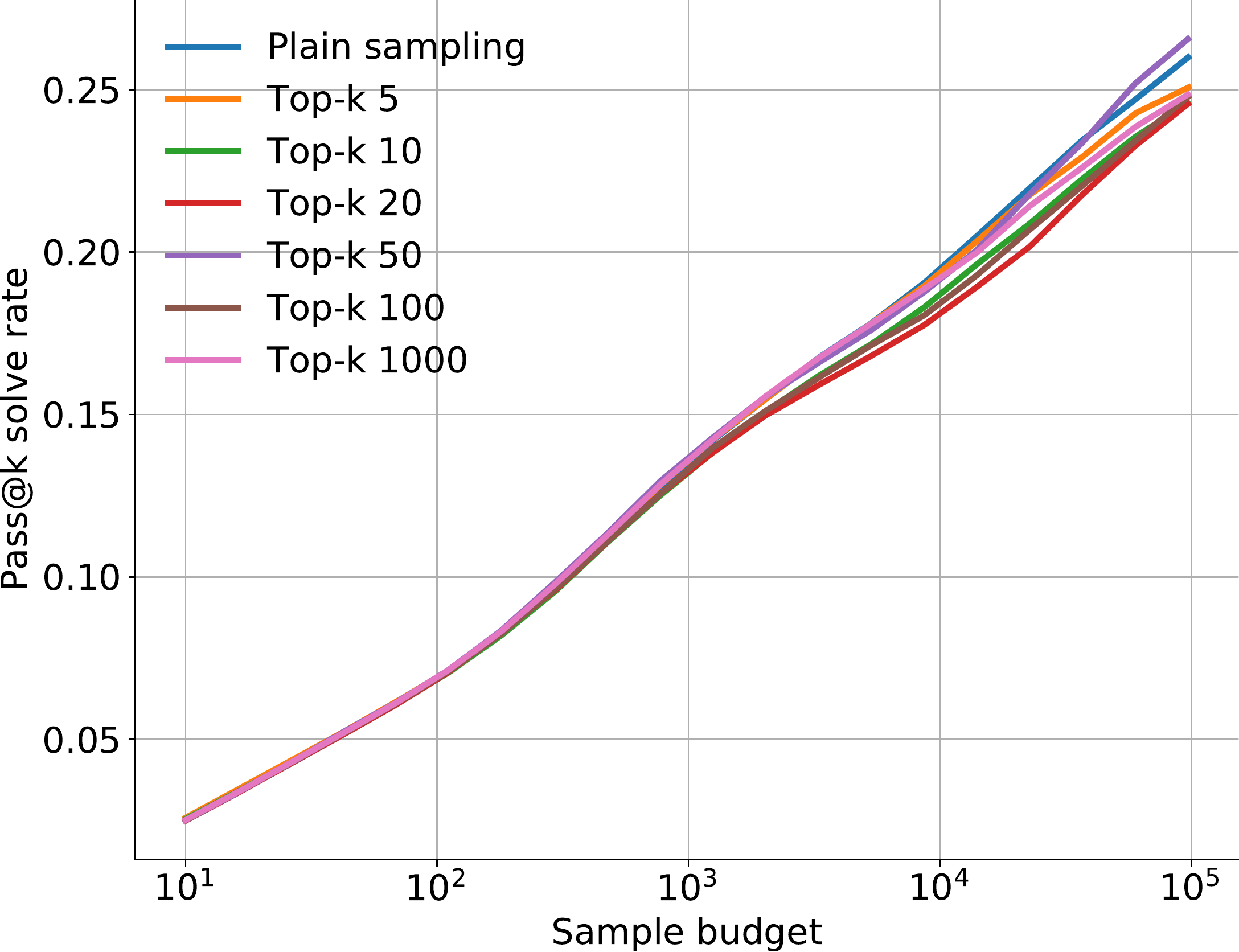} \\
        (c) Top-k sampling: 10@k & (d) Top-k sampling: pass@k \\
        \includegraphics[width=0.48\textwidth]{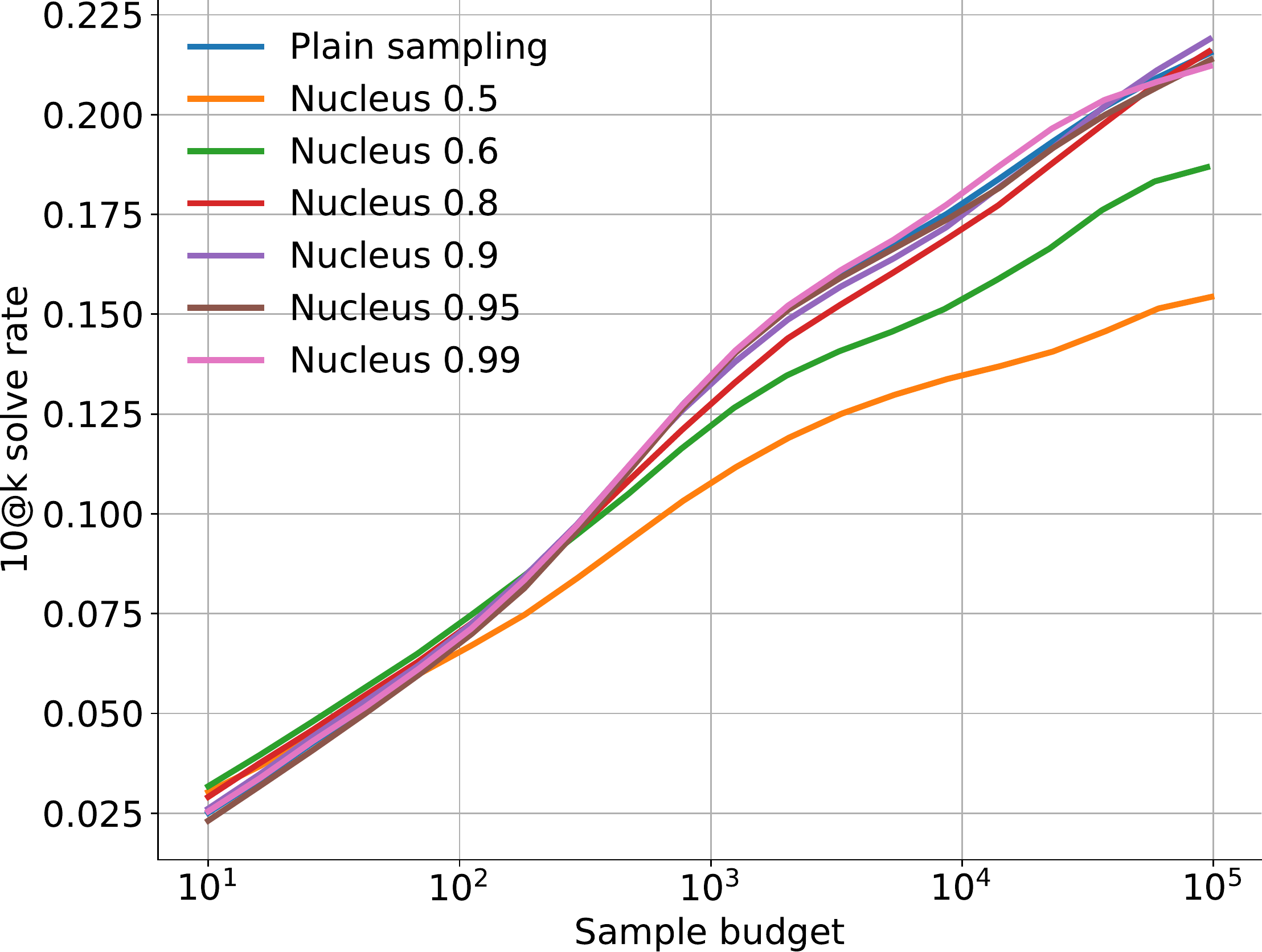} &
        \includegraphics[width=0.48\textwidth]{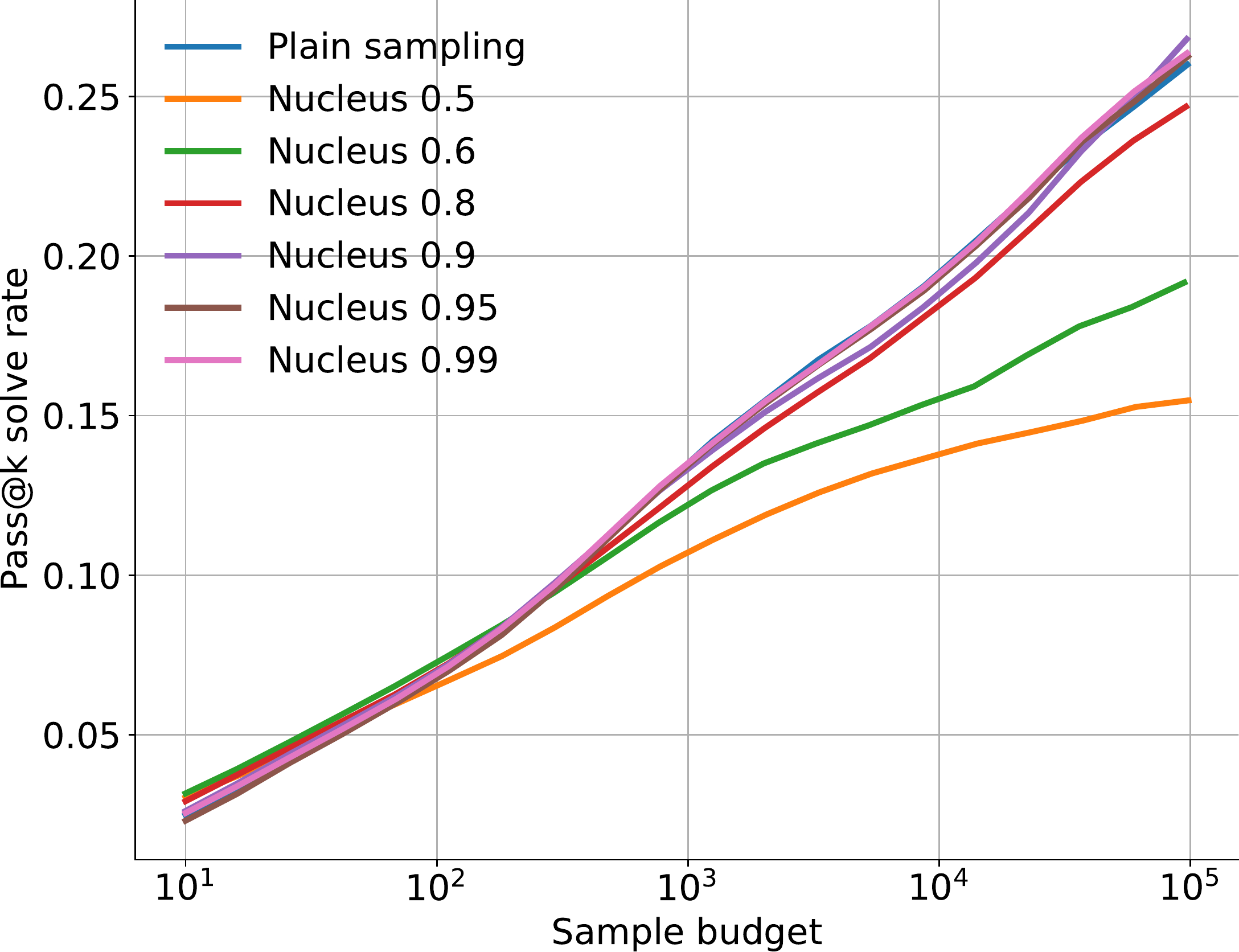} \\
        (e) Nucleus sampling: 10@k & (f) Nucleus sampling: pass@k.
    \end{tabular}
    \caption{\textbf{Sampling with temperature, top-k sampling and nucleus sampling}.
    (a-b) 
    Lower temperatures are better for small sample budgets, and higher temperatures are better for large sample budgets. However, the model is relatively tolerant to a wide range of temperatures. (c-d) Top-k sampling does not outperform the sampling with temperature baseline.
    (e-f) Nucleus sampling does not significantly improve the sampling baseline.
    Performance increases with the nucleus size, with an optimal size close to 1.0 (i.e. plain sampling).
    }
    \label{fig:different-sampling-settings}
\end{figure}

As we generate a large amount ($\ge$ 1M) of samples for each problem, the exact settings to use for sampling can potentially have a large impact on the model performance.

In \figref{fig:different-sampling-settings}(a-b), we show that sampling temperature does have an impact on the solve rate, but the temperature of $T=0.25$ works well across a wide range of sample budgets.
\figref{fig:different-sampling-settings}(c-e) shows the results for top-k sampling and nucleus sampling.  Both of them turn out to be not better than the simple sampling with temperature.

\subsection{Scaling with dataset size}\label{appendix:scaling-with-dataset-size}

\begin{figure}[t]
    \centering
    \begin{tabular}{cc}
        \includegraphics[width=0.47\textwidth]{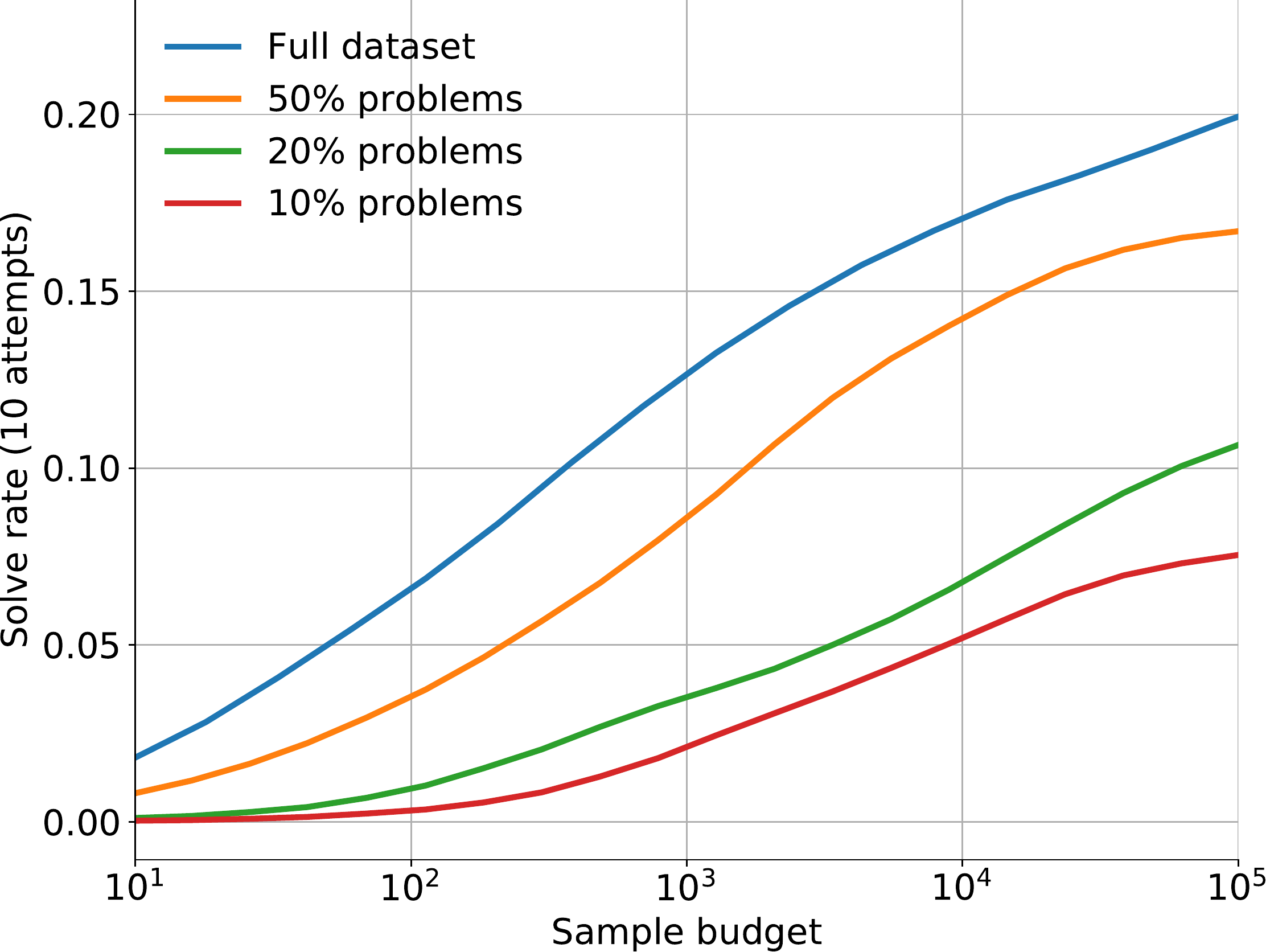} & \includegraphics[width=0.47\textwidth]{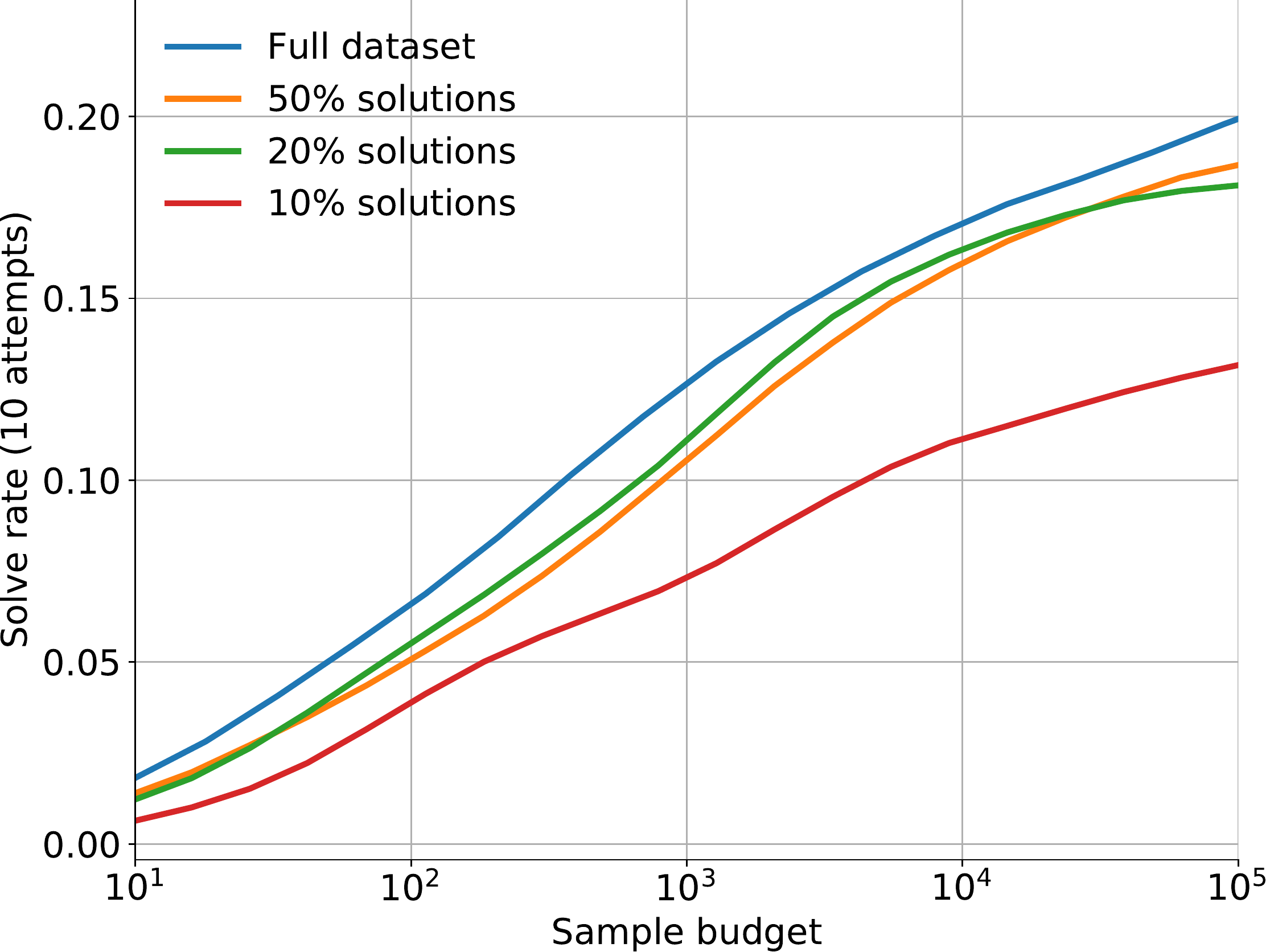} \\
        (a) Varying numbers of problems. & (b) Varying numbers of solutions.
    \end{tabular}
    \caption{\textbf{Dataset scaling}. Increasing either the number of problems (a) or solutions (b) in the finetuning dataset improves solve rate of an early 1B parameter model, although increasing the number of solutions has a smaller effect.}
    \label{fig:dataset-scaling}
\end{figure}

Besides scaling with model size, compute, and number of samples presented in \secref{sec:large-scale-sampling-and-models}, we also observe that, as expected, the model performance scales with the dataset size.
\figref{fig:dataset-scaling} shows how the model performance scales with larger datasets.  Increasing the number of problems has a more significant positive impact than increasing the number of solutions for each problem.

\section{Codeforces contest evaluation}
\label{sec:codeforces-eval}

In this section we describe the procedure and detailed settings we used for the evaluation on Codeforces contests, as well as detailed results of that evaluation. Summarized evaluation results are presented in \secref{sec:codeforces-results}.

\subsection{Simulation}
Contest scoring is based not only on whether or not a problem is solved, but also on when in the contest it is solved and how many incorrect submissions there were before a correct one. Our procedure takes all three into account.

For each contest, we simulated running live with 3750 TPUv4 and 3750 TPUv4i~\citep{jouppi2021ten} chips for the duration of the contest. These devices continuously draw samples, using a pool of workers to evaluate samples against example tests. Because clustering runs best when there are enough samples that pass example tests, we cannot continuously submit submissions. Instead, for each problem in the contest we clustered and submitted up to a total of 10 samples at three points (\tabref{tab:codeforces-submission}), where the point was either based on the fraction of the time remaining in the competition or based on the relative number of samples that pass example tests available compared to the default number. When computing these points, all contests were assumed to be two hours for simplicity even though some were slightly longer. Clustering and a specified number of submissions were done when either of these conditions were reached. The time of correct submission was then the time in the contest that corresponds to when the condition for each row was reached, plus 120 seconds for clustering. Because submissions were submitted in order, we also know the number of incorrect submissions. Additionally, we did not submit solutions to Codeforces that were obviously incorrect (by quick manual inspection) and instead automatically counted them as incorrect, though note that this can only decrease model performance.
 
\begin{table}[t]
    \centering
    \begin{tabular}{p{0.122\textwidth}p{0.155\textwidth}p{0.45\textwidth}}
    \toprule
    Number of submissions & Fraction of time left in contest & Relative number of samples that pass example tests, compared to the clustering default \\
    \midrule
    1 & 0.1 & 0.005 \\
    5 & 0.5 & 0.05 \\
    Remaining & 0.9 & 1.0 \\
    \bottomrule
    \end{tabular}
    \caption{\textbf{Codeforces submission points}. Each row specifies the conditions for when submissions happen, and the number of submissions at that point.}
    \label{tab:codeforces-submission}
\end{table}

In practice, however, we only simulated running live and instead submitted our solutions after the contest already ended. One consequence of this is that we do not fully consider the ``hacking'' phase of the competition, where competitors can get points for finding vulnerabilities in the code of others. This is only a factor for some competitions, hacking is not generally a major concern, and since the model itself does not hack others the only consequence is that the system receives more timely feedback about correctness on edge cases, so we do not believe this omission is significant.

A more significant consequence is that, as described in \appendixref{sec:program-evaluation}, we filter multiple output problems based on the consensus example output produced by human solutions, rather than the example output itself (which may be intentionally designed to cover multiple possible ways of solving the problem, and therefore not be output by any particular human solution). This example output change affected approximately five problems \OurApproachShort{} solved.

\subsection{Multiple evaluations}

After the first evaluation, we decided to run the evaluation procedure multiple times to measure variance.  Variance in this evaluation process can come from the  (1) trained model, (2) set of samples, (3) ordering of samples, and (4) clustering process.
Due to compute limitations, we did not retrain or resample the models for each evaluation, and instead tried to measure variance from (3) and (4). We used the samples from the first evaluation, but reshuffled them to simulate drawing them in a different order. We also used a different set of inputs generated by the input generation model for clustering, and selected samples from clusters with a different random seed.

\subsection{Results}

After submissions we computed our score on each contest (including penalties) using the contests' scoring system, and found where the model would have placed on the contests' official scoreboards. Per-problem contest results can be found in \tabref{tab:codeforces-result}. Overall contest results can be found in \tabref{tab:codeforces-contest-result}. In the second and third evaluations, we submitted more than 10 submissions per problem. We found that there were some problems we only solved with many samples. 
\begin{table}[tp]
\begin{center}
\begin{tabular}{l c c l l l c}
\toprule 
Contest & Problem & Problem & Solved & Incorrect & Submission & Problem \\
& Rating & Letter &  & Submissions & Time (minutes) & Link \\
\midrule
1623 & 800 & A & 1,1,1 & 0,0,0 & 14,14,14 & \href{https://codeforces.com/contest/1623/problem/A}{link} \\
1623 & 1100 & B & 1,1,1 & 0,0,0 & 14,14,14 & \href{https://codeforces.com/contest/1623/problem/B}{link}\\
1622 & 800 & A & {\color{Mahogany}0},1,1 & {\color{Mahogany}-},{\color{Mahogany}345},7 & {\color{Mahogany}-},{\color{Mahogany}110},110 & \href{https://codeforces.com/contest/1622/problem/A}{link}\\
1622 & 1000 & B & 1,1,1 & 0,0,1 & 14,14,62 & \href{https://codeforces.com/contest/1622/problem/B}{link}\\
1615 & 800 & A & 1,1,1 & 4,0,0 & 10,2,3 & \href{https://codeforces.com/contest/1615/problem/A}{link}\\
1615 & 1300 & B & {\color{Mahogany}0},{\color{Mahogany}0},{\color{Mahogany}1} & {\color{Mahogany}-},{\color{Mahogany}-},{\color{Mahogany}129} & {\color{Mahogany}-},{\color{Mahogany}-},{\color{Mahogany}110} & \href{https://codeforces.com/contest/1615/problem/B}{link}\\
1615 & 1600 & C & {\color{Mahogany}0},{\color{Mahogany}1},{\color{Mahogany}0} & {\color{Mahogany}-},{\color{Mahogany}170},{\color{Mahogany}-} & {\color{Mahogany}-},{\color{Mahogany}110},{\color{Mahogany}-} & \href{https://codeforces.com/contest/1615/problem/C}{link}\\
1619 & 800 & A & 1,1,1 & 0,0,0 & 2,2,2 & \href{https://codeforces.com/contest/1619/problem/A}{link}\\
1619 & 800 & B & 1,1,1 & 0,0,0 & 3,3,3 & \href{https://codeforces.com/contest/1619/problem/B}{link}\\
1619 & 1800 & D & {\color{Mahogany}0},{\color{Mahogany}0},{\color{Mahogany}1} & {\color{Mahogany}-},{\color{Mahogany}-},{\color{Mahogany}91} & {\color{Mahogany}-},{\color{Mahogany}-},{\color{Mahogany}110} & \href{https://codeforces.com/contest/1619/problem/D}{link}\\
1620 & 800 & A & {\color{Mahogany}0},1,{\color{Mahogany}1} & {\color{Mahogany}-},7,{\color{Mahogany}10} & {\color{Mahogany}-},110,{\color{Mahogany}110} & \href{https://codeforces.com/contest/1620/problem/A}{link}\\
1620 & 1000 & B & 1,1,1 & 0,0,0 & 14,14,14 & \href{https://codeforces.com/contest/1620/problem/B}{link}\\
1617 & 900 & B & 1,1,1 & 2,3,2 & 62,62,62 & \href{https://codeforces.com/contest/1617/problem/B}{link}\\
1618 & 800 & A & 1,1,1 & 0,0,0 & 4,4,3 & \href{https://codeforces.com/contest/1618/problem/A}{link}\\
1618 & 800 & B & {\color{Mahogany}0},1,{\color{Mahogany}1} & {\color{Mahogany}-},5,{\color{Mahogany}14} & {\color{Mahogany}-},110,{\color{Mahogany}110} & \href{https://codeforces.com/contest/1618/problem/B}{link}\\
1618 & 1300 & D & 1,1,1 & 8,4,8 & 110,110,110 & \href{https://codeforces.com/contest/1618/problem/D}{link}\\
1618 & 1700 & E & 1,1,1 & 1,0,1 & 62,110,62 & \href{https://codeforces.com/contest/1618/problem/E}{link}\\
1591 & 800 & A & 1,1,1 & 3,0,0 & 9,3,2 & \href{https://codeforces.com/contest/1591/problem/A}{link}\\
1591 & 900 & B & 1,1,1 & 0,0,0 & 2,3,2 & \href{https://codeforces.com/contest/1591/problem/B}{link}\\
1608 & 800 & A & 1,1,1 & 0,0,0 & 2,2,2 & \href{https://codeforces.com/contest/1608/problem/A}{link}\\
1613 & 900 & A & {\color{Mahogany}0},{\color{Mahogany}0},{\color{Mahogany}1} & {\color{Mahogany}-},{\color{Mahogany}-},{\color{Mahogany}359} & {\color{Mahogany}-},{\color{Mahogany}-},{\color{Mahogany}110} & \href{https://codeforces.com/contest/1613/problem/A}{link}\\
1613 & 1000 & B & {\color{Mahogany}0},1,1 & {\color{Mahogany}-},0,0 & {\color{Mahogany}-},19,14 & \href{https://codeforces.com/contest/1613/problem/B}{link}\\
1613 & 1200 & C & 1,1,1 & 3,8,4 & 18,110,19 & \href{https://codeforces.com/contest/1613/problem/B}{link}\\
\bottomrule
\end{tabular}
\end{center}
\vspace{-0.5cm}
\caption{
\textbf{Codeforces per-problem results}. Problems on Codeforces are grouped into contests, and uniquely identified by their contest and problem letter. For each solved problem, we specify whether they were solved (1 for solved), how many incorrect submissions there were before a correct one, and the simulated submission time in minutes. Unsolved problems are not listed. Results corresponding to the three evaluations are separated by commas. ``-'' indicates that the problem in the evaluation was unsolved. Submitted programs can be found on our 3 accounts on Codeforces: \href{https://codeforces.com/submissions/SelectorUnlimited}{SelectorUnlimited},  \href{https://codeforces.com/submissions/WaggleCollide}{WaggleCollide}, and  \href{https://codeforces.com/submissions/AngularNumeric}{AngularNumeric}. Note that for the first evaluation, we only submitted at most 10 times per problem. Unsolved problems and submissions beyond 10 (for the second and third evaluations) are marked in {\color{Mahogany}red} to indicate that they are not included when measuring performance in the setting with a maximum of 10 submissions per problem. 
}
\label{tab:codeforces-result}
\end{table}

\begin{table}[t]
\begingroup
\setlength{\tabcolsep}{5pt}
\begin{center}
\begin{tabular}{l c c c c}
\toprule 
Contest & 10 Submissions & Unlimited & 10 Submissions & 10 Submissions \\
&  & Submissions & Min Penalty Time & Max Penalty Time \\
\midrule
1623 & 20.9\%,20.9\%,20.9\% & -,20.9\%,20.9\% & 20.6\%,20.6\%,20.6\% & 55.3\%,55.3\%,55.3\% \\
1622 & 68.6\%,68.6\%,61.5\% & -,61.5\%,61.5\% & 61.5\%,61.5\%,49.5\% & 91.2\%,91.6\%,61.5\% \\
1615 & 91.2\%,47.2\%,48.9\% & -,41.5\%,44.9\% & 91.2\%,45.1\%,45.2\% & 92.6\%,89.2\%,89.2\% \\
1619 & 47.2\%,47.3\%,47.2\% & -,37.3\%,47.1\% & 47.1\%,47.1\%,45.2\% & 88.1\%,88.1\%,88.1\% \\
1620 & 64.4\%,61.1\%,64.4\% & -,61.1\%,62.9\% & 63.7\%,65.6\%,63.7\% & 88.8\%,82.9\%,80.8\% \\
1617 & 72.9\%,75.9\%,72.9\% & -,75.9\%,72.9\% & 64.9\%,65.6\%,64.9\% & 82.0\%,82.9\%,82.0\% \\
1618 & 62.3\%,32.1\%,62.3\% & -,32.1\%,35.1\% & 43.0\%,10.6\%,43.0\% & 62.7\%,35.1\%,62.7\% \\
1591 & 53.6\%,39.7\%,39.7\% & -,39.7\%,39.7\% & 51.1\%,39.7\%,39.7\% & 74.9\%,74.4\%,74.4\% \\
1608 & 46.3\%,46.3\%,46.3\% & -,46.3\%,46.3\% & 43.6\%,43.6\%,43.6\% & 95.7\%,95.7\%,95.7\% \\
1613 & 75.6\%,68.2\%,54.5\% & -,68.2\%,50.1\% & 72.9\%,55.3\%,51.1\% & 84.6\%,70.2\%,70.1\% \\
\midrule
Average & 60.3\%,50.8\%,51.9\% & -,49.5\%,48.2\% & 55.9\%,42.4\%,46.8\% & 84.6\%,70.2\%,70.1\% \\
Average all & 54.3\% & 48.8\% & 48.4\% & 77.2\% \\
\bottomrule
\end{tabular}
\end{center}
\vspace{-0.5cm}
\caption{
\textbf{Codeforces per-contest results}. Results of running \OurApproachShort{} on all Codeforces competitions. Top XX\%  percentile ranks are given, where the number indicates the percentage of competitors who performed better than our system. Results for the three evaluations are separated by commas. We did not track unlimited attempts numbers for the first evaluation.
}
\label{tab:codeforces-contest-result}
\endgroup
\end{table}

We also computed our estimated Codeforces Elo score by tracking what our Elo would have been if we started with the first contest, and competed in each contest in the order they were released, placing according to our calculated placement in \tabref{tab:codeforces-contest-result}. This was done separately for all three evaluations, and then averaged. 

Our Elo estimation is based on our reproduction of the Codeforces Elo method, as we didn't compete live. We checked correctness by reproducing other participants' Elo scores. Our approach largely matches the Codeforces Elo (differing by $<15$ points), but our Elo score is still only an estimation.

\section{Additional analysis of \OurApproachShort{}'s capabilities and limitations}
\label{appendix:capabilities-and-limitations}

\subsection{Model sample statistics} \label{sec:syntax-table}
\tabref{tab:syntax} shows the percentage of syntactically correct samples from the models. That is, samples that compile for \CC{}, and samples that do not generate a SyntaxError for Python.

\begin{table}[tp]
    \centering
    \begin{tabular}{cccccc}
    \toprule
    Language & 300M & 1B & 3B & 9B & 41B  \\
    \midrule
    \CC{} & $67.5\%$ & $63.7\%$ & $64.1\%$ & $61.6\%$ & $66.8\%$ \\ 
    Python & $90.4\%$ & $88.6\%$ & $90.1\%$ & $89.5\%$ & $88.8\%$ \\ 
    \bottomrule
    \end{tabular}
    \caption{\textbf{Percentage of syntactically correct model samples}. We report results on the validation set, for each language and model size.
    }
    \label{tab:syntax}
\end{table}

\subsection{Solve rate for different problem difficulty ratings}
\label{appendix:solve-rate-for-different-ratings}

\tabref{tab:solve-rate-in-difficulty-buckets} shows the 10@100k solve rates for our 9B and 41B models on the validation and test sets in different problem difficulty ratings.  The validation and test sets contain 117 and 165 problems respectively, and the number of problems in each bucket can be quite small. Therefore, the numbers in this table may have large variance.  Also note that due to the lack of long test cases, our evaluation is susceptible to accepting correct but slow solutions (see \secref{sec:false-positives}).  The high difficulty rating problems are particularly affected by this.

As expected, we see that overall our models perform significantly better on problems with low difficulty ratings, and the performance quickly degrades when the problem difficulty increases.  High solve rates at high difficulty buckets are caused by our slow positive rate being quite high (46\% as reported in \secref{sec:false-positives}), and by the buckets being quite small.

\begin{table}[th]
\begingroup
\setlength{\tabcolsep}{3pt}
    \centering
    \begin{tabular}{ccccccccc}
\toprule
Split & Bucket & 800-1100 & 1200-1500 & 1600-1900 & 2000-2300 & 2400-2700 & 2800-3100 & 3200-3500 \\
\midrule
\multirow{3}{*}{Valid} & \#Probs. & 29 & 18 & 20 & 19 & 15 & 8 & 8 \\
& 9B & 60.5\% & 18.0\% & 10.8\% & 26.6\% & 8.8\% & 12.5\% & 21.9\% \\
& 41B & 62.4\% & 17.3\% & 15.9\% & 19.7\% & 7.8\% & 14.9\% & 31.6\% \\
\midrule
\multirow{3}{*}{Test} & \#Probs. & 37 & 21 & 25 & 31 & 26 & 13 & 12 \\
& 9B & 64.7\% & 34.1\% & 17.7\% & 15.7\% & 8.4\% & 0.0\% & 0.0\% \\
& 41B & 69.5\% & 35.4\% & 20.6\% & 13.8\% & 7.8\% & 7.5\% & 0.0\% \\
\bottomrule
\end{tabular}
    \caption{\textbf{10@100k Solve rates of our models in different difficulty buckets.} Also shown is the number of problems in each difficulty bucket.}
    \label{tab:solve-rate-in-difficulty-buckets}
    \endgroup
\end{table}

\subsection{Sensitivity to the problem descriptions}
\label{sec:sensitivity-to-rewordings-appendix}
We measured model performance in response to assorted changes in the problem description, to see if the model responds appropriately and makes use of the description. Because of compute limitations, we were unable to retrain models on these modifications, but instead sampled already  trained models. The main metric used for most analyses was the 10@1024 solve rate, that is, solve rate using 10 submissions from 1024 samples.  We found overall that the problem description is critical, and \OurApproachShort{} does not simply brute force possible solutions.

\subsubsection{Simplification of problem descriptions}

Understanding what to implement is a key component of competitive programming problems. It involves parsing the problem statement (which is often phrased as a story), and coming up with the insights needed to solve it. If our model makes use of this statement, a simplified statement should improve its solve rate as this helps with problem understanding.

\begin{figure}[t]
\footnotesize
\begin{center}
\fbox{
\begin{minipage}[t]{.63\textwidth}
\textbf{Mocha and Match}\\
Mocha is a young girl from high school. She has learned so much interesting knowledge from her teachers, especially her math teacher. Recently, Mocha is learning about binary system and very interested in bitwise operation.
\\ \\
This day, Mocha got a sequence $a$ of length $n$. In each operation, she can select an arbitrary interval $[l,r]$ and for all values $i$ $(0 \leq i \leq r - l)$, replace $a_{l+i}$ with $a_{l+i} \& a_{r-i}$ at the same time, where $\&$ denotes the \href{https://en.wikipedia.org/wiki/Bitwise_operation\#AND}{bitwise AND operation}. This operation can be performed \textbf{any number of times}.
\\ \\
For example, if $n=5$, the array is $[a_1,a_2,a_3,a_4,a_5]$, and Mocha selects the interval $[2,5]$, then the new array is $[a_1, a_2 \& a_5, a_3 \& a_4, a_4 \& a_3, a_5 \& a_2]$.
\\ \\
Now Mocha wants to minimize the maximum value in the sequence. As her best friend, can you help her to get the answer?
\\ \\
\textbf{Input}\\
Each test contains multiple test cases.
\\
The first line contains a single integer $t$ $(1 \leq t \leq 100)$ -- the number of test cases. Each test case consists of two lines.
\\
The first line of each test case contains a single integer $n$ $(1 \leq n \leq 100)$ -- the length of the sequence.
\\
The second line of each test case contains $n$ integers $a_1, a_2, ..., a_n (0 \leq a_i \leq 10^9)$.
\\ \\
\textbf{Output}\\
For each test case, print one integer -- the minimal value of the maximum value in the sequence.
\end{minipage}
}
\hspace{0.1cm}
\begin{minipage}[t]{.31\textwidth}
\footnotesize
\textbf{Example Input}
\begin{lstlisting}
4
2
1 2
3
1 1 3
4
3 11 3 7
5
11 7 15 3 7
\end{lstlisting}
\textbf{Example Output}
\begin{lstlisting}
0
1
3
3
\end{lstlisting}
\textbf{Explanation}\\
In the first test case, Mocha can choose the interval $[1,2]$, then the sequence becomes $[0,0]$, where the first element is $1\&2$, and the second element is $2\&1$.
\\ \\
In the second test case, Mocha can choose the interval $[1,3]$, then the sequence becomes $[1,1,1]$, where the first element is $1\&3$, the second element is $1\&1$, and the third element is $3\&1$.
\end{minipage}
\end{center}
\vspace{-0.5cm}
\caption{\textbf{Example problem statement}. Problem statement of \textbf{\href{https://codeforces.com/problemset/problem/1559/A}{Mocha and Math}}, a Codeforces problem~\citep{codeforces2020}. This is an easy problem, with a rating of 900.
}
\label{fig:mocha-and-math}
\end{figure}

For example, consider Codeforces problem 1559A, \emph{Mocha and Math}, shown in full in~\figref{fig:mocha-and-math}. The key observation to solve this problem is to note that the optimal solution is achieved when every element is bitwise ANDed with every other element. We can therefore simplify this problem by changing the description part of the problem statement to a single sentence:

\begin{quote}
Given a sequence of integers, compute the bitwise AND of all of its elements.
\end{quote}

\begin{table}[h]
    \centering

    \begin{tabular}{lrr}
    \toprule
    Problem
    & Original & Simplified \\
    \midrule
1554A Cherry & 2.98\% & 15.74\% \\
1559A Mocha and Math & 12.25\% & 55.53\% \\
1569A Balanced Substring & 10.61\% & 31.97\% \\
No consecutive zeros & 0.17\% & 1.25\% \\
Nim & 0.95\% & 85.38\% \\
\bottomrule
    \end{tabular}
    \caption{\textbf{Performance on original vs. simplified problems}. The percentage of correct samples from a total of 100k samples, for original problem wordings and rewordings which make the required algorithm more explicit.  This result was obtained using a 1B parameter model.}
    \label{tab:rewording-solves}
\end{table}

We find that as expected, simplifying problem descriptions this way significantly improves our model's performance (\tabref{tab:rewording-solves}). The solve rate for our base 1B parameter model on this particular problem increased from 12\% to 55\%. We performed this analysis for five problems: three from our validation set and two hand-written ourselves. The full wordings of these problems are included in  \appendixref{sec:simplified-rewordings}.

\subsubsection{Incorrect or irrelevant rewordings}

To investigate what parts of the problem description the model pays attention to and how strongly it conditions on the description, we investigated how the solve rate for a problem changes when we add irrelevant information to the description, or reword it to be under-specified or incorrect. We performed this analysis on the simplified \emph{Cherry} problem, measuring solve rate of different rewordings. Full problem descriptions can be found in \appendixref{sec:incorrect-rewordings}.

\begin{table}[t]
    \centering
    \begin{tabular}{lr}
    \toprule
    Rewording & Solve rate  \\
    \midrule
    Original (maximum product of two consecutive array elements) & 17.1\% \\
    Opposite (minimum product of two consecutive array elements) & 0.1\% \\
    Related (maximum product of two any array elements) & 3.2\% \\
    Underspecified (maximum function of two consecutive array elements) & 0.03\% \\
    Verbose & 19.4\% \\
    Algorithm described in words & 19.7\% \\
    \bottomrule
    \end{tabular}
    \caption{\textbf{Rewording the \textit{Cherry} problem}. The percentage of solutions in 50000 samples from the 1B parameter model when attempting the simplified version of the \emph{Cherry} problem with different rewordings.}
    \label{tab:rewording-simplified}
\end{table}

The results in~\tabref{tab:rewording-simplified} continue to suggest that the model is strongly conditioning on the description (rather than, for example, brute forcing all possible solutions related to the problem domain). The model is also able to parse algorithm descriptions from either symbols or from natural language, and can ignore irrelevant natural language explanations; indeed, the model actually does better with more language-heavy descriptions, perhaps because of the verbose nature of the training distribution.

\subsubsection{Capturing variables and their relations}
\label{sec:appendix-var-relations}
Problem descriptions include multiple variable names to denote objects and quantities of interest, and to specify relationships between them.
For example, a problem might feature an array, $a$, of length $n$, or two integers called $n$ and $m$ with $n<m$.
Understanding these variables and their relations is necessary for solving problems.

\begin{figure}[h]
\footnotesize
\begin{center}
\fbox{
\begin{minipage}[ht]{1.0\textwidth}
\textbf{Gregor and Cryptography}\\
Gregor is learning about RSA cryptography, and although he doesn't understand how RSA works, he is now fascinated with prime numbers and factoring them. Gregor's favorite prime number is \textbf{\textcolor{blue}{[P]}\textcolor{ForestGreen}{[H]}\textcolor{red}{[b]}}. Gregor wants to find two bases of \textbf{\textcolor{blue}{[P]}\textcolor{ForestGreen}{[H]}\textcolor{red}{[y]}}. Formally, Gregor is looking for two integers a and b which satisfy both of the following properties. \\ 
* \textbf{\textcolor{blue}{[P]}\textcolor{ForestGreen}{[H]}\textcolor{red}{[b]}} mod a = \textbf{\textcolor{blue}{[P]}\textcolor{ForestGreen}{[H]}\textcolor{red}{[b]}} mod b, where x mod \textbf{\textcolor{blue}{[y]}\textcolor{ForestGreen}{[f]}\textcolor{red}{[b]}} denotes the remainder when x is divided by \textbf{\textcolor{blue}{[y]}\textcolor{ForestGreen}{[f]}\textcolor{red}{[x]}}, and \\
 * 2 $\leq$ a < b $\leq$ \textbf{\textcolor{blue}{[P]}\textcolor{ForestGreen}{[H]}\textcolor{red}{[t]}}. 
\\
Help Gregor find two bases of his favorite prime number!
\\ \\
\textbf{Input}\\
Each test contains multiple test cases. The first line contains the number of test cases \textbf{\textcolor{blue}{[t]}\textcolor{ForestGreen}{[m]}\textcolor{red}{[b]}} (1 $\leq$ \textbf{\textcolor{blue}{[t]}\textcolor{ForestGreen}{[m]}\textcolor{red}{[b]}} $\leq$ 1000). Each subsequent line contains the integer \textbf{\textcolor{blue}{[P]}\textcolor{ForestGreen}{[H]}\textcolor{red}{[b]}} (5 $\leq$ \textbf{\textcolor{blue}{[P]}\textcolor{ForestGreen}{[H]}\textcolor{red}{[x]}} $\leq$ ${10}^9$), with \textbf{\textcolor{blue}{[P]}\textcolor{ForestGreen}{[H]}\textcolor{red}{[b]}} guaranteed to be prime.
\\ \\
\textbf{Output}\\
Your output should consist of \textbf{\textcolor{blue}{[t]}\textcolor{ForestGreen}{[m]}\textcolor{red}{[t]}} lines. Each line should consist of two integers a and b (2 $\leq$ a < b $\leq$ \textbf{\textcolor{blue}{[P]}\textcolor{ForestGreen}{[H]}\textcolor{red}{[b]}}). If there are multiple possible solutions, print any.
\end{minipage}
}\hspace{0.5cm}
\end{center}
\caption{\textbf{Consistent / inconsistent variable renaming example}. This example shows the original description that has variable names $P$, $y$, and $t$, along with versions with consistent and inconsistent replacement. The original description's variables are the \textbf{\textcolor{blue}{blue}} variables in each triplet \textbf{\textcolor{blue}{[A]}\textcolor{ForestGreen}{[B]}\textcolor{red}{[C]}}, and similarly the consistent and inconsistent replacements are \textbf{\textcolor{ForestGreen}{green}} ($A$ replaced with $B$ everywhere) and \textbf{\textcolor{red}{red}} ($A$ replaced with a random name $C$ independently for each appearance) respectively.
Consistent replacement does not change the problem, but random replacement introduces noise and renders the problem ill-posed. Model deterioration with such modifications allows analysis of how well a model captures the given variable structure. Problem sourced from \href{https://codeforces.com}{Codeforces}.
}
\label{fig:varname_replacement_example}
\end{figure}

We investigated two changes: random variable name substitutions either consistently applied throughout a problem, or applied at random for each instance such that consistency is not guaranteed (which renders the problem ill-posed). Perturbations were done on up to 6 different variables, maintaining consistent character case, and variables were replaced by other existing variable names from the same problem (see \figref{fig:varname_replacement_example} for a concrete example).

\begin{figure}[t]
    \centering
    \includegraphics[scale=1]{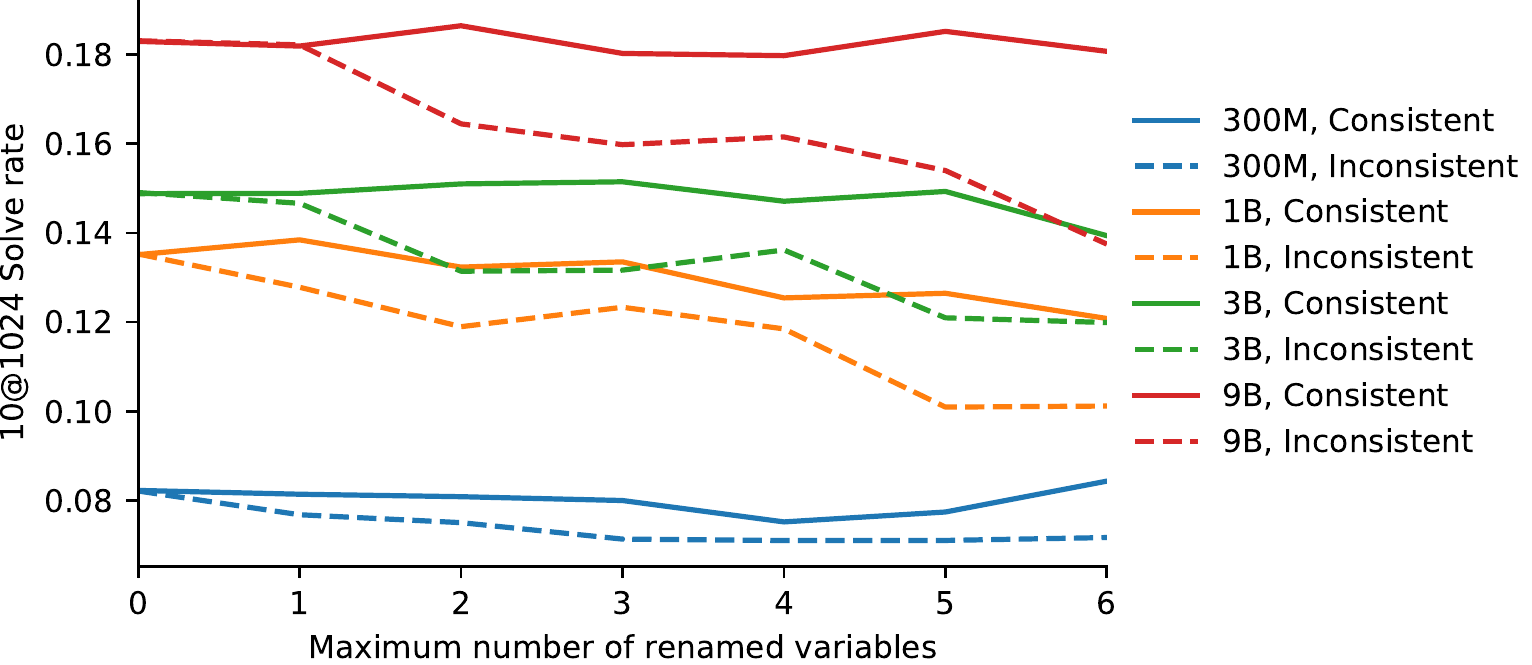}
    \caption{\textbf{Sensitivity to variable renaming}. The 10@1024 solve rate for consistent and inconsistent (i.e.~ill-posed) variable renaming, for different model sizes. Larger models are robust to invariant variable renaming, but deteriorate with greater amounts of inconsistent renaming.}
    \label{fig:variable-renaming-scaled}
\end{figure}

\figref{fig:variable-renaming-scaled} shows the results of this evaluation.
The small (300M) model is largely unaffected by both consistent and inconsistent renaming, suggesting that it does not model the different variables well. 
As model size increases however, the relative performance drop observed with ill-formed inputs becomes more and more pronounced, while sensitivity to consistent variable renaming decreases.
This suggests that as models get larger, they are increasingly able to capture relevant relationships between variables described in the problem formulation.
The non-trivial solve rate for ill-posed problems, on the other hand, suggests that other parts in the problem description provide important cues for a solution, which models can learn to leverage.

\subsubsection{Sensitivity to word-level changes}

\begin{figure}[t]
    \centering
    \begin{tabular}{ll}
    (a) Transposing characters & (b) Replacing words with synonyms \\
    \vspace{0.05cm} \\
    \includegraphics[width=0.48\textwidth]{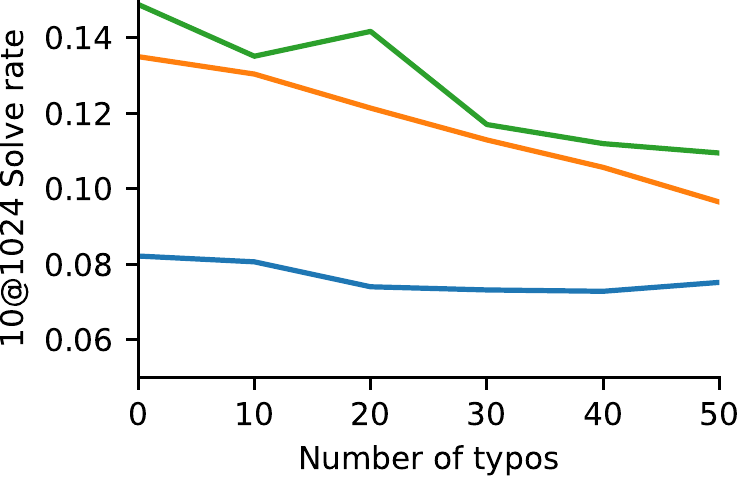} &
    \includegraphics[width=0.48\textwidth]{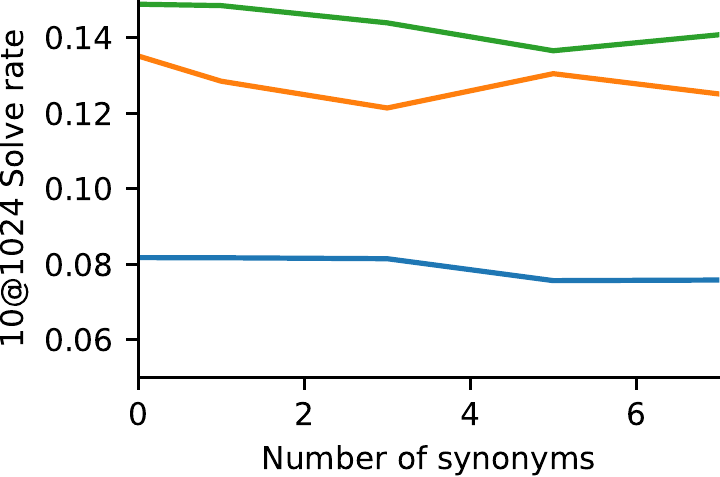} \\
    \vspace{0.1cm} \\
    (c) Permuting words & (d) Deleting words probabilistically \\
    \vspace{0.05cm} \\
    \includegraphics[width=0.48\textwidth]{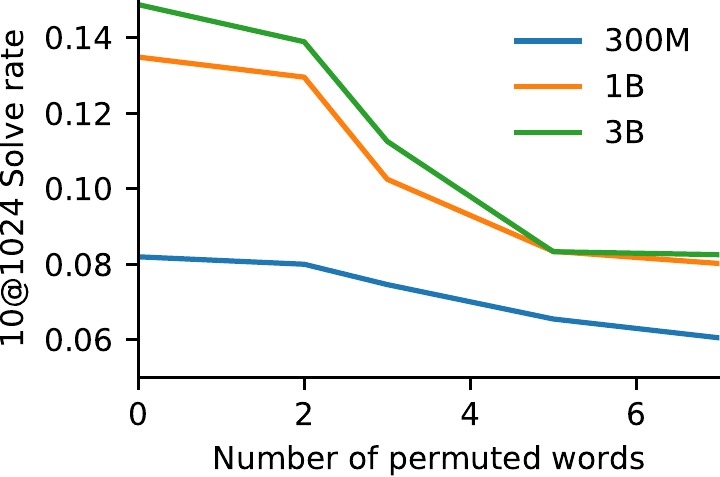} &
    \includegraphics[width=0.48\textwidth]{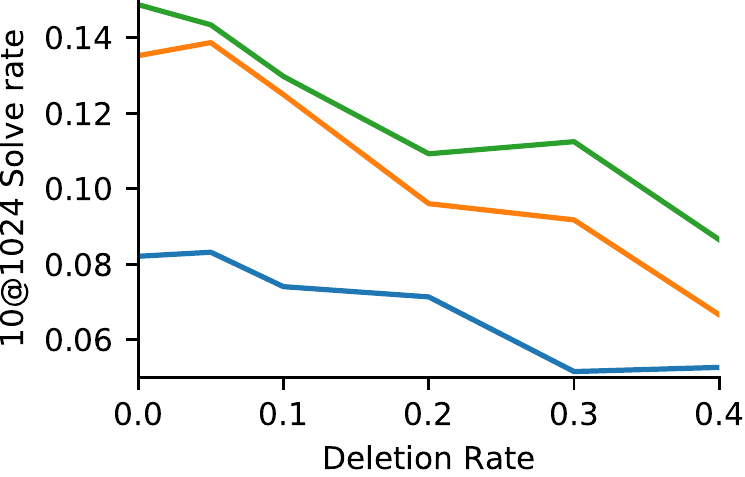} \\
    \end{tabular}
    \caption{\textbf{Solve rates under different word-level change regimes}.}
    \label{fig:word-level-changes}
\end{figure}

\textbf{Typing.~~}We analysed whether the model is sensitive to the implicit type information contained in problem descriptions. In particular, we replaced the words integer, array, and string with the more generic terms number, sequence, and sequence of characters, respectively. This process allows us to determine whether the model pays undue attention to specific output and input types to generate viable solutions. Overall, we observe no significant differences in solve rates with and without type information (\tabref{tab:type-removal}).

\begin{table}[ht]
\centering
\begin{tabular}{lrrr}
\toprule
 & 300M & 1B & 3B \\
\midrule
With type information & $8.23\%$ & $13.54\%$ & $14.89\%$\\
Without type information & $8.27\%$ & $13.30\%$ & $14.89\%$\\
\bottomrule
\end{tabular}
\caption{\textbf{Solve rate sensitivity to type information}. 10@1024 solve rates with different model sizes, with and without type information.}
\label{tab:type-removal}
\end{table}

\textbf{Typos.~~}
Typographical errors are another type of word changes.  To measure the model performance in this setting, we introduced a number of typos to the description, where each typo is a swap of adjacent characters of a randomly chosen English word (to avoid modifying nonsensical strings that are relevant to the solution). For example, the word "middle" could become "midlde". The solve rate deteriorates roughly linearly with the number of introduced typos, as shown in \figref{fig:word-level-changes} (a).

\textbf{Synonym substitution.~~}
We also analysed sensitivity to synonym substitutions in the problem description and specification, using synonym pairs in \citet{huang-2019-achieving} relying on the PPDB database \citep{ganitkevitch-etal-2013-ppdb}. 
\figref{fig:word-level-changes} (b) shows model performance with different numbers of synonym changes. 
Overall, we observe very little degradation in solve rate as we increase the number of substitutions. %

\textbf{Word-level perturbations.~~}
We studied the impact of word-level perturbations as done in \citet{edunov-etal-2018-understanding}, specifically swapping words and deleting words in the problem description and specification.
Words were swapped by randomly permuting words no more than $N$ positions apart (\figref{fig:word-level-changes} (c)), and words were deleted with probability $p$ (\figref{fig:word-level-changes} (d)). 
With both permutations and deletions, we observe stronger word level noise has a negative impact on the model performance. However, the model is relatively robust for low levels of words deletion ($p=0.05$) and swapping ($N=2$).

\subsubsection{Description section ablations}

We performed an ablation study on the importance of the different parts of the problem description: task description, specification, and input/output examples.
In \tabref{tab:prompt_ablation}, we see that removing any of the three sections impacts performance. Removing the IO examples has the least impact, followed by the description, and then the specification.
This is as we might expect, as without the specification it is difficult to know how to parse the problem input.
We also studied the impact of reordering sections, however different permutations have only a small impact on the solve rates of the models.

\begin{table}[h]
\centering
\begin{tabular}{lrrr}
\toprule
Prompt & 300M & 1B & 3B \\
\midrule
Description + Specification + IO & $8.33$\%  & $13.75$\%  & $14.90$\% \\
\midrule
Description + IO + Specification & $8.20$\%  & $14.35$\%  & $14.22$\% \\
Specification + Description + IO & $7.74$\%  & $12.88$\%  & $14.62$\% \\
Specification + IO + Description & $8.06$\%  & $13.08$\%  & $12.45$\% \\
IO + Description + Specification & $7.51$\%  & $12.94$\%  & $13.76$\% \\
IO + Specification + Description & $7.48$\%  & $13.41$\%  & $13.08$\% \\
\midrule
Description + Specification & $5.92$\%  & $10.42$\%  & $11.75$\% \\
Description + IO & $1.70$\%  & $4.81$\%  & $4.98$\% \\
Specification + IO & $5.43$\%  & $6.87$\%  & $7.95$\% \\
\bottomrule
\end{tabular}
\caption{\textbf{Description section ablations}. We report 10@1024 solve rates for different model sizes and different description ablations. `Description + Specification + IO' is the original prompt, the middle rows are different orderings, and the bottom rows remove sections. %
}
\label{tab:prompt_ablation}
\end{table}

\subsection{Sensitivity to problem metadata}\label{sec:sensitivity-to-metadata-appendix}
\subsubsection{Problem ratings}

\begin{figure}[h]
    \centering
    \includegraphics{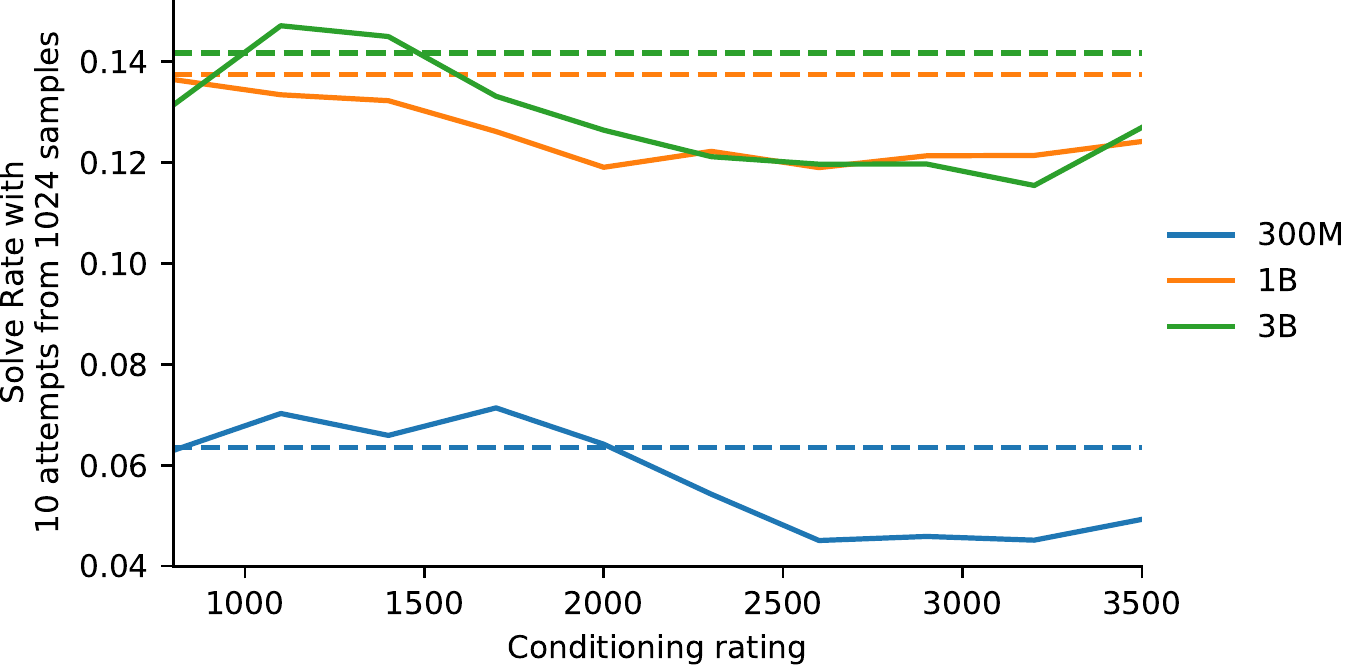}
    \caption{\textbf{10@1024 solve rates for samples conditioned on different ratings}. Dotted lines shows solve rates when conditioning on a uniform random rating.}
    \label{fig:rating-overall-solve}
\end{figure}

Problems are often giving ratings that indicate how difficult they are, though ratings are typically not available during the competition.
We investigated how the ratings provided to the model change the samples it produces. \figref{fig:rating-overall-solve} plots the solve rate when conditioning on various ratings. Specifying an easier rating is generally better than a harder one (although there are concerns that this could lead to more algorithmically inefficient solutions), and specifying a uniform random rating is competitive with any fixed rating.

\begin{figure}[h]
    \centering
    \includegraphics[scale=0.45]{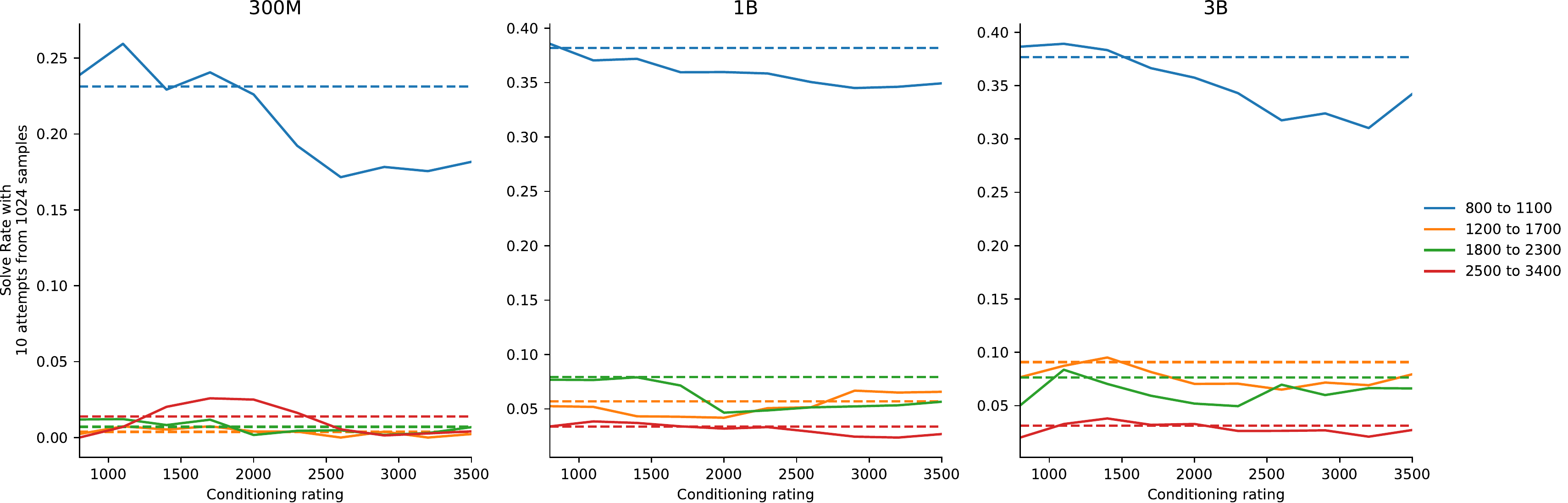}
    \caption{\textbf{10@1024 solve rates by problem difficulty for samples conditioned on different ratings}. Dotted lines shows solve rates when conditioning on a uniformly random rating.}
    \label{fig:rating-solve-rate-buckets}
\end{figure}

Next, we might expect that conditioning on a rating close to the true one could increase model solve rate. \figref{fig:rating-solve-rate-buckets} shows the solve rate for the four quartiles of problem difficulty in the validation set, conditioning on different ratings. This indicates that for more difficult problems it is relatively better to condition the model on a harder rating, while for the easier problems there is a larger negative impact.

\subsubsection{Solution correctness}
\begin{figure}[th]
    \centering
    \begin{tabular}{cc}
        \includegraphics[width=0.48\textwidth]{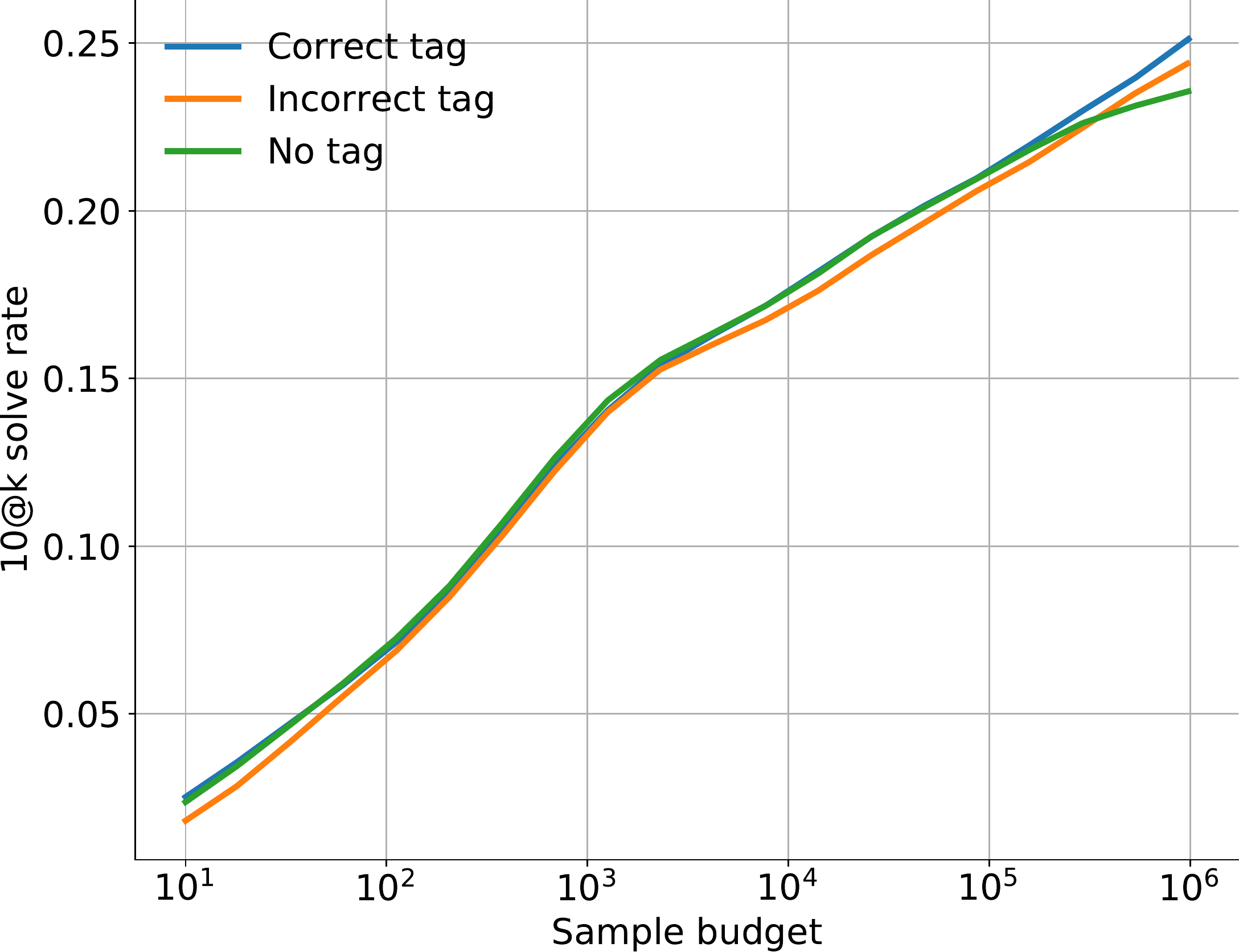} &
        \includegraphics[width=0.48\textwidth]{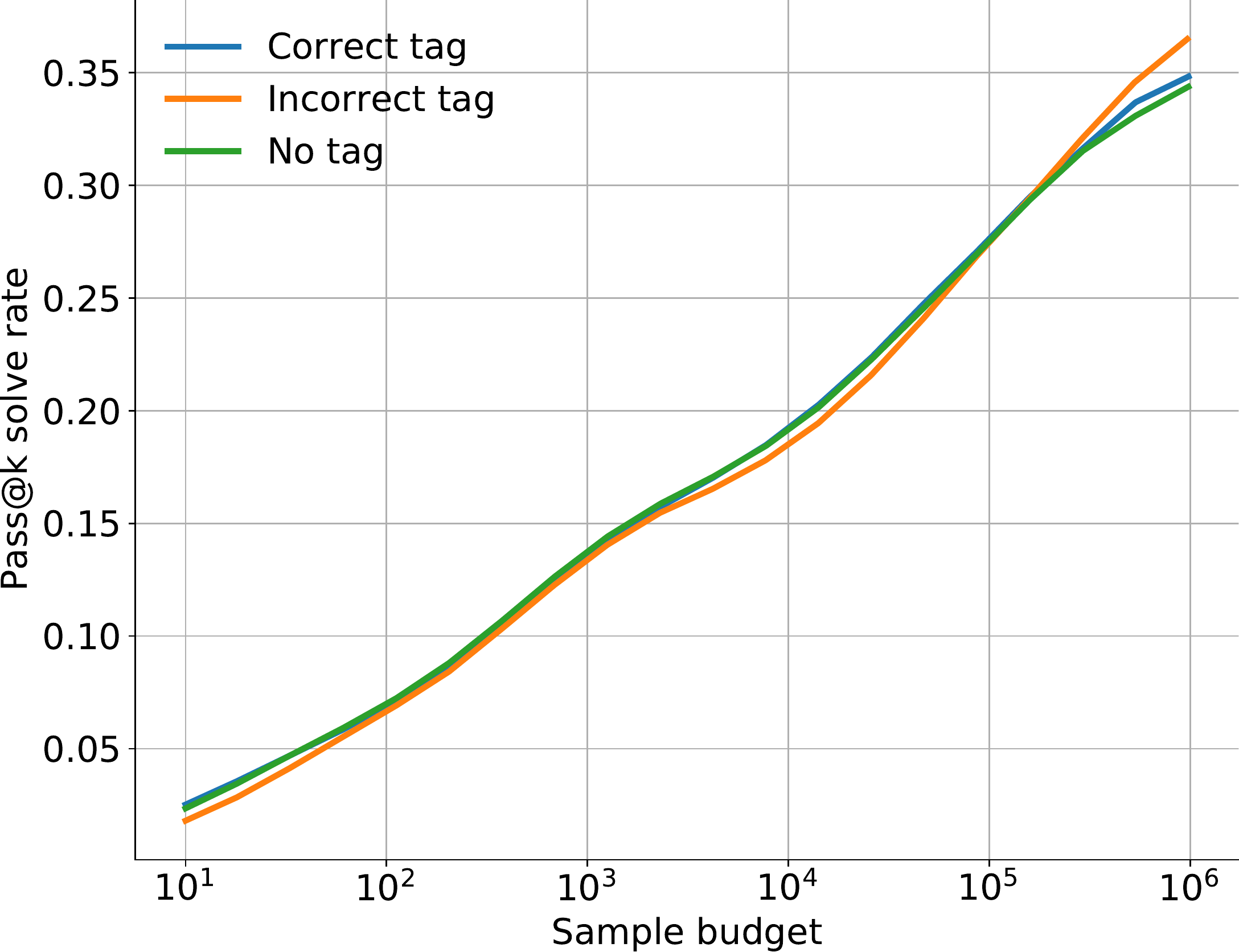} \\
        (a) 10 attempts per problem & (b) Unlimited attempts per problem
    \end{tabular}
    \caption{\textbf{Conditioning on the \texttt{CORRECT SOLUTION}, the \texttt{INCORRECT SOLUTION}, or no tag}.}
    \label{fig:incorrect-sampling}
\end{figure}

Value conditioning conditions the model with either a \texttt{CORRECT SOLUTION} or an \texttt{INCORRECT SOLUTION} tag at training time depending on the solution, and always with a \texttt{CORRECT SOLUTION} tag during sampling. \tabref{tab:fine-tuning-ablations} shows that it results in a performance improvement. Here, we investigate whether our models are strongly conditioned on this tag by supplying the  \texttt{INCORRECT SOLUTION} tag or even no tag at all, instead of the \texttt{CORRECT SOLUTION}.
The comparison is illustrated in \figref{fig:incorrect-sampling}.
Conditioning on the \texttt{INCORRECT SOLUTION} tag hurts in the $10@k$ metric, although not the $pass@k$ metric, indicating that the model may produce more solutions that pass example tests but not hidden tests.
Removing conditioning hurts in both metrics, although not significantly.

\clearpage

\section{Complete prompt and model examples}\label{sec:complete-examples}
All problems and human solutions in this section are sourced from \href{https://codeforces.com}{Codeforces}.
{
\lstset{
    inputencoding = utf8,  %
    extendedchars = true,  %
    literate      =        %
      {∈}{{$\in$}}1  {⋅}{{$\cdot$}}1  {≤}{{$\leq$}}1 {—}{{--}}1  {ú}{{\'u}}1
      {Á}{{\'A}}1  {É}{{\'E}}1  {Í}{{\'I}}1 {Ó}{{\'O}}1  {Ú}{{\'U}}1
      {à}{{\`a}}1  {è}{{\`e}}1  {ì}{{\`i}}1 {ò}{{\`o}}1  {ù}{{\`u}}1     
  }
\begin{figure}[h!]
\begin{center}
\begin{minipage}[t]{.63\textwidth}
Encoder Input $X$:
\begin{lstlisting}[language=C++,basicstyle=\fontsize{4.5}{5}\selectfont\ttfamily]
// RATING: 1200
// TAGS: math
// LANGUAGE IS cpp
// CORRECT SOLUTION
// n towns are arranged in a circle sequentially. The towns are numbered from 1
// to n in clockwise order. In the i-th town, there lives a singer with a
// repertoire of a_i minutes for each i ∈ [1, n].
// 
// Each singer visited all n towns in clockwise order, starting with the town he
// lives in, and gave exactly one concert in each town. In addition, in each
// town, the i-th singer got inspired and came up with a song that lasts a_i
// minutes. The song was added to his repertoire so that he could perform it in
// the rest of the cities.
// 
// Hence, for the i-th singer, the concert in the i-th town will last a_i
// minutes, in the (i + 1)-th town the concert will last 2 ⋅ a_i minutes, ...,
// in the ((i + k) mod n + 1)-th town the duration of the concert will be (k +
// 2) ⋅ a_i, ..., in the town ((i + n - 2) mod n + 1) — n ⋅ a_i minutes.
// 
// You are given an array of b integer numbers, where b_i is the total duration
// of concerts in the i-th town. Reconstruct any correct sequence of positive
// integers a or say that it is impossible.
// 
// Input
// 
// The first line contains one integer t (1 ≤ t ≤ 10^3) — the number of test
// cases. Then the test cases follow.
// 
// Each test case consists of two lines. The first line contains a single
// integer n (1 ≤ n ≤ 4 ⋅ 10^4) — the number of cities. The second line contains
// n integers b_1, b_2, ..., b_n (1 ≤ b_i ≤ 10^{9}) — the total duration of
// concerts in i-th city.
// 
// The sum of n over all test cases does not exceed 2 ⋅ 10^5.
// 
// Output
// 
// For each test case, print the answer as follows:
// 
// If there is no suitable sequence a, print NO. Otherwise, on the first line
// print YES, on the next line print the sequence a_1, a_2, ..., a_n of n
// integers, where a_i (1 ≤ a_i ≤ 10^{9}) is the initial duration of repertoire
// of the i-th singer. If there are multiple answers, print any of them.
// 
// Example
// 
// Input
// 
// 
// 4
// 3
// 12 16 14
// 1
// 1
// 3
// 1 2 3
// 6
// 81 75 75 93 93 87
// 
// 
// Output
// 
// 
// YES
// 3 1 3
// YES
// 1
// NO
// YES
// 5 5 4 1 4 5
// 
// Note
// 
// Let's consider the 1-st test case of the example:
// 
//   1. the 1-st singer in the 1-st city will give a concert for 3 minutes, in
// the 2-nd — for 6 minutes, in the 3-rd — for 9 minutes;
//   2. the 2-nd singer in the 1-st city will give a concert for 3 minutes, in
// the 2-nd — for 1 minute, in the 3-rd - for 2 minutes;
//   3. the 3-rd singer in the 1-st city will give a concert for 6 minutes, in
// the 2-nd — for 9 minutes, in the 3-rd — for 3 minutes.



\end{lstlisting}
\end{minipage}
\begin{minipage}[t]{.35\textwidth}
Decoder Output $Y$:
\begin{lstlisting}[language=C++,basicstyle=\tiny\selectfont\ttfamily]
#include <bits/stdc++.h>
using namespace std;
void solve() {
  long long int n, i, j;
  cin >> n;
  vector<long long int> a(n);
  for (i = 0; i < n; i++) cin >> a[i];
  long long int sum = 0;
  for (i = 0; i < n; i++) sum += a[i];
  if (sum %
    cout << "NO" << '\n';
    return;
  }
  long long int k = sum / (n * (n + 1) / 2);
  vector<long long int> ans(n);
  for (i = 0; i < n; i++) {
    long long int t = 0;
    if (i == 0)
      t = a[n - 1];
    else
      t = a[i - 1];
    long long int p = k - (a[i] - t);
    if (p %
      cout << "NO" << '\n';
      return;
    }
    ans[i] = p / n;
    if (ans[i] <= 0) {
      cout << "NO" << '\n';
      return;
    }
  }
  cout << "YES" << '\n';
  for (i = 0; i < n; i++) cout << ans[i] << " ";
  cout << '\n';
}
int main() {
  ios_base::sync_with_stdio(false);
  cin.tie(NULL);
  long long int t;
  cin >> t;
  while (t--) {
    solve();
  }
}
\end{lstlisting}
\end{minipage}
\vspace{-0.4cm}
\caption{\textbf{Complete model \protect\CC{} sample}. The tags, rating, and language are sampled randomly per problem at test time. See \url{https://alphacode.deepmind.com/} for more examples.
}
\label{fig:example-sample-cpp}
\end{center}
\end{figure}

\begin{figure}[h!]
\begin{center}
\begin{minipage}[t]{0.63\textwidth}
Encoder Input $X$:
\begin{lstlisting}[language=Python,basicstyle=\fontsize{4.5}{5}\selectfont\ttfamily]
# RATING: 3100
# TAGS: binary search,math
# LANGUAGE IS python3
# CORRECT SOLUTION
# n towns are arranged in a circle sequentially. The towns are numbered from 1
# to n in clockwise order. In the i-th town, there lives a singer with a
# repertoire of a_i minutes for each i ∈ [1, n].
# 
# Each singer visited all n towns in clockwise order, starting with the town he
# lives in, and gave exactly one concert in each town. In addition, in each
# town, the i-th singer got inspired and came up with a song that lasts a_i
# minutes. The song was added to his repertoire so that he could perform it in
# the rest of the cities.
# 
# Hence, for the i-th singer, the concert in the i-th town will last a_i
# minutes, in the (i + 1)-th town the concert will last 2 ⋅ a_i minutes, ..., in
# the ((i + k) mod n + 1)-th town the duration of the concert will be (k + 2) ⋅
# a_i, ..., in the town ((i + n - 2) mod n + 1) — n ⋅ a_i minutes.
# 
# You are given an array of b integer numbers, where b_i is the total duration
# of concerts in the i-th town. Reconstruct any correct sequence of positive
# integers a or say that it is impossible.
# 
# Input
# 
# The first line contains one integer t (1 ≤ t ≤ 10^3) — the number of test
# cases. Then the test cases follow.
# 
# Each test case consists of two lines. The first line contains a single integer
# n (1 ≤ n ≤ 4 ⋅ 10^4) — the number of cities. The second line contains n
# integers b_1, b_2, ..., b_n (1 ≤ b_i ≤ 10^{9}) — the total duration of
# concerts in i-th city.
# 
# The sum of n over all test cases does not exceed 2 ⋅ 10^5.
# 
# Output
# 
# For each test case, print the answer as follows:
# 
# If there is no suitable sequence a, print NO. Otherwise, on the first line
# print YES, on the next line print the sequence a_1, a_2, ..., a_n of n
# integers, where a_i (1 ≤ a_i ≤ 10^{9}) is the initial duration of repertoire
# of the i-th singer. If there are multiple answers, print any of them.
# 
# Example
# 
# Input
# 
# 
# 4
# 3
# 12 16 14
# 1
# 1
# 3
# 1 2 3
# 6
# 81 75 75 93 93 87
# 
# 
# Output
# 
# 
# YES
# 3 1 3
# YES
# 1
# NO
# YES
# 5 5 4 1 4 5
# 
# Note
# 
# Let's consider the 1-st test case of the example:
# 
#   1. the 1-st singer in the 1-st city will give a concert for 3 minutes, in
# the 2-nd — for 6 minutes, in the 3-rd — for 9 minutes;
#   2. the 2-nd singer in the 1-st city will give a concert for 3 minutes, in
# the 2-nd — for 1 minute, in the 3-rd - for 2 minutes;
#   3. the 3-rd singer in the 1-st city will give a concert for 6 minutes, in
# the 2-nd — for 9 minutes, in the 3-rd — for 3 minutes.



\end{lstlisting}
\end{minipage}
\begin{minipage}[t]{0.35\textwidth}
Decoder Output $Y$:
\begin{lstlisting}[language=Python,basicstyle=\tiny\selectfont\ttfamily]
import sys
input = sys.stdin.readline

t=int(input())
for tests in range(t):
    n=int(input())
    B=list(map(int,input().split()))

    A=[-1<<100]*n
    MIN=float("inf")
    MAX=-1

    for i,b in enumerate(B):
        MIN=min(MIN,b)
        MAX=max(MAX,b)

    SUM=sum(B)

    if SUM%
        print("NO")
        continue

    SUM//=(n*(n+1)//2)

    for i in range(n):
        j=i-1
        if j==-1:
            j=n-1
        a=(SUM-B[i]+B[j])//n
        if a>MIN or a>MAX or a<1:
            print("NO")
            break
        A[i]=a
    else:
        print("YES")
        print(*A)
\end{lstlisting}
\end{minipage}
\vspace{-0.4cm}
\caption{\textbf{Complete model Python sample}. The tags, rating, and language are sampled randomly per problem at test time. See \url{https://alphacode.deepmind.com/} for more examples.}
\label{fig:example-sample-py}
\end{center}
\end{figure}
}
\clearpage

\clearpage
\subsection{Solution duplication} \label{appendix:verbatim-copying}

\subsubsection{Solution decompositions}
\begin{figure}[h]
\begin{center}

\begin{tabular}{ll}
    Human solution & Model solution \\
    \\
    \raisebox{3.6cm}{
    \includegraphics[width=0.45\textwidth]{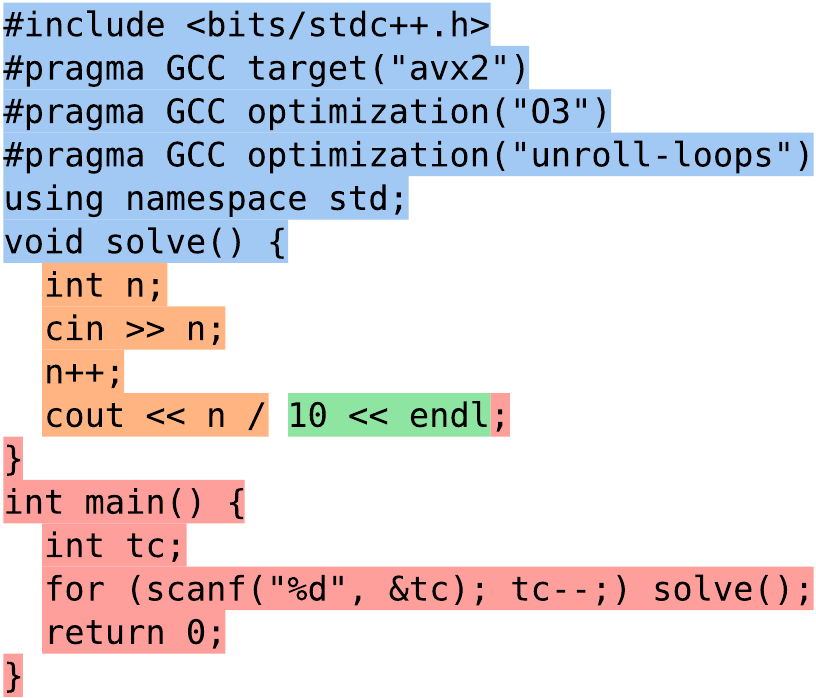}} & 
    \includegraphics[width=0.3\textwidth]{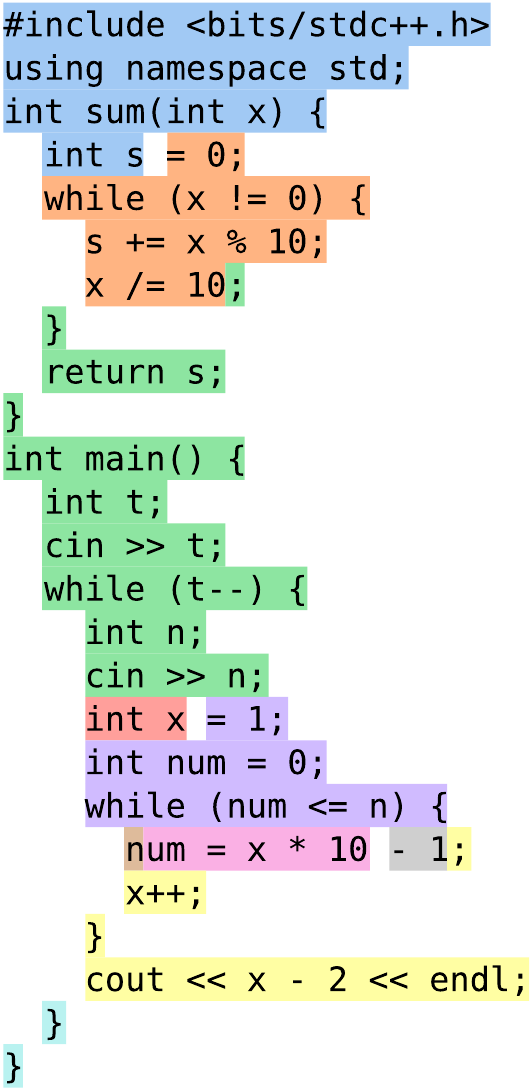}
\end{tabular}

\caption{\textbf{Example decomposition of human and model solutions to the `Digits Sum' problem into substrings from the finetuning dataset}. Each color identifies one substring, but repetition of any color is not meaningful, nor is there a relationship between substrings in the human and model solutions with the same color.}
\label{fig:lcs_decomposition_digits_sum_cpp}
\end{center}
\end{figure}
\begin{figure}[h]
\begin{center}

\begin{tabular}[t]{ll}
    Human solution & Model solution  \\
    \\
    \includegraphics[width=0.4\textwidth]{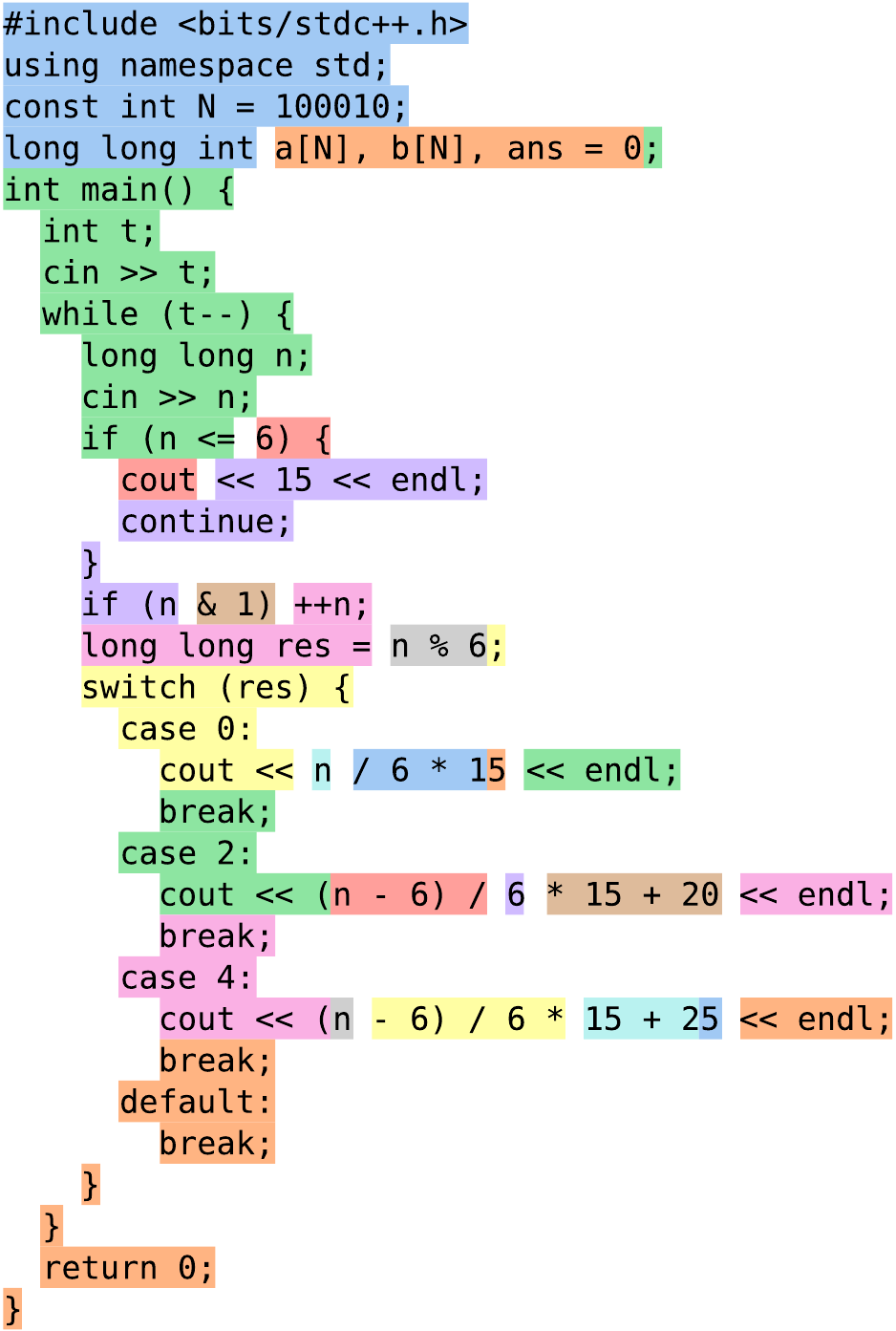} &
    \raisebox{1.7cm}{
    \includegraphics[width=0.5\textwidth]{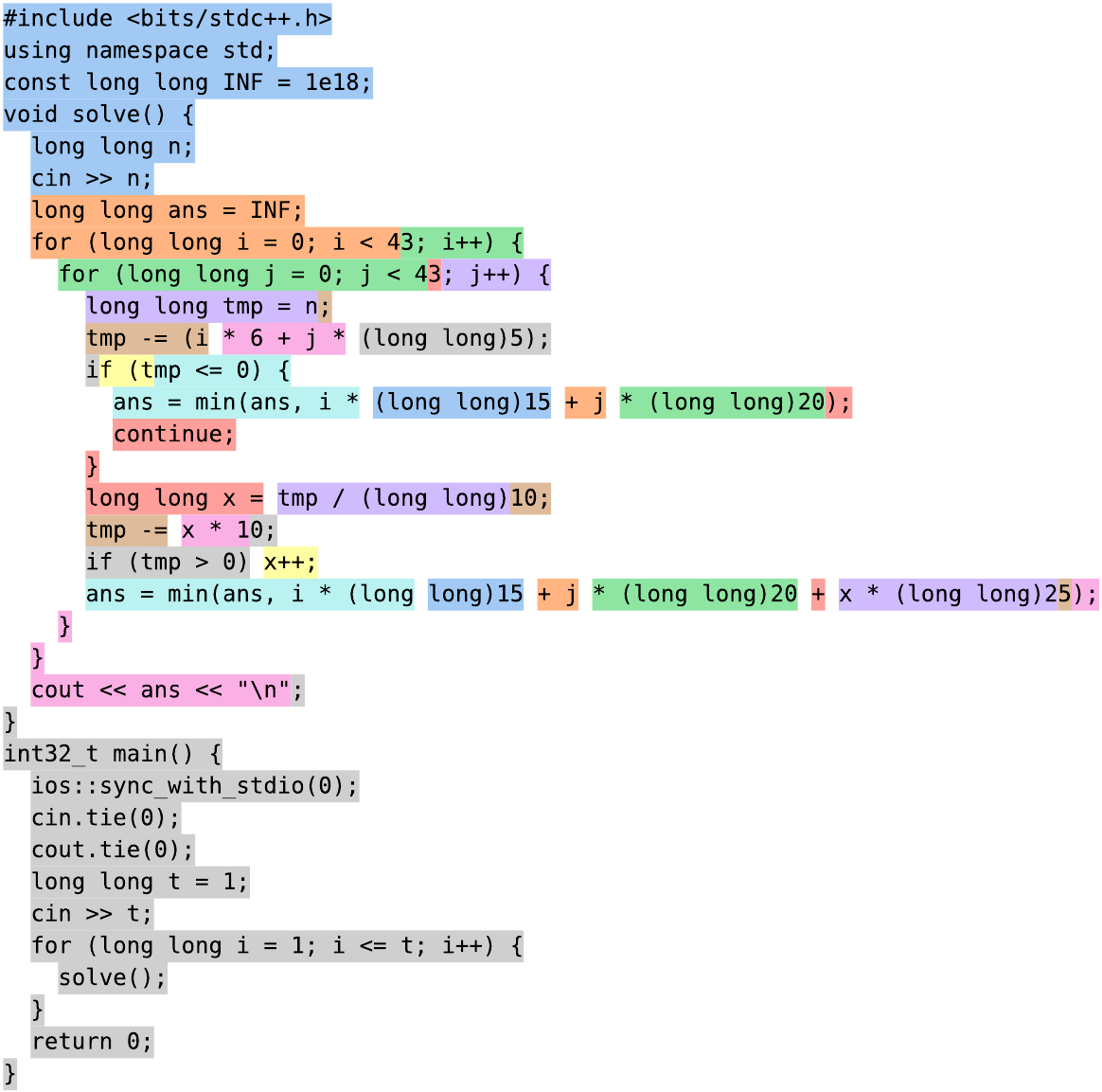}}
\end{tabular}

\caption{\textbf{Example decomposition of human and model solutions to the `Pizzaforces' problem into substrings from the finetuning dataset}. Each color identifies one substring, but repetition of any color is not meaningful, nor is there a relationship between substrings in the human and model solutions with the same color.}
\label{fig:lcs_decomposition_pizzaforces_cpp}
\end{center}
\end{figure}
\begin{figure}[h]
\begin{center}

\begin{tabular}[t]{ll}
    Human solution & Model solution \\
    \\
    \raisebox{2cm}{
    \includegraphics[width=0.4\textwidth]{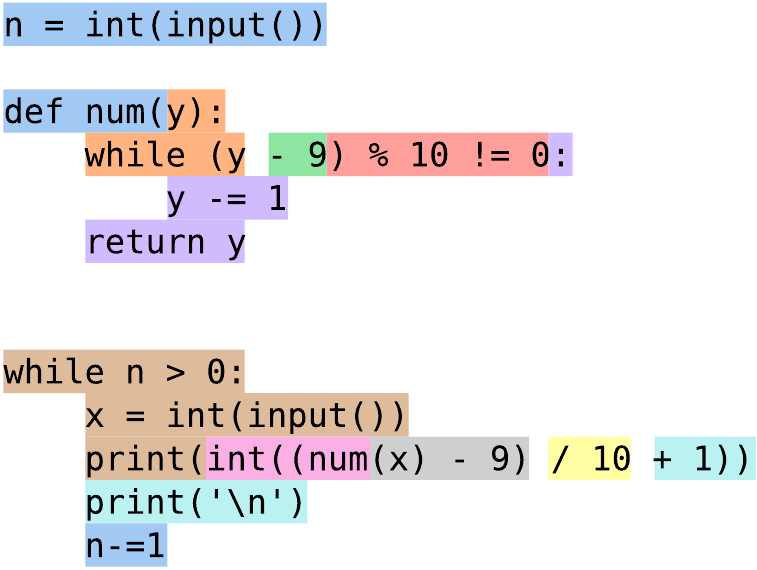}} & 
    \includegraphics[width=0.5\textwidth]{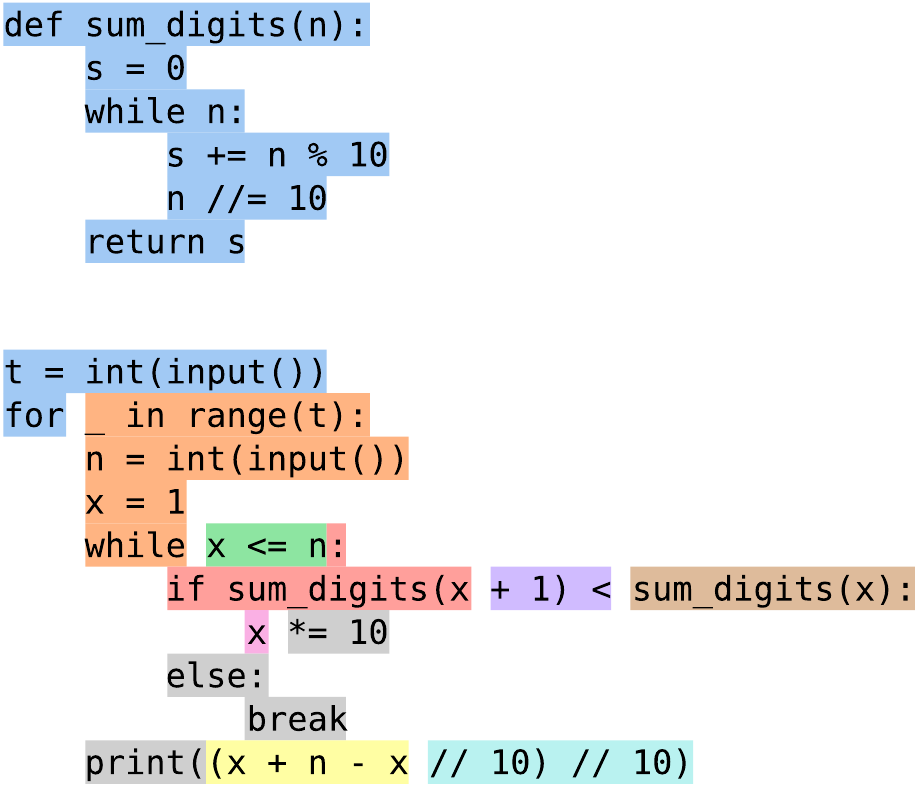}
\end{tabular}

\caption{\textbf{Example decomposition of human and model solutions to the `Digits Sum' problem into substrings from the finetuning dataset}. Each color identifies one substring, but repetition of any color is not meaningful, nor is there a relationship between substrings in the human and model solutions with the same color.}
\label{fig:lcs_decomposition_digits_sum_py}
\end{center}
\end{figure}
\begin{figure}[h]
\begin{center}

\begin{tabular}[t]{ll}
    Human solution  & Model solution \\
    \\
    \raisebox{5.6cm}{
    \includegraphics[width=0.3\textwidth]{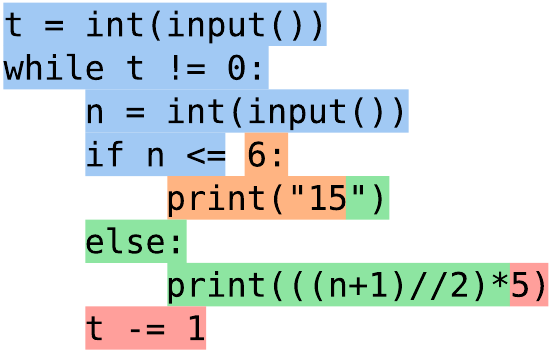}} & 
    \includegraphics[width=0.5\textwidth]{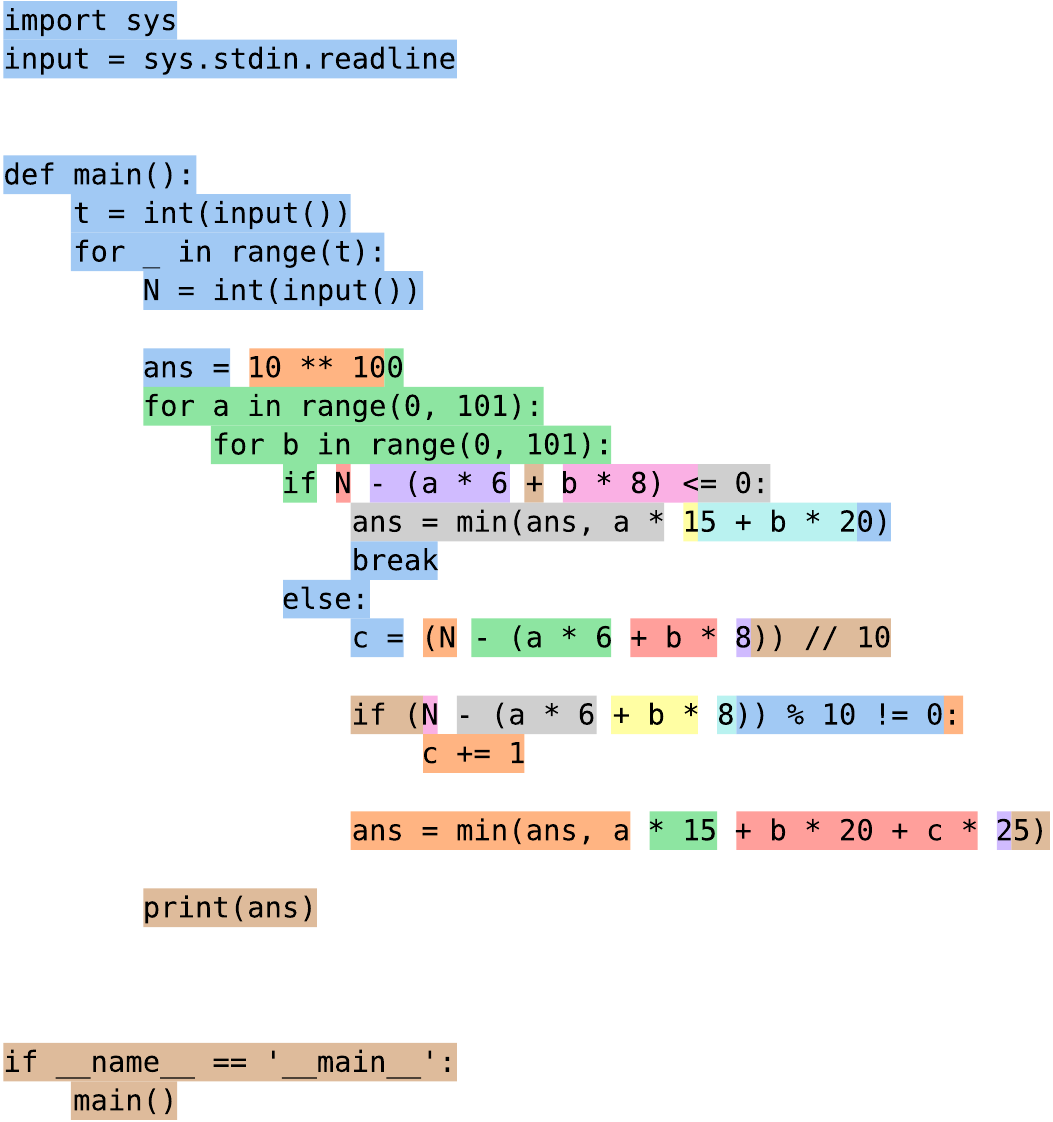}
\end{tabular}

\caption{\textbf{Example decomposition of human and model solutions to the `Pizzaforces' problem into substrings from the finetuning dataset}. Each color identifies one substring, but repetition of any color is not meaningful, nor is there a relationship between substrings in the human and model solutions with the same color.}
\label{fig:lcs_decomposition_pizzaforces_py}
\end{center}
\end{figure}
\clearpage

\subsubsection{Very long common subsequences between human solutions and finetuning data}
\begin{figure}[H]
\begin{center}
\includegraphics[width=0.6\textwidth]{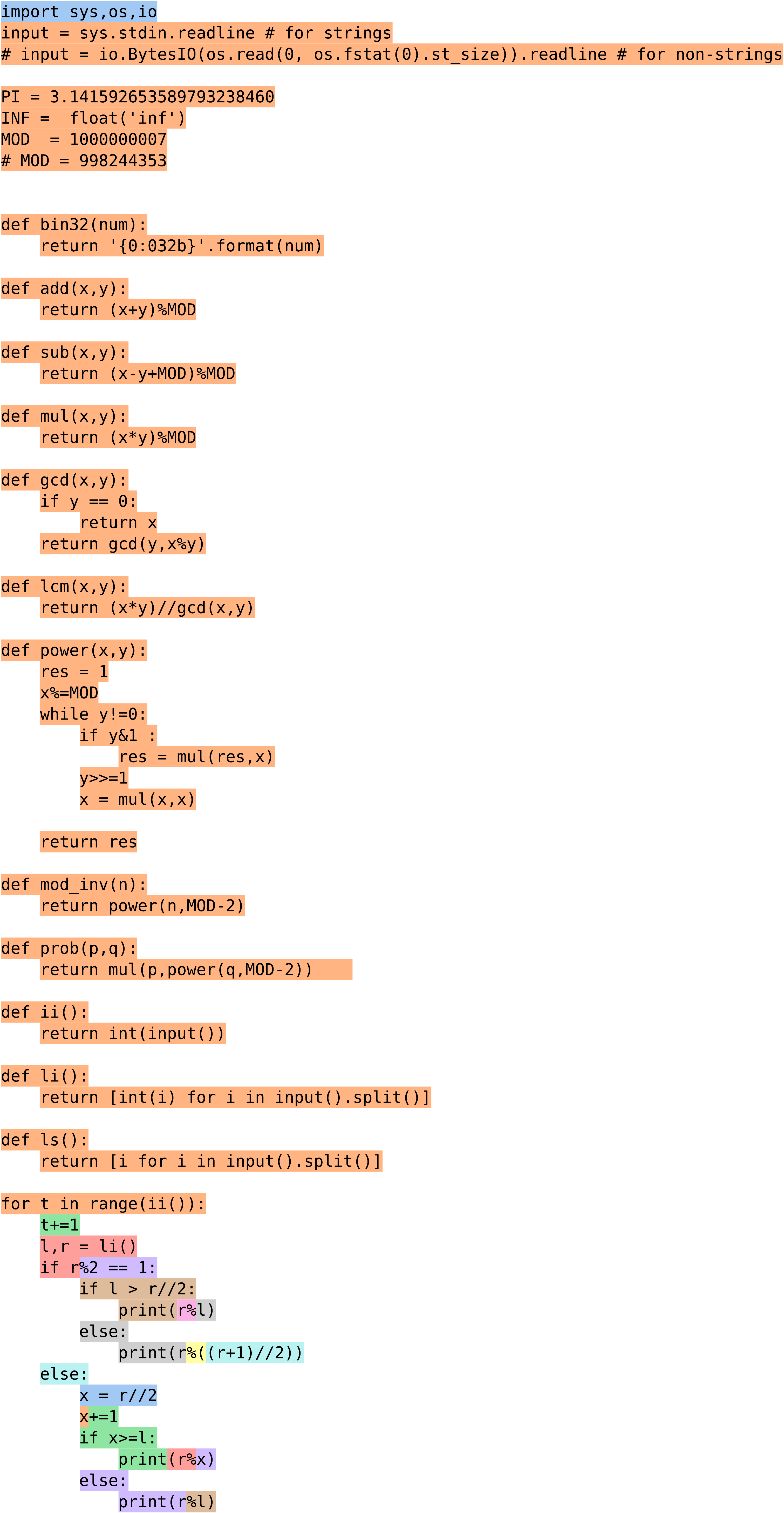} 
\caption{\textbf{A human validation solution to the `The Miracle and the Sleeper' problem with a very long LCS with the finetuning dataset (length 914)}. The remaining part of the solution is composed of much smaller substrings. Each color identifies one substring, but repetition of any color is not meaningful.}
\label{fig:human_long_lcs_py}
\end{center}
\end{figure}

\begin{figure}[h]
\begin{center}
\includegraphics[width=0.5\textwidth]{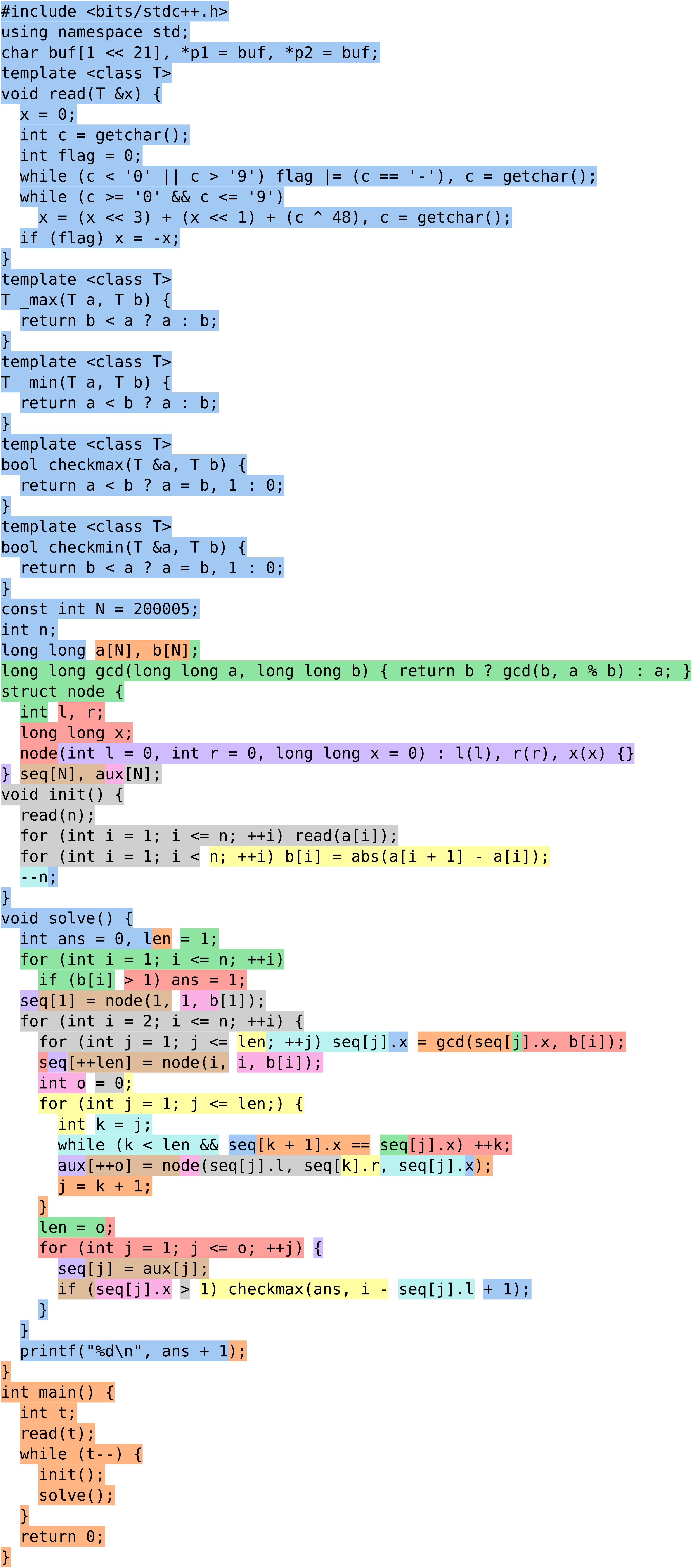} 
\caption{\textbf{A human validation solution to the 'Integers Have Friends' problem with a very long LCS with the finetuning dataset (length 666)}. The remaining part of the solution is composed of much smaller substrings. Each color identifies one substring, but repetition of any color is not meaningful.}
\label{fig:human_long_lcs_cpp}
\end{center}
\end{figure}

\clearpage

\subsection{Problem description rewordings}
\subsubsection{Simplified rewordings} \label{sec:simplified-rewordings}
\begin{center}
\begin{minipage}[t]{.48\textwidth}
Nim -- Original
\begin{lstlisting}[language={},basicstyle=\tiny\selectfont\ttfamily,breaklines=true,breakautoindent=false,breakindent=0pt]
Nim is a game in which 2 players take turns removing objects from heaps of different sizes. On each turn, a player must remove at least one object, and may remove any number of objects provided they all come from the same heap. The player to remove the last object is the winner.

Formally there are n heaps, with integer values a_1, ..., a_n. A turn consists of reducing the value of some a_i to a value between zero and a_i - 1.

Given the list of heap sizes you need to figure out which player wins if both play optimally.

Input

The first line contains a single integer t (1 <= t <= 10 000) - the number of test cases.

The first line of each test case contains a single integer n (2 <= n <= 10^5).

The second line of each test case contains n integers a_1, a_2, ..., a_n (1 <= a_i <= 10^6).

Output

If the first player wins print "1", otherwise print "2"

Example

Input

3
2
10 10
3
1 2 3
2
1 2

Output

2
2
1
\end{lstlisting}
\end{minipage}
\begin{minipage}[t]{.48\textwidth}
Nim -- Simplified
\begin{lstlisting}[language={},basicstyle=\tiny\selectfont\ttfamily,breaklines=true,breakautoindent=false,breakindent=0pt]
Given an array a, of length n, with values a_1, ..., a_n, compute the xor of all of the a_i.

If the xor is zero, output "2", else output "1".

Input

The first line contains a single integer t (1 $\leq$ t <= 10 000) - the number of test cases.

The first line of each test case contains a single integer n (2 <= n <= 10^5).

The second line of each test case contains n integers a_1, a_2, ..., a_n (1 <= a_i <= 10^6).

Output

If the xor of all the a_i is zero, print "2", otherwise print "1"

Example

Input

3
2
10 10
3
1 2 3
2
1 2

Output

2
2
1
\end{lstlisting}
\end{minipage}

\begin{minipage}[t]{.48\textwidth}
No Consecutive Zeros -- Original
\begin{lstlisting}[language={},basicstyle=\tiny\selectfont\ttfamily,breaklines=true,breakautoindent=false,breakindent=0pt]
Find the number of binary strings of length n that have no two consecutive zeros.

Consider all possible binary strings of length n. Many of these have two consecutive zeros, such as 101001. But some, such as 11010, do not. Find the number which do not have two consecutive zeros.

Input

The first line contains a single integer t (1 <= t <= 10 000) - the number of test cases.

The first line of each test case contains a single integer n (2 <= n <= 10^5).

Output

For each test case, print a single integer - the number of binary strings of length n which do not contain two consecutive zeros.

Example

Input

2
1
2

Output

2
3
\end{lstlisting}
\end{minipage}
\begin{minipage}[t]{.48\textwidth}
No Consecutive Zeros -- Simplified
\begin{lstlisting}[language={},basicstyle=\tiny\selectfont\ttfamily,breaklines=true,breakautoindent=false,breakindent=0pt]
Given an integer n, find the (n+2)th fibonacci number.

Consider the 0th fibonacci number to be 0, and the 1st fibonacci number to be 1.

Input

The first line contains a single integer t (1 <= t <= 10 000) - the number of test cases.

The first line of each test case contains a single integer n (2 <= n <= 10^5).

Output

For each test case, print a single integer - the (n+2)th fibonacci number.

Example

Input

2
1
2

Output

2
3
\end{lstlisting}
\end{minipage}

\begin{minipage}[t]{.48\textwidth}
1554A Cherry -- Original
\begin{lstlisting}[language={},basicstyle=\tiny\selectfont\ttfamily,breaklines=true,breakautoindent=false,breakindent=0pt]
You are given n integers a_1, a_2, ..., a_n. Find the maximum value of max(a_l, a_{l + 1}, ..., a_r) . min(a_l, a_{l + 1}, ..., a_r) over all pairs (l, r) of integers for which 1 <= l < r <= n.

Input

The first line contains a single integer t (1 <= t <= 10 000) - the number of test cases.

The first line of each test case contains a single integer n (2 <= n <= 10^5).

The second line of each test case contains n integers a_1, a_2, ..., a_n (1 <= a_i <= 10^6).

It is guaranteed that the sum of n over all test cases doesn't exceed 3 . 10^5.

Output

For each test case, print a single integer - the maximum possible value of the product from the statement.

Example

Input


4
3
2 4 3
4
3 2 3 1
2
69 69
6
719313 273225 402638 473783 804745 323328


Output


12
6
4761
381274500335

Note

Let f(l, r) = max(a_l, a_{l + 1}, ..., a_r) . min(a_l, a_{l + 1}, ..., a_r).

In the first test case, 

  * f(1, 2) = max(a_1, a_2) . min(a_1, a_2) = max(2, 4) . min(2, 4) = 4 . 2 = 8. 
  * f(1, 3) = max(a_1, a_2, a_3) . min(a_1, a_2, a_3) = max(2, 4, 3) . min(2, 4, 3) = 4 . 2 = 8. 
  * f(2, 3) = max(a_2, a_3) . min(a_2, a_3) = max(4, 3) . min(4, 3) = 4 . 3 = 12. 
\end{lstlisting}
\end{minipage}
\begin{minipage}[t]{.48\textwidth}
1554A Cherry -- Simplified
\begin{lstlisting}[language={},basicstyle=\tiny\selectfont\ttfamily,breaklines=true,breakautoindent=false,breakindent=0pt]
You are given n integers a_1, a_2, ..., a_n. Find the maximum value of a_l times a_{l + 1} for an integer l for which 1 <= l < n.

Input

The first line contains a single integer t (1 <= t <= 10 000) - the number of test cases.

The first line of each test case contains a single integer n (2 <= n <= 10^5).

The second line of each test case contains n integers a_1, a_2, ..., a_n (1 <= a_i <= 10^6).

It is guaranteed that the sum of n over all test cases doesn't exceed 3 . 10^5.

Output

For each test case, print a single integer - the maximum possible value of the product from the statement.

Example

Input


4
3
2 4 3
4
3 2 3 1
2
69 69
6
719313 273225 402638 473783 804745 323328


Output


12
6
4761
381274500335
\end{lstlisting}
\end{minipage}

\begin{minipage}[t]{.48\textwidth}
1559A Mocha and Math -- Original
\begin{lstlisting}[language={},basicstyle=\tiny\selectfont\ttfamily,breaklines=true,breakautoindent=false,breakindent=0pt]
Mocha is a young girl from high school. She has learned so much interesting knowledge from her teachers, especially her math teacher. Recently, Mocha is learning about binary system and very interested in bitwise operation.

This day, Mocha got a sequence a of length n. In each operation, she can select an arbitrary interval [l, r] and for all values i (0<= i <= r-l), replace a_{l+i} with a_{l+i}  \&  a_{r-i} at the same time, where \& denotes the [bitwise AND operation](https://en.wikipedia.org/wiki/Bitwise_operation#AND). This operation can be performed any number of times.

For example, if n=5, the array is [a_1,a_2,a_3,a_4,a_5], and Mocha selects the interval [2,5], then the new array is [a_1,a_2 \&  a_5, a_3 \&  a_4, a_4 \&  a_3, a_5 \&  a_2].

Now Mocha wants to minimize the maximum value in the sequence. As her best friend, can you help her to get the answer?

Input

Each test contains multiple test cases. 

The first line contains a single integer t (1 <= t <= 100) - the number of test cases. Each test case consists of two lines.

The first line of each test case contains a single integer n (1 <= n <= 100) - the length of the sequence.

The second line of each test case contains n integers a_1, a_2, ..., a_n (0 <= a_i <= 10^9).

Output

For each test case, print one integer - the minimal value of the maximum value in the sequence.

Example

Input


4
2
1 2
3
1 1 3
4
3 11 3 7
5
11 7 15 3 7


Output


0
1
3
3

Note

In the first test case, Mocha can choose the interval [1,2], then the sequence becomes [ 0, 0], where the first element is 1 \& 2, and the second element is 2 \& 1.

In the second test case, Mocha can choose the interval [1,3], then the sequence becomes [ 1,1,1], where the first element is 1 \& 3, the second element is 1 \& 1, and the third element is 3 \& 1.
\end{lstlisting}
\end{minipage}
\begin{minipage}[t]{.48\textwidth}
1559A Mocha and Math -- Simplified
\begin{lstlisting}[language={},basicstyle=\tiny\selectfont\ttfamily,breaklines=true,breakautoindent=false,breakindent=0pt]
Given a sequence of integers, compute the bitwise AND of all of its elements.

Input

Each test contains multiple test cases. 

The first line contains a single integer t (1 <= t <= 100) - the number of test cases. Each test case consists of two lines.

The first line of each test case contains a single integer n (1 <= n <= 100) - the length of the sequence.

The second line of each test case contains n integers a_1, a_2, ..., a_n (0 <= a_i <= 10^9).

Output

For each test case, print one integer - the bitwise AND of all elements of a.


Example

Input


4
2
1 2
3
1 1 3
4
3 11 3 7
5
11 7 15 3 7


Output


0
1
3
3

Note

In the first test case, Mocha can choose the interval [1,2], then the sequence becomes [ 0, 0], where the first element is 1 \& 2, and the second element is 2 \& 1.

In the second test case, Mocha can choose the interval [1,3], then the sequence becomes [ 1,1,1], where the first element is 1 \& 3, the second element is 1 \& 1, and the third element is 3 \& 1.
\end{lstlisting}
\end{minipage}

\begin{minipage}[t]{.48\textwidth}
1569A Balanced Substring -- Original
\begin{lstlisting}[language={},basicstyle=\tiny\selectfont\ttfamily,breaklines=true,breakautoindent=false,breakindent=0pt]
You are given a string s, consisting of n letters, each letter is either 'a' or 'b'. The letters in the string are numbered from 1 to n.

s[l; r] is a continuous substring of letters from index l to r of the string inclusive. 

A string is called balanced if the number of letters 'a' in it is equal to the number of letters 'b'. For example, strings "baba" and "aabbab" are balanced and strings "aaab" and "b" are not.

Find any non-empty balanced substring s[l; r] of string s. Print its l and r (1 <= l <= r <= n). If there is no such substring, then print -1 -1.

Input

The first line contains a single integer t (1 <= t <= 1000) - the number of testcases.

Then the descriptions of t testcases follow.

The first line of the testcase contains a single integer n (1 <= n <= 50) - the length of the string.

The second line of the testcase contains a string s, consisting of n letters, each letter is either 'a' or 'b'.

Output

For each testcase print two integers. If there exists a non-empty balanced substring s[l; r], then print l r (1 <= l <= r <= n). Otherwise, print -1 -1.

Example

Input


4
1
a
6
abbaba
6
abbaba
9
babbabbaa


Output


-1 -1
1 6
3 6
2 5

Note

In the first testcase there are no non-empty balanced subtrings.

In the second and third testcases there are multiple balanced substrings, including the entire string "abbaba" and substring "baba".
\end{lstlisting}
\end{minipage}
\begin{minipage}[t]{.48\textwidth}
1569A Balanced Substring -- Simplified
\begin{lstlisting}[language={},basicstyle=\tiny\selectfont\ttfamily,breaklines=true,breakautoindent=false,breakindent=0pt]
You are given a string s, consisting of n letters, each letter is either 'a' or 'b'. The letters in the string are numbered from 1 to n.

Find two adjacent letters which are not equal and print their indexes. If there is no such pair, print -1, -1.

Input

The first line contains a single integer t (1 <= t <= 1000) - the number of testcases.

Then the descriptions of t testcases follow.

The first line of the testcase contains a single integer n (1 <= n <= 50) - the length of the string.

The second line of the testcase contains a string s, consisting of n letters, each letter is either 'a' or 'b'.

Output

For each testcase print two integers. If there is an adjacent pair of non-identical letters at indexes l and r, print l, r. Otherwise, print -1 -1.

Example

Input


4
1
a
6
abbaba
6
abbaba
9
babbabbaa


Output


-1 -1
1 6
3 6
2 5

Note

In the first testcase there are no non-identical pairs.

In the second and third testcases there are non-identical pairs.
\end{lstlisting}
\end{minipage}
\end{center}

\clearpage

\subsubsection{Incorrect and verbose rewordings} \label{sec:incorrect-rewordings}
\begin{center}
\begin{minipage}[t]{.48\textwidth}
Original
\begin{lstlisting}[language={},basicstyle=\tiny\selectfont\ttfamily,breaklines=true,breakautoindent=false,breakindent=0pt]
You are given n integers a_1, a_2, ..., a_n. Find the maximum value of a_l times a_{l + 1} for an integer l for which 1 <= l < n.

Input

The first line contains a single integer t (1 <= t <= 10 000) -- the number of test cases.

The first line of each test case contains a single integer n (2 <= n <= 10^5).

The second line of each test case contains n integers a_1, a_2, ..., a_n (1 <= a_i <= 10^6).

It is guaranteed that the sum of n over all test cases doesn't exceed 3 . 10^5.

Output

For each test case, print a single integer -- the maximum possible value of the product from the statement.

Example

Input


4
3
2 4 3
4
3 2 3 1
2
69 69
6
719313 273225 402638 473783 804745 323328


Output


12
6
4761
381274500335

Note

Let f(l) =a_l . a_{l+1}

In the first test case, 

  * f(1) =  a_1 . a_2  = 2 . 4 = 8. 
  * f(2) = a_2 . a_3  = 4 . 3 = 12. 



So the maximum is f(2, 3) = 12.

In the second test case, the maximum is f(1) = f(2) = 6.
\end{lstlisting}
\end{minipage}
\begin{minipage}[t]{.48\textwidth}
Opposite
\begin{lstlisting}[language={},basicstyle=\tiny\selectfont\ttfamily,breaklines=true,breakautoindent=false,breakindent=0pt]
You are given n integers a_1, a_2, ..., a_n. Find the minimum value of a_l times a_{l + 1} for an integer l for which 1 <= l < n.

Input

The first line contains a single integer t (1 <= t <= 10 000) -- the number of test cases.

The first line of each test case contains a single integer n (2 <= n <= 10^5).

The second line of each test case contains n integers a_1, a_2, ..., a_n (1 <= a_i <= 10^6).

It is guaranteed that the sum of n over all test cases doesn't exceed 3 . 10^5.

Output

For each test case, print a single integer -- the minimum possible value of the product from the statement.

Example

Input


4
3
2 4 3
4
3 2 3 1
2
69 69
6
719313 273225 402638 473783 804745 323328


Output


8
3
4761
88341292800

Note

Let f(l) =a_l . a_{l+1}

In the first test case, 

  * f(1) =  a_1 . a_2  = 2 . 4 = 8. 
  * f(2) = a_2 . a_3  = 4 . 3 = 12. 



So the minimum is f(2, 3) = 8.

In the second test case, the minimum is f(3) = 3.
\end{lstlisting}
\end{minipage}

\begin{minipage}[t]{.48\textwidth}
Related
\begin{lstlisting}[language={},basicstyle=\tiny\selectfont\ttfamily,breaklines=true,breakautoindent=false,breakindent=0pt]
You are given n integers a_1, a_2, ..., a_n. Find the maximum value of (a_l . a_r) over all pairs (l, r) of integers for which 1 <= l < r <= n.

Input

The first line contains a single integer t (1 <= t <= 10 000) -- the number of test cases.

The first line of each test case contains a single integer n (2 <= n <= 10^5).

The second line of each test case contains n integers a_1, a_2, ..., a_n (1 <= a_i <= 10^6).

It is guaranteed that the sum of n over all test cases doesn't exceed 3 . 10^5.

Output

For each test case, print a single integer -- the maximum possible value of the product from the statement.

Example

Input


4
3
2 4 3
4
3 2 3 1
2
69 69
6
719313 273225 402638 473783 804745 323328


Output


12
6
4761
381274500335

Note

Let f(l, r) = (a_l . a_r).

In the first test case, 

  * f(1, 2) = 2 . 4 = 8. 
  * f(1, 3) = 2 . 3 = 8. 
  * f(2, 3) = 4 . 3 = 12. 



So the maximum is f(2, 3) = 12.

In the second test case, the maximum is f(1, 3) = 9.
\end{lstlisting}

\end{minipage}
\begin{minipage}[t]{.48\textwidth}
Underspecified
\begin{lstlisting}[language={},basicstyle=\tiny\selectfont\ttfamily,breaklines=true,breakautoindent=false,breakindent=0pt]
You are given n integers a_1, a_2, ..., a_n. Find the maximum value of f(a_l, a_r) over all pairs (l, r) of integers for which 1 <= l < r <= n.


Input

The first line contains a single integer t (1 <= t <= 10 000) -- the number of test cases.

The first line of each test case contains a single integer n (2 <= n <= 10^5).

The second line of each test case contains n integers a_1, a_2, ..., a_n (1 <= a_i <= 10^6).

It is guaranteed that the sum of n over all test cases doesn't exceed 3 . 10^5.

Output

For each test case, print a single integer -- the maximum value of the function from the statement.


Example

Input


4
3
2 4 3
4
3 2 3 1
2
69 69
6
719313 273225 402638 473783 804745 323328


Output


12
9
4761
578863540185
\end{lstlisting}
\end{minipage}

\begin{minipage}[t]{.48\textwidth}
Verbose
\begin{lstlisting}[language={},basicstyle=\tiny\selectfont\ttfamily,breaklines=true,breakautoindent=false,breakindent=0pt]
William has been given an array for his birthday, which consists of n integers a_1, a_2, ..., a_n. He is very proud of his array, but naturally his friend Mary is curious about it. Mary would like to know a certain function of consecutive elements of the array. Concretely, Mary would like to know the maximum value of a_l times a_{l + 1} for an integer l for which 1 <= l < n. Can you help William by calculating this value for him?

Input

The first line contains a single integer t (1 <= t <= 10 000) -- the number of test cases.

The first line of each test case contains a single integer n (2 <= n <= 10^5).

The second line of each test case contains n integers a_1, a_2, ..., a_n (1 <= a_i <= 10^6).

It is guaranteed that the sum of n over all test cases doesn't exceed 3 . 10^5.

Output

For each test case, print a single integer -- the maximum possible value of the product from the statement.

Example

Input


4
3
2 4 3
4
3 2 3 1
2
69 69
6
719313 273225 402638 473783 804745 323328


Output


12
6
4761
381274500335

Note

Let f(l) =a_l . a_{l+1}

In the first test case, 

  * f(1) =  a_1 . a_2  = 2 . 4 = 8. 
  * f(2) = a_2 . a_3  = 4 . 3 = 12. 



So the maximum is f(2, 3) = 12.

In the second test case, the maximum is f(1) = f(2) = 6.
\end{lstlisting}

\end{minipage}
\begin{minipage}[t]{.48\textwidth}
Algorithm described in words
\begin{lstlisting}[language={},basicstyle=\tiny\selectfont\ttfamily,breaklines=true,breakautoindent=false,breakindent=0pt]
You are given n integers a_1, a_2, ..., a_n. Find the maximum value of the product of two consecutive members of the array.



Input

The first line contains a single integer t (1 <= t <= 10 000) -- the number of test cases.

The first line of each test case contains a single integer n (2 <= n <= 10^5).

The second line of each test case contains n integers a_1, a_2, ..., a_n (1 <= a_i <= 10^6).

It is guaranteed that the sum of n over all test cases doesn't exceed 3 . 10^5.

Output

For each test case, print a single integer -- the maximum possible value of the product from the statement.

Example

Input


4
3
2 4 3
4
3 2 3 1
2
69 69
6
719313 273225 402638 473783 804745 323328


Output


12
6
4761
381274500335

Note

Let f(l) =a_l . a_{l+1}

In the first test case, 

  * f(1) =  a_1 . a_2  = 2 . 4 = 8. 
  * f(2) = a_2 . a_3  = 4 . 3 = 12. 



So the maximum is f(2, 3) = 12.

In the second test case, the maximum is f(1) = f(2) = 6.
\end{lstlisting}
\end{minipage}

\end{center}

\clearpage

\end{document}